\documentclass[fleqn,usenatbib]{mnras}

\usepackage{newtxtext,newtxmath}

\usepackage[T1]{fontenc}

\DeclareRobustCommand{\VAN}[3]{#2}
\let\VANthebibliography\thebibliography
\def\thebibliography{\DeclareRobustCommand{\VAN}[3]{##3}\VANthebibliography}

\usepackage{graphicx}	
\usepackage{amsmath}	

\newcommand{\oii}{[O~{\sc ii}]}
\newcommand{\oiii}{[O~{\sc iii}]}
\newcommand{\lya}{Ly$\alpha$}
\newcommand{\ha}{H$\alpha$}
\newcommand{\hb}{H$\beta$}
\newcommand{\nii}{[N~{\sc ii}]}

\title[$\xi_{ion,0}$ for 30 LAEs using JEMS and MUSE]{The ionising photon production efficiency at $z \sim 6$ for Lyman-alpha emitters using JEMS and MUSE}

\author[C. Simmonds et al.]{
C. Simmonds,$^{1,2}$\thanks{E-mail: cs2210@cam.ac.uk}
S. Tacchella,$^{1,2}$
M. Maseda,$^{3}$
C. C. Williams,$^{4,5}$
W. M. Baker,$^{1,2}$
C. E. C. Witten,$^{6,1}$
\newauthor
B. D. Johnson,$^{17}$
B. Robertson,$^{7}$
A. Saxena,$^{8,9}$
F. Sun,$^{5}$
J. Witstok,$^{1,2}$
R. Bhatawdekar,$^{10,11}$
K. Boyett,$^{12,13}$
\newauthor
A. J. Bunker,$^{8}$
S. Charlot,$^{14}$
E. Curtis-Lake,$^{15}$
E. Egami,$^{5}$
D. J. Eisenstein,$^{17}$
Z. Ji,$^{5}$
R. Maiolino,$^{1,2,9}$
\newauthor
L. Sandles,$^{1,2}$
R. Smit,$^{18}$
H. \"Ubler,$^{1,2}$
and C. J. Willott$^{16}$
\\
$^{1}$The Kavli Institute for Cosmology (KICC), University of Cambridge, Madingley Road, Cambridge, CB3 0HA\\
$^{2}$Cavendish Laboratory, University of Cambridge, 19 JJ Thomson Avenue, Cambridge, CB3 0HE, UK\\
$^{3}$Department of Astronomy, University of Wisconsin-Madison, 475 N. Charter Street, Madison, WI 53706, USA\\
$^{4}$NSF’s National Optical-Infrared Astronomy Research
Laboratory, 950 North Cherry Avenue, Tucson, AZ 85719, USA\\
$^{5}$Steward Observatory, University of Arizona, 933 North Cherry Avenue, Tucson, AZ 85721, USA\\
$^{6}$Institute of Astronomy, University of Cambridge, Madingley Road, Cambridge CB3 0HA, United Kingdom\\
$^{7}$Department of Astronomy and Astrophysics University of California, Santa Cruz, 1156 High Street, Santa Cruz CA 96054, USA\\ 
$^{8}$Department of Physics, University of Oxford, Denys Wilkinson Building, Keble Road, Oxford OX1 3RH, UK\\
$^{9}$Department of Physics and Astronomy, University College London, Gower Street, London WC1E 6BT, UK\\
$^{10}$European Space Agency (ESA), European Space Astronomy Centre (ESAC), Camino Bajo del Castillo s/n, 28692 Villanueva de la Cañada, Madrid, Spain\\
$^{11}$European Space Agency, ESA/ESTEC, Keplerlaan 1, 2201 AZ Noordwijk, NL
$^{12}$School of Physics, University of Melbourne, Parkville 3010, VIC, Australia\\
$^{13}$ARC Centre of Excellence for All Sky Astrophysics in 3 Dimensions (ASTRO 3D), Australia\\
$^{14}$Sorbonne Universit\'e, CNRS, UMR 7095, Institut d'Astrophysique de Paris, 98 bis bd Arago, 75014 Paris, France\\
$^{15}$Centre for Astrophysics Research, Department of Physics, Astronomy and Mathematics, University of Hertfordshire, Hatfield AL10 9AB, UK\\
$^{16}$NRC Herzberg, 5071 West Saanich Rd, Victoria, BC V9E 2E7, Canada\\
$^{17}$Center for Astrophysics $|$ Harvard \& Smithsonian, 60 Garden St., Cambridge MA 02138 USA\\
$^{18}$Astrophysics Research Institute, Liverpool John Moores University, 146 Brownlow Hill, Liverpool L3 5RF, UK
}

\begin{document}
\label{firstpage}
\pagerange{\pageref{firstpage}--\pageref{lastpage}}
\maketitle

\begin{abstract}
We study the ionising photon production efficiency at the end of the Epoch of Reionisation ($z \sim 5.4 - 6.6$) for a sample of 30 Lyman-$\alpha$ emitters. This is a crucial quantity to infer the ionising photon budget of the Universe. These objects were selected to have reliable spectroscopic redshifts, assigned based on the profile of their Lyman-$\alpha$ emission line, detected in the MUSE deep fields. We exploit medium-band observations from the JWST Extragalactic Medium-band Survey (JEMS) to find the flux excess corresponding to the redshifted \ha\ emission line. We estimate the ultra-violet (UV) luminosity by fitting the full JEMS photometry, along with several HST photometric points, with \texttt{Prospector}. We find a median UV continuum slope of $\beta = -2.09^{+0.23}_{-0.21}$ for the sample, indicating young stellar populations with little-to-no dust attenuation. Supported by this, we derive $\xi_{ion,0}$ with no dust attenuation and find a median value of log$\frac{\xi_{ion,0}}{\text{Hz erg}^{-1}} = 25.44^{+0.21}_{-0.15}$. If we perform dust attenuation corrections and assume a Calzetti attenuation law, our values are lowered by $\sim 0.1$ dex. Our results suggest Lyman-$\alpha$ emitters at the Epoch of Reionisation have slightly enhanced $\xi_{ion,0}$ compared to previous estimations from literature, in particular, when compared to the non-Lyman-$\alpha$ emitting population. This initial study provides a promising outlook on the characterisation of ionising photon production in the early Universe. In the future, a more extensive study will be performed on the entire dataset provided by the JWST Advanced Deep Extragalactic Survey (JADES). Thus, for the first time, allowing us to place constraints on the wider galaxy populations driving reionisation.
\end{abstract}

\begin{keywords}
Galaxies: high-redshift -- Galaxies: evolution -- Galaxies: general 
\end{keywords}



\section{Introduction}
The Epoch of Reionisation (EoR) is one of the major phase changes of the universe. It corresponds to the transition between a dark and neutral universe to an ionised one, where the intergalactic medium (IGM) became transparent to Lyman continuum (LyC) radiation. Observational evidence places the end of this epoch at redshift $z \sim 6$ \citep{Becker2001,Fan2006,Yang2020}. Understanding the sources responsible for ionising the IGM is one of the most significant, unsolved questions in modern-day astronomy. Young massive stars in galaxies are currently believed to be the main responsible culprit, producing copious amounts of hydrogen-ionising photons (E $\geq$ 13.6 eV) that escape the interstellar medium (ISM) to then ionise the IGM \citep{Hassan2018,Rosdahl2018,Trebitsch2020}. Until recently, average escape fractions ($f_{esc}$) of 10-20\% were believed to be a key threshold for galaxies to be the main responsible sources of reionisation  \citep{Ouchi2009,Robertson2013,Robertson2015,Finkelstein2019,Naidu2020}. 
High escape fractions have been seen in some LyC leaking galaxies \citep[e.g. ][]{Borthakur2014,Bian2017,Vanzella2018,Fletcher2019,Ji2020,Izotov2021}. These are galaxies from which we observe (or infer through indirect probes) LyC radiation. Understanding how LyC photons escape into the IGM and subsequently ionise it during the EoR is of utmost importance. Strong LyC leakers are not generally observed in large samples \citep{Leitet2013,Leitherer2016,Steidel2018,Flury2022a}. Based on population-wide studies, it is currently uncertain which types of galaxies dominate the budget of reionisation \citep{Finkelstein2019,Naidu2020}. Promisingly, studies targeting galaxies at high redshift (up to $z \sim 9$) have found that as we look further into the past of the universe, galaxies seem to be more efficient in producing ionising photons. This can be measured as the ratio between the production rate of ionising photons over the non-ionising luminosity density, called the ionising photon production efficiency ($\xi_{ion}$), which has been shown to increase with redshift \citep[e.g. ][]{Bouwens2016,Faisst2019,Endsley2021,Stefanon2022,Tang2023}. An increase of $\xi_{ion}$ implies that smaller $f_{esc}$ are required in galaxies to explain the reionisation of the universe.
   \begin{figure*}
   \centering
   \includegraphics[width=1\textwidth]{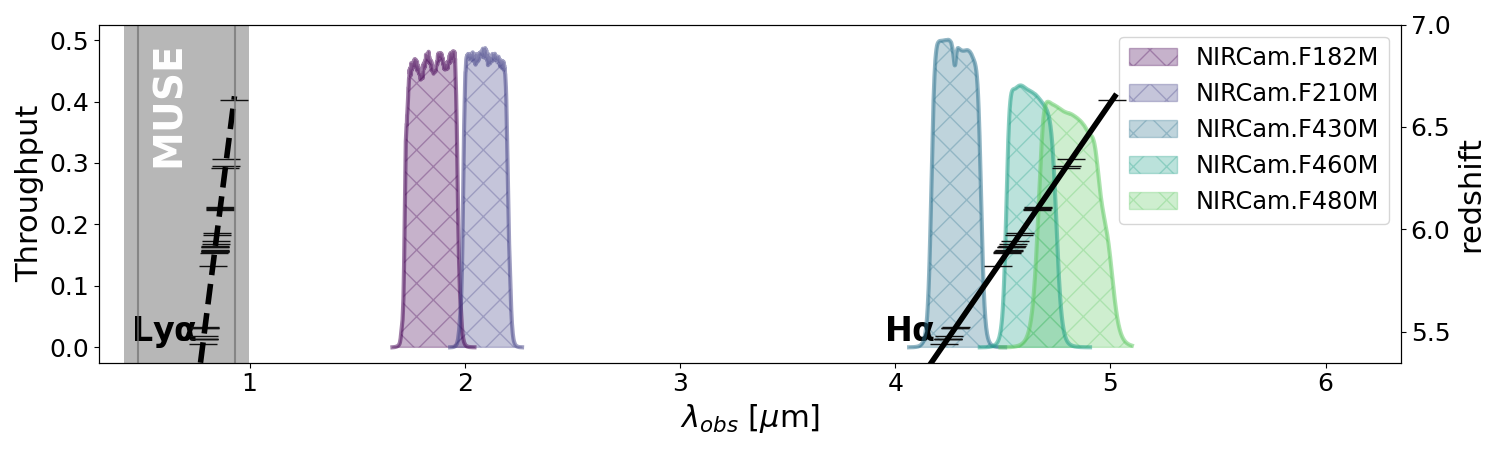}
   \caption{Horizontal lines show the systemic redshift of the 30 selected LAEs from the MUSE UDF DR2 \citep{Bacon2022}. They were selected because their redshift makes it possible to retrieve \ha\ information using JEMS. The throughput of each medium band is shown as filled hatched areas. The exposure time, effective wavelength and 5$\sigma$ sensitivity in AB magnitudes for each band is, from left to right: F182M ($t_{\rm{exp}} = 27,830$ s, $\lambda_{\rm{eff}} = 1.829\text{ }\mu$m, $m_{\rm{AB}}$ = 29.3), F210M ($t_{\rm{exp}} = 13,915$ s, $\lambda_{\rm{eff}} = 2.091\text{ }\mu$m, $m_{\rm{AB}}$ = 29.2), F430M ($t_{\rm{exp}} = 13,915$ s, $\lambda_{\rm{eff}} = 4.287\text{ }\mu$m, $m_{\rm{AB}}$ = 28.5), F460M ($t_{\rm{exp}} = 13,915$ s, $\lambda_{\rm{eff}} = 4.627\text{ }\mu$m, $m_{\rm{AB}}$ = 28.5), and F480M ($t_{\rm{exp}} = 27,830$ s, $\lambda_{\rm{eff}} = 4.814\text{ }\mu$m, $m_{\rm{AB}}$ = 28.6). The details are provided in \citet{Williams2023}.}
              \label{fig:filters}%
    \end{figure*}

Direct observations of LyC radiation are impossible at the EoR due to the increasing absorption by neutral hydrogen in the IGM along the line of sight. At $z \sim 6$ the average IGM transmission of photons with $\lambda_{rest-frame} = 900$ \AA\ is virtually zero \citep{Inoue2014}. However, hydrogen recombination lines can provide indirect evidence of ionising photons, since they are produced after photoionisation has taken place. The strongest of such lines is \lya\/, and due to its resonant nature, the shape of its profile is to date one of the most reliable indirect tracers of LyC leakage. In particular, when double peaks are detected, the separation between them can be related to the neutral hydrogen column density and consequently, to the escape of LyC photons \citep{Verhamme2015,Verhamme2017,Hogarth2020,Izotov2021,Naidu2021}. Moreover, LyC leakers tend to show strong \lya\ emission \citep{Nakajima2020}. These kind of objects have high \lya\ equivalent widths \citep[EW(\lya\/) > 20 \AA\/; ][]{Ajiki2003,Ouchi2005,Ouchi2008}, in some extreme cases reaching hundreds of angstroms \citep{Kerutt2022,Saxena2023}, and are appropriately called \lya\ emitters (LAEs) \citep[see ][and references within]{Ouchi2020}. We note that the \lya\/-LyC correlation has scatter, and that strong LyC leakage has also been observed in galaxies with relatively low \lya\ emission \citep[e.g.][]{Ji2020,Flury2022b}.

The second strongest hydrogen recombination line is \ha\/. Unlike \lya\/, it does not scatter resonantly under normal star-forming conditions. It is commonly used as a measure of star formation rate (SFR), because its luminosity is directly related to the amount of ionising photons that are scattered or absorbed by the medium, and thus, do not escape the ISM. If used in combination with a measure of the ultra-violet (UV) continuum, \ha\ can constrain $\xi_{ion}$ \citep[e.g.][]{Bouwens2016,Chisholm2022,Stefanon2022}. We note that $\xi_{ion}$ has a strong degeneracy with LyC escape fractions, thus, it is common to report the values obtained under the assumption of zero LyC escape ($\xi_{ion,0}$).

In this work we aim to estimate $\xi_{ion}$ for a sample of bright LAEs identified with the ground-based Multi Unit Spectroscopic Explorer \citep[MUSE; ][]{Bacon2010}, by measuring \ha\ and the UV continuum through observations obtained with the Near-Infrared Camera \citep[NIRCam; ][]{Rieke2005} onboard the James Webb Space Telescope  \citep[JWST; ][]{Gardner2006}. In this first study we focus on a sample of bright LAEs, the advantage of using confirmed LAEs is the availability of reliable systemic redshifts, which can be derived from their \lya\ profiles. In addition, LAEs could be the main producers of ionising photons at the EoR, due to their connection to LyC leakage \citep{Gazagnes2020,Nakajima2020,Izotov2022,Matthee2022,Naidu2022}. In the future, this study will be expanded upon with larger samples using the JWST Advanced Deep Extragalactic Survey \citep[JADES; ][Eisenstein in prep.]{Eisenstein2017,rieke_jwst_2019} and a stellar-mass selected sample, to understand how ionising radiation varies among all known galaxy types at the EoR.

The structure of this paper is the following. In $\S$2 we describe the datasets and catalogues used in this work, followed by the methods used to measure \ha\ and the UV continuum luminosity in $\S$3. In $\S$4 we present our \texttt{Prospector} fitting method, along with the dust attenuation prescription used in this work. Using these results, we place constraints on $\xi_{ion}$ in $\S$5, and discuss them in $\S$6. Finally, we provide concluding remarks in $\S$7. Throughout this work we assume $\Omega_0 = 0.315$ and $H_0 = 67.4$ km s$^{-1}$ Mpc$^{-1}$, following \cite{Planck2020}.

\begin{table*}
    \small
        \centering
        \begin{tabular}{ccccccccc}
        \hline
        \noalign{\smallskip}
        N & Name & z$_{\rm{sys}}$ & f(Ly$\alpha$) & F182M  & F210M  & F430M & F460M & F480 \\ 
         & & & [10$^{-20}$ cgs]& [nJy] &  [nJy] & [nJy] & [nJy] & [nJy] \\
        \noalign{\smallskip}
        \hline
        \noalign{\smallskip}
        1$^{\rm{H23}}$ & JADES-GS+53.16408-27.79971 & 5.443 & 1008.2 (15.7) &  91.6$\pm$1.1 & 91.8$\pm$1.3 & 303.4$\pm$2.2 & 77.6$\pm$2.7 & 77.6$\pm$2.2 \\ 
2$^{\rm{H23}}$ & JADES-GS+53.16612-27.78576 & 5.467 & 717.2 (44.1) & 48.5$\pm$1.6 & 50.9$\pm$1.8 & 263.7$\pm$2.5 & 67.5$\pm$3.2 & 60.3$\pm$2.4 \\ 
3 & JADES-GS+53.16904-27.78769 & 5.468 & 200.0 (15.4) & 12.8$\pm$1.4 & 14.2$\pm$1.6 & 60.1$\pm$2.6 & 16.1$\pm$3.2 & 12.0$\pm$2.8 \\ 
4$^{\rm{H23}}$ & JADES-GS+53.16571-27.78494 & 5.468 & 1093.9 (70.3) & 16.8$\pm$1.6 & 15.5$\pm$1.8 & 54.5$\pm$2.5 & 18.9$\pm$3.1 & 15.6$\pm$2.7 \\ 
5$^{\rm{H23}}$ & JADES-GS+53.13859-27.79024 & 5.479 & 4127.7 (52.2) & 41.7$\pm$1.4 & 44.9$\pm$1.6 & 1153.4$\pm$3.9 & 394.8$\pm$4.1 & 393.9$\pm$3.0 \\ 
6 & JADES-GS+53.15337-27.77119 & 5.520 & 183.9 (8.8) & 4.6$\pm$1.3 & 3.4$\pm$1.6 & 21.5$\pm$2.2 & 7.9$\pm$3.2 & 2.1$\pm$2.6 \\ 
7 & JADES-GS+53.14669-27.78620 & 5.520 & 384.5 (9.1) & 52.1$\pm$1.2 & 53.7$\pm$1.3 & 115.5$\pm$2.3 & 54.1$\pm$2.8 & 61.9$\pm$2.2 \\ 
8 & JADES-GS+53.14023-27.78709 & 5.522 & 315.9 (7.8) & 6.5$\pm$1.2 & 6.9$\pm$1.3 & 19.1$\pm$2.1 & 10.2$\pm$2.7 & 8.6$\pm$2.0 \\ 
9$^{\rm{B04}}$ & JADES-GS+53.16674-27.80424 & 5.822 & 2423.2 (25.4) & 211.1$\pm$1.5 & 195.1$\pm$1.7 & 189.2$\pm$2.9 & 239.7$\pm$3.9 & 205.2$\pm$3.1 \\ 
10 & JADES-GS+53.16576-27.80344 & 5.885 & 308.7 (10.8) & 12.8$\pm$1.1 & 12.6$\pm$1.3 & 10.8$\pm$1.9 & 37.5$\pm$2.7 & 10.7$\pm$1.9 \\ 
11 & JADES-GS+53.17009-27.77512 & 5.889 & 234.3 (8.7) & 6.8$\pm$1.2 & 6.0$\pm$1.4 & 6.0$\pm$2.6 & 19.0$\pm$3.4 & 2.2$\pm$2.5 \\ 
12 & JADES-GS+53.17829-27.77728 & 5.893 & 325.1 (7.9) & 3.6$\pm$1.3 & 6.4$\pm$1.4 & -0.8$\pm$2.4 & 19.2$\pm$3.2 & 4.0$\pm$2.4 \\ 
13 & JADES-GS+53.16677-27.78592 & 5.895 & 98.5 (8.1) & 8.8$\pm$1.4 & 7.0$\pm$1.6 & 3.0$\pm$2.5 & 15.2$\pm$3.2 & 3.6$\pm$2.4 \\ 
14 & JADES-GS+53.16248-27.79273 & 5.897 & 246.4 (24.1) & 11.1$\pm$1.1 & 8.3$\pm$1.2 & 12.6$\pm$2.3 & 35.5$\pm$2.9 & 0.7$\pm$2.4 \\ 
15 & JADES-GS+53.15900-27.79268 & 5.899 & 142.0 (8.3) & 7.9$\pm$1.0 & 13.8$\pm$1.2 & 9.6$\pm$2.3 & 24.5$\pm$2.9 & 9.8$\pm$2.2 \\ 
16 & JADES-GS+53.16229-27.77746 & 5.915 & 219.5 (18.6) & 2.0$\pm$1.3 & 1.7$\pm$1.6 & 5.3$\pm$2.5 & 21.6$\pm$3.3 & 3.9$\pm$2.5 \\ 
17$^{\rm{B04}}$ & JADES-GS+53.16279-27.76084 & 5.915 & 1640.1 (32.0) & 38.5$\pm$1.7 & 40.6$\pm$2.0 & 18.1$\pm$11.0 & 103.5$\pm$15.4 & 14.3$\pm$9.7\\ 
18$^{\rm{B04}}$ & JADES-GS+53.14208-27.77985 & 5.919 & 465.2 (9.1) & 96.8$\pm$1.2 & 96.3$\pm$1.4 & 122.0$\pm$2.1 & 448.3$\pm$3.0 & 152.6$\pm$2.3 \\ 
19 & JADES-GS+53.13836-27.77878 & 5.930 & 459.8 (9.2) & 17.4$\pm$1.4 & 14.6$\pm$1.6 & 12.0$\pm$2.8 & 34.7$\pm$4.0 & 19.1$\pm$3.5 \\ 
20 & JADES-GS+53.15591-27.79382 & 5.946 & 263.5 (6.4) & 6.5$\pm$1.0 & 8.7$\pm$1.1 & 4.0$\pm$2.3 & 21.4$\pm$2.7 & 8.9$\pm$2.1 \\ 
21 & JADES-GS+53.15705-27.77272 & 5.972 & 127.3 (9.7) & 7.7$\pm$1.5 & 8.2$\pm$1.6 & 4.4$\pm$2.6 & 13.0$\pm$3.4 & 2.7$\pm$2.6 \\ 
22 & JADES-GS+53.17010-27.79694 & 5.981 & 330.6 (7.7) & 6.4$\pm$1.0 & 6.1$\pm$1.2 & 5.6$\pm$2.0 & 18.7$\pm$2.8 & 2.2$\pm$2.1 \\ 
23$^{\rm{B04}}$ & JADES-GS+53.16441-27.76214 & 6.096 & 442.1 (5.7) & 17.5$\pm$1.7 & 17.5$\pm$1.9 & 12.2$\pm$4.2 & 36.9$\pm$5.0 & 24.3$\pm$4.0 \\ 
24$^{\rm{B04}}$ & JADES-GS+53.17929-27.77331 & 6.102 & 326.0 (6.6) & 23.1$\pm$1.3 & 20.4$\pm$1.5 & 9.2$\pm$3.4 & 103.9$\pm$4.6 & 58.0$\pm$3.4\\ 
25 & JADES-GS+53.16853-27.77567 & 6.106 & 214.0 (9.9) & 1.4$\pm$1.1 & 3.5$\pm$1.3 & 1.8$\pm$2.4 & 8.0$\pm$3.2 & 10.5$\pm$2.8 \\ 
26 & JADES-GS+53.18744-27.77804 & 6.109 & 519.9 (7.3) & 2.6$\pm$1.5 & 6.5$\pm$1.7 & -3.0$\pm$3.5 & 23.9$\pm$4.2 & 6.2$\pm$3.6 \\ 
27$^{\rm{B04}}$ & JADES-GS+53.16611-27.77204 & 6.302 & 796.9 (21.2) & 22.7$\pm$1.4 & 25.9$\pm$1.7 & 55.1$\pm$2.6 & 79.1$\pm$3.2 & 257.1$\pm$2.7 \\ 
28 & JADES-GS+53.15603-27.78098 & 6.310 & 161.9 (4.5) & 18.7$\pm$1.4 & 19.3$\pm$1.7 & 20.1$\pm$2.3 & 15.8$\pm$3.2 & 38.4$\pm$2.4 \\ 
29 & JADES-GS+53.16912-27.76591 & 6.342 & 233.0 (7.0) & 4.4$\pm$1.8 & 7.0$\pm$2.0 & 7.4$\pm$3.5 & 12.3$\pm$4.2 & 19.2$\pm$3.3 \\ 
30 & JADES-GS+53.15803-27.78632 & 6.631 & 160.2 (8.1) & 17.7$\pm$1.3 & 21.9$\pm$1.5 & 31.1$\pm$2.5 & 28.5$\pm$3.2 & 39.5$\pm$2.7\\ 

        \end{tabular}
        \caption{General properties of galaxies in the sample, in order of increasing redshift. Medium band survey counterparts are within a 0.5$''$ radii of the MUSE coordinates (after correction for astrometric offset).  \textsl{Column 1:} sequential identifier from this work. "B04" highlights the LAEs that were identified as LBGs in \citet{Bunker2004}, while "H23" highlights the galaxies which are part of the sample studied in \citet{Helton2023}. \textsl{Column 2:} JADES identifier, composed of the coordinates of the centroid in units of degrees, rounded to the fifth decimal point. \textsl{Column 3:} systemic redshift from \citet{Bacon2022}, the error associated to this measurement is z$_{\rm{sys,e}} = 0.002$ for this sample. \textsl{Column 4:} Ly$\alpha$ fluxes from \citet{Bacon2022} in units of erg s$^{-1}$ cm$^{-2}$, rounded to the first decimal point. The Ly$\alpha$ signal-to-noise ratio is given in parenthesis. \textsl{Columns 5-9:} JEMS flux density in F182M, F210M, F430M, F460M and F480M, respectively, with corresponding errors in units of nJy. Circular apertures with radius 0.15$''$ are used throughout. All photometry has been aperture corrected.} 
        \label{tab:LAEs_properties}
    \end{table*}

\section{Data}

\subsection{JEMS}

The JWST Extragalactic Medium-band Survey \citep[JEMS; ][PID=1963]{Williams2023} is a JWST imaging program whose primary goal is to target \ha\ and the UV continuum at the EoR ($z \sim 5.4 - 6.6$). Many galaxies have been identified in this redshift range through the Lyman break technique, using the original HST/ACS images \citep[e.g. ][]{Bunker2004,Bouwens2015}. Including several with spectroscopic confirmation \citep[e.g. ][]{Bunker2003,Stanway2004}. JEMS covers the Hubble Ultra Deep Field \citep[HUDF; ][]{Beckwith2006} region in the GOODS-South field, which has been extensively studied with (among others) deep HST and MUSE observations. The survey consists of single visit observations with three filter pairs on the NIRCam: F210M-F430M, F210M-F460M, and F182M-F480M. In each pair, one module covers the entirety of the UDF while the second one points to the surrounding region,  both pointings have publicly available ancillary data. Figure~\ref{fig:filters} shows the throughputs of each filter as a function of wavelength, along with the wavelength evolution of the observed \lya\ and \ha\ with redshift (dashed and full black lines, respectively) in the redshift range of interest. The spectral coverage of MUSE is shown as a grey shaded area. The effective wavelength, exposure time and 5$\sigma$ sensitivity limit for each band is given in the caption.

The source detection and photometry leverage both the JEMS NIRCam medium band and  JWST Advanced Deep Extragalactic Survey (JADES; Eisenstein et al., in preparation) NIRCam broad and medium band imaging. Mosaics of the NIRCam long wavelength filter images (F277W, F335M, F356W, F410M, F430M, F444W, F460M, F480M) are combined into an inverse variance-weighted signal-to-noise ratio (SNR) image. Detection is performed using the {\it photutils} \citep{bradley2022a} software package, identifying sources with contiguous regions of the SNR mosaic with signal $>3\sigma$ and five or more contiguous pixels. We
also use {\it photutils} to perform circular aperture
photometry with filter-dependent aperture corrections based on empirical point-spread-functions measured from stars in the mosaic. The object flux within a range of aperture sizes is measured, ranging from $r=0.1"$ to $r=0.5"$ as well as the PSF 80\% encircled energy radius in each filter. \citet{kron1980a} photometry is also performed using Kron parameters of 1.2 and 2.5. Details of the JADES source catalogue generation and photometry will be presented in Robertson et al., (in prep.).

\subsection{\lya\ emitters from the MUSE Ultra-Deep Field surveys}
The Multi Unit Spectroscopic Explorer (MUSE) is a large integral field unit on the VLT, in Chile, with a wavelength coverage of $\lambda = 4800-9300$ \AA\/. The MUSE Hubble Ultra-Deep Field surveys (MUSE UDF) consist of deep observations in the HUDF area, in this work we use these surveys to find \lya\ emitters at the tail end of the EoR. The first release \citep{Bacon2017,Inami2017} was based on two datasets: (1) a $3 \times 3$ arcmin$^2$ mosaic of 9 MUSE fields with an exposure of 10 hours, called "MOSAIC", and (2) a single $1 \times 1$ arcmin$^2$ field with a depth of 31 hours, called "UDF-10". The second release \citep{Bacon2022} extended this survey to unprecedented depths by including an adaptive optics assisted survey in the same area, reaching 141 hours of exposure time. This new survey is called the MUSE eXtremely Deep-Field (MXDF) and is the deepest spectroscopic survey in existence. The astrometry was matched to the Hubble ACS astrometry, however, there is an offset between the HST and Gaia DR2 catalogues \citep{Dunlop2017,Franco2018} amounting to $\Delta \text{RA}= +0.094 \pm 0.042$ arcsec and $\Delta \text{DEC}= -0.26 \pm 0.10$ arcsec. The average offset for each dataset is given in \cite{Bacon2022}, we take this offset into account when assigning counterparts to the LAEs in JEMS.

To assemble our sample, we use the catalogues presented in \cite{Bacon2022}, in which the datasets from the first release were reprocessed with the same tools as the MXDF. The redshifts assignment and confidence are of particular interest for this work. Briefly, at $z > 2.9$ the redshift is estimated through the \lya\ profile. The redshift is based on the peak of the \lya\ line, known to be offset from the systemic redshift due to its resonant nature \citep{Shapley2003,Song2014}. In order to estimate the systemic redshift, statistical corrections from \cite{Verhamme2018} are applied in the catalogue. In summary, two corrections are taken into account: one based on the separation of the peaks when double peaks are detected, and one based on the full-width-half-maximum (FWHM) of the line. For the redshifts range of interest, the confidence of the detected line being \lya\ (ZCONF) is a number between zero and three, where
\begin{itemize}
    \item ZCONF = 0. No redshift solution can be found
    \item ZCONF = 1. Low confidence arising for example from low S/N of the lines or the existence of other redshift solutions
    \item ZCONF = 2. Good confidence. The \lya\ line is detected with a S/N $>$ 5, and has a width and asymmetry that are compatible with \lya\ line profiles
    \item ZCONF = 3. High confidence. If \lya\ is the only detected line, then it has to have a S/N $>$ 7 with the expected line shape (i.e. a red asymmetrical profile and/or a blue bump, or double peaked profile)
\end{itemize}
A group of experts iterates until they converge on a classification for each object. Often the difference between "good" and "high" confidence are subtle and objects in either group can be considered as having a certain redshift assignment. 

\subsection{Sample}
We focus on LAEs at $z \sim 5.4 - 6.6$ that have ZCONF = 2 or ZCONF = 3, and for which \ha\ falls in one of the following JEMS filters: F430M, F460M, or F480M. This leaves us with a sample of 30 LAEs with spectroscopic redshift, shown as horizontal line markers on Figure~\ref{fig:filters}. After correcting for the astrometry offset we find counterparts in JEMS by searching within a $0.5''$ radius of the coordinates, this radius is favourable due to the known possible offset between the \lya\ and the stellar continuum emission \citep{Hoag2019,Claeyssens2022}. We then visually examine each source to confirm the counterpart assignment. Table~\ref{tab:LAEs_properties} shows the coordinates of JEMS survey counterparts, along with their systemic redshift, \lya\ flux, and JEMS photometry. We note that we see a flux excess that corresponds to the redshifted \ha\ location in these objects. All the \lya\ profiles and medium band cutouts are shown in the Appendix~\ref{appendix}. For reference, seven of these galaxies were identified previously as Lyman Break Galaxies (LBGs) in \cite{Bunker2004}, and four of them are part of the \ha\ emitting sample studied in \cite{Helton2023} they have been highlighted in Table~\ref{tab:LAEs_properties}.

   \begin{figure*}
   \centering
   \includegraphics[width=1\textwidth]{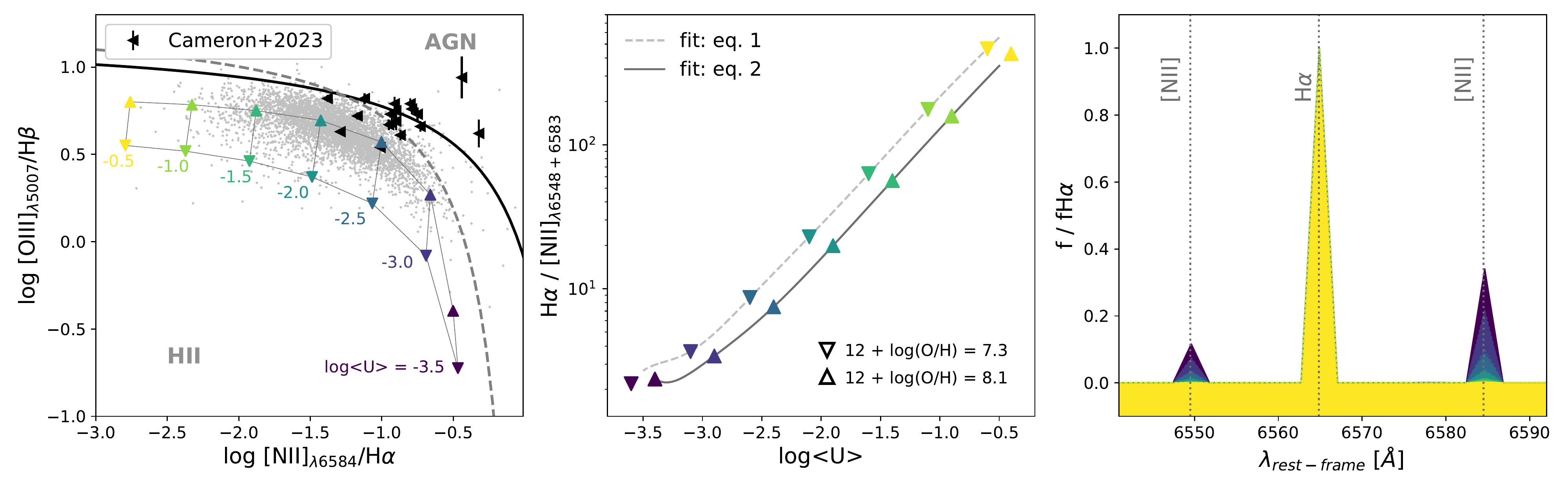}
   \caption{Summary of the Cloudy models used in this work, showing the relative intensity of \nii\ in comparison to \ha\/. As the ionisation parameter (log<U>) increases, the \nii\ emission decreases. Specifically, for the parameter space populated by high redshift galaxies (log<U> $\gtrapprox -2.5$), \nii\ is negligible compared to \ha\/. The coloured triangles represent two different stellar and gas-phase metallicities, as labelled. $\textsl{Left panel:}$ BPT diagram showing the SDSS sample (grey points) and the evolution of Cloudy models with varying ionisation parameter values. Also shown are the \oiii\//\hb\ measurements and \nii\//\ha\ upper limits for galaxies at $z \sim 6$ from \citet{Cameron2023} (black sideways triangles). $\textsl{Middle panel:}$ evolution of the \nii\//\ha\ ratio with increasing log<U>. $\textsl{Right panel:}$ zoom-in of spectral region enclosing \ha\ and \nii\/, normalised to the \ha\ peak. The \nii\ emission is colour-coded by log<U>. }
              \label{fig:cloudy}%
    \end{figure*}

\section{\ha\ measurement}
There has been extensive work in the literature to retrieve \ha\ fluxes by combining multi-band photometry to estimate both the line flux and the local continuum \citep[e.g.  ][]{Bunker1995,Shim2011,Labbe2013,Stark2013}. It is a common approach to estimate \ha\ emission from photometry, and then use models to interpret the results  \citep[e.g. ][]{Sobral2012,Vilella-Rojo2015,Faisst2016}. The JEMS survey has proven to be useful to measure emission lines in high redshift galaxies \citep[$7 < z < 9$; ][]{Laporte2022}, in this work we use JEMS photometry to measure \ha\ luminosities. Specifically, depending on the redshift of the sources, we interpret colour excess in the F430M, F460M or F480M bands as the presence of \ha\ emission. The luminosity can then be estimated by using the remaining two filters to measure the local continuum and subtract it. We first study the possible contamination of \nii\ in the photometry, then describe the steps that were followed to measure \ha\/, and correct it for dust attenuation.

\subsection{\nii\ contribution to \ha\ photometry}
\label{section:nii_contamination}
Extracting \ha\ emission from broad-band photometry can be troublesome, mainly due to the nearby \nii\ emission lines at rest-frame 6548 and 6583 \AA\/, although also potentially affected by other metal lines in the vicinity of the continuum band measurement. To study the effect this contamination might have in our medium band observations, we create appropriate photoionisation models and simulate photometry for F182M, F210M, F430M, F460M and F480M, for the redshifts of the LAEs in the sample. It is important to note that at the redshifts of interest for this work (i.e. $z \sim 5.4 - 6.6$) we expect metal-poor galaxies with high ionisation parameters (log<U>) to dominate \citep{Harikane2020,Katz2022,Sugahara2022}. Although recently some tentative evidence has been shown for a galaxy with solar metallicity at $z > 7$ \citep{Killi2022}, this is not the norm. Promisingly, \cite{Cameron2023} produce emission line diagnostic diagrams for 26 Lyman break galaxies at $z \sim 5.5-9.5$ using spectroscopy obtained with the JWST/NIRSpec micro-shutter assembly \citep{Ferruit2022,Jakobsen2022}, and find that these galaxies populate a parameter region comparable to extreme local ones, having low metallicity and high ionisation parameters \citep{Curti2023,Nakajima2023}.  

To investigate the relative intensity of \nii\ compared to \ha\ we produce, as in \cite{Simmonds2021}, photoionisation models following a sample of 5607 star-forming galaxies \citep{Izotov2014} from the Sloan Digital Sky Survey (SDSS) Data Release 14 \citep{Abolfathi_2018}, compiled by Y. Izotov and collaborators. The properties of these galaxies are discussed in previous literature \citep{Guseva2020, Ramambason2020} but of main importance to this work are their high \oiii\//\hb\ ratios, and their low metallicities (mean metallicity, 12 + log(O/H) = 7.97$^{+0.14}_{-0.17}$). 

We then use the version 17.03 of the photoionisation code Cloudy \citep{Ferland2017}, combined with stellar population models from the Binary Population and Spectral Synthesis version 2.2.1 \citep[BPASS;][]{Eldridge2017}. We run a grid of simple Cloudy photoionisation models to convergence, using BPASSv2.2.1 models with binary interactions. The initial mass function (IMF) only slightly changes the results, therefore, we choose to use a Salpeter IMF ($\alpha = -2.35$, with M$_{*,max} = 100$ M$_\odot$). At redshifts close to the EoR, we expect galaxies to have young and metal-poor stellar populations, so we decide to use 10 Myr old stellar population models with a constant star-formation history (SFH). We use two different metallicities, one that follows the bulk of the the SDSS sample ($Z = 0.006$; 12 + log(O/H) = 8.1), and one to explore lower metallicity galaxies ($Z = 0.001$; 12 + log(O/H) = 7.3), shown as upwards and downwards triangles in Figure~\ref{fig:cloudy}, respectively. Most importantly, we vary the ionisation parameter (log<U>), in a range between $-3.5$ and $-0.5$. This is a dimensionless ratio of ionising over non-ionising hydrogen photon densities. A spherical cloud with the same gas-phase metallicity as the stellar SEDs is assumed in each case. The left panel of Figure~\ref{fig:cloudy} shows one of the classical low-ionisation emission line diagrams \citep[BPT diagram;][]{Baldwin1981,Kewley2006}. The star-forming SDSS sample is plotted as grey dots. These galaxies populate virtually the same region of the BPT diagram that is expected of galaxies at $z > 6$, i.e. log<U> $> -3.0$ and log \oiii\//\hb\ $> 0.4$ \citep[see Figures 5 and 6 of][]{Sugahara2022}, which has recently been confirmed by observations of $z \sim 6$ Lyman break galaxies in \cite{Cameron2023}, making them appropriate low-redshift analogues to galaxies at the EoR. 

The relation between the \nii\//\ha\ intensities ratio and log<U> is shown for our Cloudy models in the middle panel of Figure~\ref{fig:cloudy}, and are described by the following cubic equations
\begin{equation}
    \frac{\text{\nii\/}}{\text{\ha\/}}\bigg|_{Z=0.006} = -0.020\text{x}^3 - 0.045\text{x}^2 - 0.028\text{x} - 0.001
\end{equation}
and
\begin{equation}
    \frac{\text{\nii\/}}{\text{\ha\/}}\bigg|_{Z=0.001} = -0.004\text{x}^3 + 0.048\text{x}^2 + 0.092\text{x} + 0.037 
\end{equation}
where the subscripts represent the intensity ratios for the Cloudy models corresponding to metallicities $Z = 0.001$ (dashed curve, 12 + log(O/H) = 7.3) and $Z = 0.006$ (filled curve, 12 + log(O/H) = 8.1), respectively, and "x" represents log<U>.

 We find that the \ha\ emission dominates over \nii\ in our photoionisation models, in agreement with previous studies \citep{Maiolino2019,Onodera2020}. In addition, \nii\ appears to be virtually non-existent in a large sample of galaxies at $z = 5.1-5.5$, for which grism spectra covering the \ha\ emission line is available (Sun in prep.). At its maximum, our models indicate \nii\ can account for $\sim 45$\% of the band flux, but this is only in cases where log<U> = -3.5. At $z \sim 6$ higher ionisation parameters are predicted \citep[e.g. ][]{Sugahara2022}. Moreover, log<U> is directly related to the \oiii\//\oii\ ratio, which has been observed to be elevated at $z > 6$ \citep{Cameron2023}, and therefore, we assume negligible \nii\ contamination in our next steps.

    \begin{figure}
          \centering
   \includegraphics[width=0.5\textwidth]{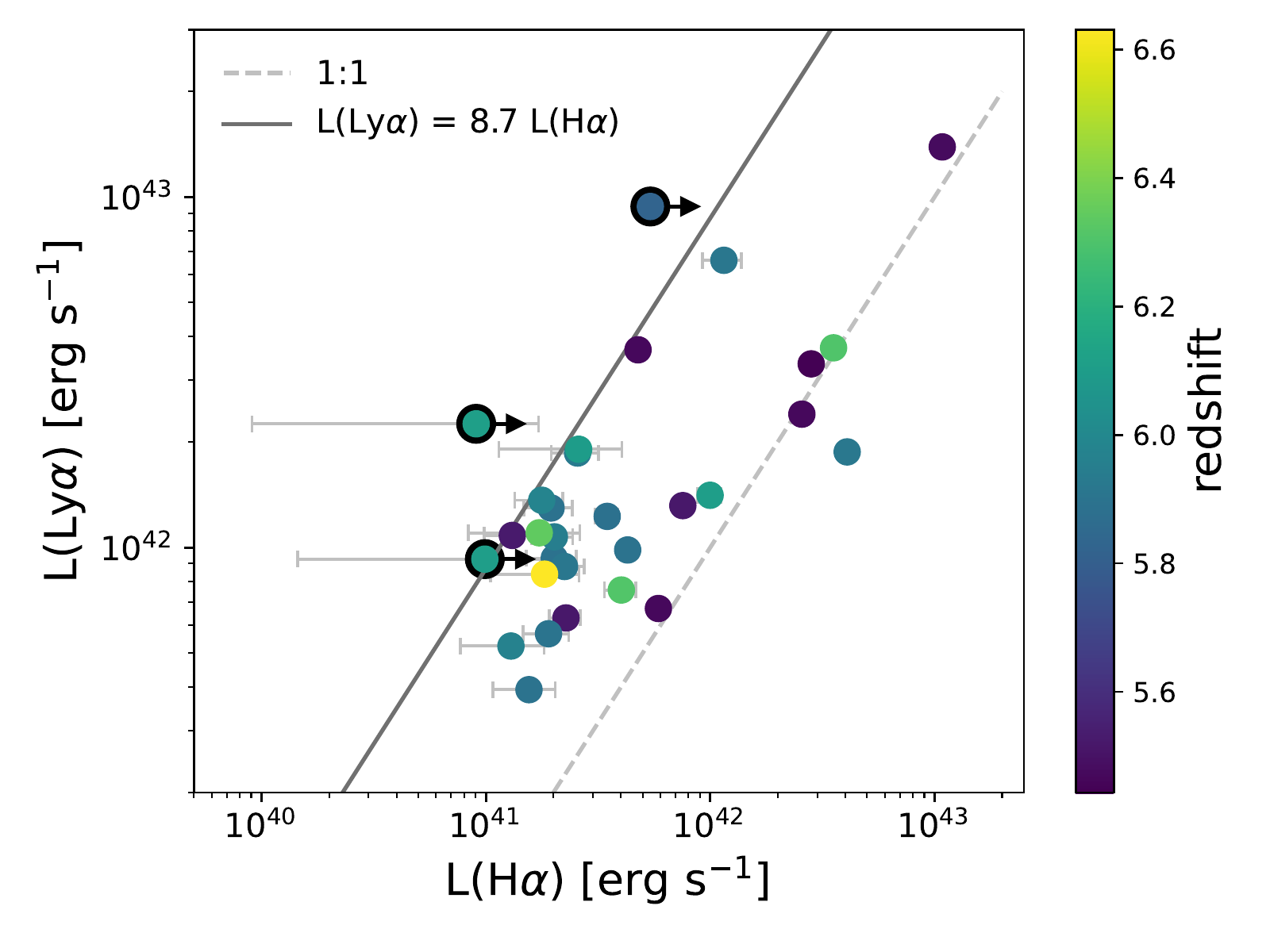}
   \caption{Comparison of measured \ha\ luminosities derived from JEMS photometry and \lya\ luminosities from MUSE. The lower limits are shown as circles with black edges. The grey dashed line shows the 1:1 relation, while the filled grey line shows the expected relation between \lya\ and \ha\ assuming Case B recombination. The observed range in \lya\//\ha\ agrees with LAEs from literature \citep[e.g ][]{Finkelstein2011,Hayes2013}, and is commonly attributed to differences in dust attenuation.} 
              \label{fig:Ha_vs_Lya}%
    \end{figure}

\subsection{\ha\ luminosity estimation}

We define four redshift bins in our sample, depending on the expected wavelength of \ha\/, and for each bin use a slightly different prescription to estimate \ha\ luminosities. We find that the local continuum is more reliably measured when we use the full JEMS photometry to fit a line in logarithmic space, and use this fit to estimate the local continuum around \ha\/. We note that for our objects there is no evidence of any strong emission lines (other than \ha\/) contaminating the JEMS photometry, and thus, for the continuum estimation we only exclude the band that contains \ha\ in each bin, as follows:
\begin{enumerate}
    \item $z \leq 5.75$: f(\ha\/) falls in F430M
    \item $5.75 < z \leq 6$: f(\ha\/) falls in F460M
    \item $6 < z \leq 6.15$: f(\ha\/) falls in both F460M and F480M
    \item $z > 6.15$: f(\ha\/) falls in F480M
\end{enumerate}

The JEMS photometry flux densities in units of nJy can be found in Table~\ref{tab:LAEs_properties}, while the derived \ha\ luminosities are presented in Table~\ref{tab:LAEs_derived_properties}. 
We have enforced a floor error of 5\% in the \ha\ measurements. Our measurements (when available) are consistent with those reported in \cite{Helton2023}. It must be noted, however, that this simple prescription in some cases underestimates the flux of \ha\ emission. In particular, there are two LAEs (JADES-GS+53.17829-27.77728 and JADES-GS+53.18744-27.77804) in which the F430M photometry yields negative fluxes. In these cases we ignore this filter when estimating \ha\/, resulting in lower limits for $\xi_{ion,0}$ (circles with black edges in Figure~\ref{fig:xi_ion}). In the case of JADES-GS+53.16674-27.80424, the \ha\ emission line falls in the wavelength range with low transmission of the F460M filter. The \ha\ line flux derived from the NIRCam grism observations through the First Reionisation Epoch Spectroscopic Complete Survey  \citep[FRESCO, PI: Oesch; ][]{Oesch2023} is approximately six times higher than our photometric measurement (private communication from F. Sun). For consistency, we estimate $\xi_{ion,0}$ using the underestimated \ha\ for this source (shown as a circle with a black edge and a plus sign in Figure~\ref{fig:xi_ion}), but note that this value is a lower limit.

   \begin{figure*}
        \centering
   \includegraphics[width=1\textwidth,trim={0 6cm 0 0}]{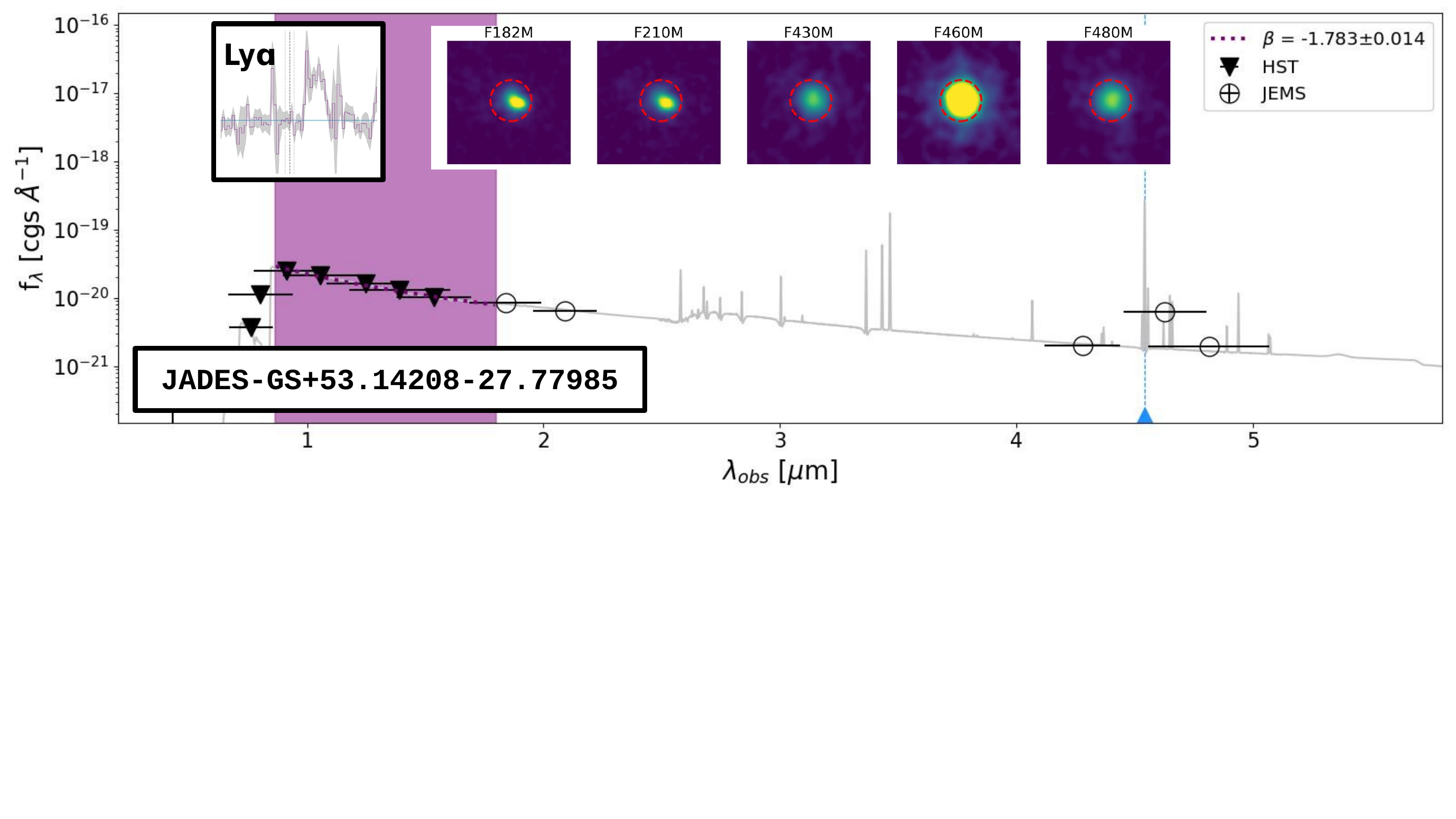}
   \caption{Representative example of the LAEs in this work (JADES-GS+53.14208-27.77985), at $z = 5.919$, assuming a continuity (i.e. non-parametric) SFH. $30 \times 30$ pixel$^2$ ($0.9 \times 0.9$ arcsecond$^2$) JEMS cutouts are shown, the aperture used to extract photometry is shown as a red dashed circle. The symbols show the photometric points for HST (triangles) and JEMS (circles), respectively. The grey curve shows the best fit provided by \texttt{Prospector}, with the spectral region used for the $\beta$ estimation shaded in purple. For reference, the shape of the \lya\ profile and the \ha\ expected location (vertical dotted line) are included.}
              \label{fig:bestfit}%
    \end{figure*}

\subsection{The \lya\//\ha\ ratio}
The synergy between MUSE and NIRCam is exceptional to study the \lya\//\ha\ ratio, since there are no slit loss effects and MUSE is currently the best instrument to observe \lya\ at the redshift range of this work. Figure~\ref{fig:Ha_vs_Lya} compares the estimated \ha\ to the measured \lya\ luminosities. If Case B recombination is assumed, \lya\//\ha\ $= 8.7$ \citep[filled grey line, ][]{Osterbrock2006}. In LAEs this ratio has been observed to fluctuate between $\sim 0.1-10$, the resonant nature of the \lya\ emission line complicates the interpretation of this fluctuation, but it has been proposed that dust attenuation plays a big role by reducing the \lya\ emission \citep[see ][]{Scarlata2009,Cowie2011,Finkelstein2011,Hayes2013,Hayes2014}. Therefore, our range of \ha\ fluxes are expected and in agreement with previous studies. 

If we assume Case-B recombination, the ratio between the \lya\ and \ha\ luminosities can give us constraints on the \lya\ escape fractions, using the simple relation given by atomic physics: f$_{\rm{esc}}$(\lya\/)$ = \frac{1}{8.7} \frac{\text{L(\lya)}}{\text{L(\ha)}}$. We provide these estimations, excluding the cases in which \ha\ is a lower limit, in Table~\ref{tab:LAEs_derived_properties}.


\section{UV luminosities and dust attenuation}
\subsection{SED fitting with \texttt{Prospector}}
To obtain the UV continuum slope, $\beta$, we use the galaxy spectral energy distribution (SED) fitting code \texttt{Prospector} \citep{Johnson2019,Johnson2021}. This tool infers stellar population parameters from the UV to IR wavelengths using photometry and/or spectroscopy as inputs. For this work we use the JEMS photometry, in addition to observations in the HST ACS bands: F435W ($\lambda_{\rm{eff}} =  0.432 \mu$m), F606W ($\lambda_{\rm{eff}} = 0.578 \mu$m), F775W ($\lambda_{\rm{eff}} = 0.762 \mu$m), F815W ($\lambda_{\rm{eff}} = 0.803 \mu$m), F850LP ($\lambda_{\rm{eff}} = 0.912 \mu$m), and the HST WFC3 IR bands: F105W ($\lambda_{\rm{eff}} = 1.055 \mu$m), F125W ($\lambda_{\rm{eff}} = 1.249 \mu$m), F140W ($\lambda_{\rm{eff}} = 1.382 \mu$m) and F160W ($\lambda_{\rm{eff}} = 1.537 \mu$m). The same circular aperture of radius 0.15$''$ is used to extract both the JEMS and HST photometry, all photometry has been aperture corrected. We fix the redshifts to the ones provided in \cite{Bacon2022}, then allow the dust attenuation and properties of the stellar populations (stellar and gas-phase metallicity, mass, ionisation parameter, and SFH) to vary following the prescription in \cite{Tacchella2022}. Specifically, for the dust attenuation we use a two component dust model that accounts for the increased effect on young stars, embedded in dust, and the nebular emission \citep[<10 Myr, as described in ][]{Conroy2009}, with a variable dust index. We allow the stellar metallicity to vary between log(Z/Z$_{\odot}$) $= -3.0$ and $0.19$, adopting a top-hat prior in logarithmic space. The continuum and line emission properties of the SEDs are provided by the Flexible Stellar Population Synthesis (FSPS) code \citep{Byler2017}, which are based on interpolated models from Cloudy \citep[version 13.03; ][]{Ferland2013}, limiting the ionisation parameter to a maximum value of -1. Regarding SFHs, we adopt a non-parametric SFH \citep[continuity SFH; ][]{Leja2019}. In brief, this option describes the SFH as a combination of six different SFR bins with the bursty-continuity prior \citep{Tacchella2022bursty}. The latter consists of five free parameters controlling the ratios of the amplitudes of the adjacent bins. Finally, we include both nebular and dust emission. The nebular emission is particularly important because we incorporate the JEMS photometry that contain the \ha\ emission line in our fits.

Once the best fit spectrum has been produced for each galaxy, we fit a straight line in logarithmic space between rest-frame $\lambda = 1250-2600$ \AA\/ \citep{Calzetti1994}, in the form $f_{\lambda} \propto \lambda^{\beta}$. The results are given in Table~\ref{tab:LAEs_derived_properties}. In addition to providing constraints on the dust attenuation, this fitted line also allows us to predict the monochromatic UV luminosity density at rest-frame $\lambda = 1500$ \AA\/. A representative example is shown on Figure~\ref{fig:bestfit}.

\subsection{Dust attenuation estimation}
The observed \ha\ luminosities must now be corrected for dust attenuation. Since we do not have access to \hb\/, and therefore, cannot measure the Balmer decrement, we can use the rest-frame UV continuum slope \citep[$\beta$; ][]{Calzetti1994} obtained with \texttt{Prospector} to estimate the effect of dust attenuation in the galaxies \citep[e.g. ][]{Reddy2018}. We note that the values obtained following this method are consistent with the dust attenuation obtained with \texttt{Prospector}.

The stellar and nebular continuum colour excess, E(B-V), can be estimated using
\begin{equation}
    E(B-V)_{stellar} = \frac{1}{4.684}[\beta + 2.616] 
\end{equation}
given in \cite{Reddy2018}. The nebular excess is then simply given by $E(B-V)_{neb} = 2.27 \times E(B-V)_{stellar}$ \citep{Calzett2000}. The optical depth of \ha\ is given as a function of $E(B-V)_{neb}$ in \cite{Reddy2020} as $\tau \sim 3 \times E(B-V)_{neb}$ assuming a Calzetti attenuation law. The attenuation law at high redshifts is not well understood \citep{Gallerani2010,Ma2019}. Moreover, the amount of dust attenuation that affects emission lines versus the stellar continuum is highly uncertain. At $z \sim 2$, nebular lines have been observed to be attenuated by the same amount as the stellar continuum \citep{Calzett2000,Erb2006,Reddy2010,Reddy2015}. In addition, the lack of understanding of the geometry and amount of dust attenuation at early galaxy phases complicate the situation even further \citep{Bowler2018,Bowler2022}. It has been shown that a steeper attenuation curve like SMC is more appropriate than a Calzetti law for young high-redshift galaxies \citep{Shivaei2020}. The choice to use Calzetti over SMC in this work is due the increased effect it has in lowering $\xi_{ion,0}$ \citep[e.g., log $\xi_{ion}$ (SMC) $\sim$ log $\xi_{ion}$ (Calzetti) + 0.3; ][]{Shivaei2018}. And to thus highlight the increased ionising photon production we find in our sample.  

Given that the galaxies in this work show evidence of blue UV slopes ($\beta \sim -2.2$) we believe the dust attenuation effects to be small. We therefore present our results uncorrected by dust in Table~\ref{tab:LAEs_derived_properties}. However, for illustrative purposes, in Figure~\ref{fig:xi_ion} we show both uncorrected (circles) and dust-corrected (crosses) values, assuming a Calzetti attenuation law, and adopting local relations for the nebular and stellar continuum attenuation. This prescription lowers our resulting $\xi_{ion,0}$ by $\sim 0.1$ dex in the few cases where $\beta$ suggests non-negligible dust attenuation. We note that if an SMC law is assumed instead, the effect is even smaller.

   \begin{table*}
    \selectfont
        \centering
        \begin{tabular}{ccccc}
        \hline
        \noalign{\smallskip}
        N & log$_{10}$ L(H$\alpha$) & $\beta$ & log$_{10}$ $\xi_{ion,0}$ & f$_{\rm{esc}}$(\lya\/)\\
         & [erg s$^{-1}$] & & [Hz erg$^{-1}$] & \\
        \noalign{\smallskip}
        \hline
        \noalign{\smallskip}
        1 & 42.45 (+0.02,-0.02) & -1.60$\pm$0.02 & 25.61 (+0.12,-0.11) & 0.14 (+0.01,-0.01)\\ 
2 & 42.41 (+0.02,-0.02) & -1.84$\pm$0.01 & 25.86 (+0.13,-0.12) & 0.11 (+0.01,-0.00)\\ 
3 & 41.77 (+0.05,-0.05) & -2.07$\pm$0.01 & 25.66 (+0.02,-0.02) & 0.13 (+0.02,-0.01)\\ 
4 & 41.68 (+0.03,-0.04) & -1.95$\pm$0.01 & 25.56 (+0.05,-0.05) & 0.88 (+0.06,-0.08)\\ 
5 & 43.03 (+0.02,-0.02) & -1.47$\pm$0.03 & 26.59 (+0.09,-0.10) & 0.15 (+0.01,-0.01)\\ 
6 & 41.36 (+0.10,-0.12) & -1.75$\pm$0.03 & 25.96 (+0.06,-0.07) & 0.32 (+0.08,-0.08)\\ 
7 & 41.88 (+0.06,-0.08) & -1.83$\pm$0.01 & 25.26 (+0.02,-0.08) & 0.20 (+0.03,-0.03)\\ 
8 & 41.12 (+0.06,-0.08) & -2.24$\pm$0.02 & 25.31 (+0.02,-0.09) & 0.94 (+0.14,-0.16)\\ 
9$^{*}$ & 41.73 (+0.11,-0.14) & -2.05$\pm$0.01 & 24.48 (+0.08,-0.10) & - \\ 
10 & 41.54 (+0.06,-0.07) & -2.26$\pm$0.02 & 25.44 (+0.03,-0.06) & 0.41 (+0.06,-0.06)\\ 
11 & 41.30 (+0.08,-0.09) & -1.93$\pm$0.01 & 25.61 (+0.11,-0.13) & 0.54 (+0.11,-0.10)\\ 
12 & 41.29 (+0.13,-0.18) & -2.43$\pm$0.02 & 25.44 (+0.11,-0.14) & 0.76 (+0.27,-0.26)\\ 
13 & 41.19 (+0.09,-0.11) & -2.03$\pm$0.01 & 25.43 (+0.07,-0.15) & 0.29 (+0.07,-0.07)\\ 
14 & 41.63 (+0.04,-0.04) & -2.35$\pm$0.01 & 25.57 (+0.04,-0.04) & 0.27 (+0.03,-0.02)\\ 
15 & 41.28 (+0.08,-0.09) & -1.95$\pm$0.01 & 25.29 (+0.11,-0.13) & 0.34 (+0.07,-0.06)\\ 
16 & 41.35 (+0.11,-0.15) & -2.66$\pm$0.02 & 26.03 (+0.08,-0.11) & 0.45 (+0.13,-0.13)\\ 
17 & 42.06 (+0.12,-0.17) & -2.24$\pm$0.02 & 25.52 (+0.10,-0.13) & 0.66 (+0.21,-0.21)\\ 
18 & 42.61 (+0.02,-0.02) & -1.78$\pm$0.01 & 25.72 (+0.11,-0.10) & 0.05 (+0.00,-0.00)\\ 
19 & 41.41 (+0.19,-0.33) & -2.28$\pm$0.02 & 25.22 (+0.23,-0.30) & 0.83 (+0.46,-0.44)\\ 
20 & 41.31 (+0.08,-0.10) & -2.22$\pm$0.02 & 25.42 (+0.10,-0.07) & 0.60 (+0.12,-0.12)\\ 
21 & 41.11 (+0.13,-0.18) & -2.36$\pm$0.01 & 25.16 (+0.11,-0.14) & 0.47 (+0.16,-0.16)\\ 
22 & 41.25 (+0.09,-0.12) & -2.42$\pm$0.02 & 25.35 (+0.12,-0.09) & 0.88 (+0.20,-0.21)\\ 
23 & 41.41 (+0.19,-0.35) & -2.31$\pm$0.01 & 25.15 (+0.24,-0.33) & 0.85 (+0.47,-0.47)\\ 
24 & 42.00 (+0.05,-0.06) & -2.16$\pm$0.01 & 25.68 (+0.01,-0.04) & 0.16 (+0.02,-0.02)\\ 
25$^{*}$ & 41.00 (+0.28,-1.00) & -2.38$\pm$0.02 & 25.53 (+0.67,-1.00) & - \\ 
26$^{*}$ & 40.96 (+0.21,-0.41) & -2.46$\pm$0.02 & 25.25 (+0.30,-0.39) & - \\ 
27 & 42.55 (+0.02,-0.02) & -1.79$\pm$0.01 & 26.21 (+0.10,-0.09) & 0.12 (+0.01,-0.01)\\ 
28 & 41.60 (+0.09,-0.11) & -2.01$\pm$0.01 & 25.32 (+0.05,-0.07) & 0.22 (+0.05,-0.05)\\ 
29 & 41.24 (+0.11,-0.15) & -2.11$\pm$0.01 & 25.33 (+0.08,-0.11) & 0.73 (+0.21,-0.21)\\ 
30 & 41.26 (+0.09,-0.12) & -1.82$\pm$0.01 & 24.92 (+0.06,-0.08) & 0.53 (+0.12,-0.13)\\ 

        \end{tabular}
        \caption{Derived properties rounded to the second decimal point assuming zero dust correction. The median values for the sample are as follows: $\beta = -2.09^{+0.23}_{-0.21}$, and log$\frac{\xi_{ion,0}}{\text{Hz erg}^{-1}} = 25.44^{+0.21}_{-0.15}$. If we assume a \citet{Calzett2000} attenuation law, $\xi_{ion,0}$ is reduced by $\sim 0.1$ dex. \textsl{Column 1:} sequential identifier from this work (see Table~\ref{tab:LAEs_properties}). LAEs where lower limits for L(\ha\/) and $\xi_{ion,0}$ are provided are marked with an asterisk. \textsl{Column 2:} logarithm of the measured \ha\ luminosities, with corresponding errors. \textsl{Column 3:} UV continuum slope, $\beta$, defined as $f_{\lambda} \propto \lambda^{\beta}$ in the region delimited by rest-frame $\lambda = 1250 - 2600$ \AA\/. \textsl{Column 4:} logarithm of the ionising photon production efficiency assuming no dust attenuation, with corresponding errors. \textsl{Column 5:} \lya\ escape fractions assuming no dust attenuation. The cases in which L(\ha\/) lower limits have been provided have been left blank.}
        \label{tab:LAEs_derived_properties}
    \end{table*}

\section{Constraints on $\xi_{ion}$}

The dust-corrected \ha\ luminosity is related to the amount of ionising photons ($N(H^0)$) that are being emitted. If we assume Case B recombination ($f_{esc} = 0$), a temperature of $10^4$ K and an electron density of 100 cm$^{-3}$, this relation is described in \cite{Osterbrock2006} as
\begin{equation}
\label{eq:xi_ion}
    N(H^0) = 7.28 \times 10^{11} L(H\alpha)    
\end{equation}
where $N(H^0)$ is in units of Hz and L(\ha\/) in units of erg s$^{-1}$. Since we expect our sources to have a non-zero LyC escape fraction, the Case B recombination assumption yields a conservative estimation of the amount of ionising photons. If these galaxies are LyC leakers, which we will study in a future work, the ionising photon production efficiency would be boosted with respect to the values presented here. We use this equation to estimate the production rate of ionising photons, $\xi_{ion,0}$, using
\begin{equation}
    \xi_{ion,0} = \frac{N(H^0)}{L_{UV}},
\end{equation}
where the zero subscript indicates the assumption of zero escape of ionising photons into the IGM, $L_{UV}$ is the observed monochromatic luminosity density in units of erg s$^{-1}$ Hz$^{-1}$ at rest-frame $\lambda$ = 1500 \AA\/.

\begin{figure*}
        \centering
   \includegraphics[width=1\textwidth]{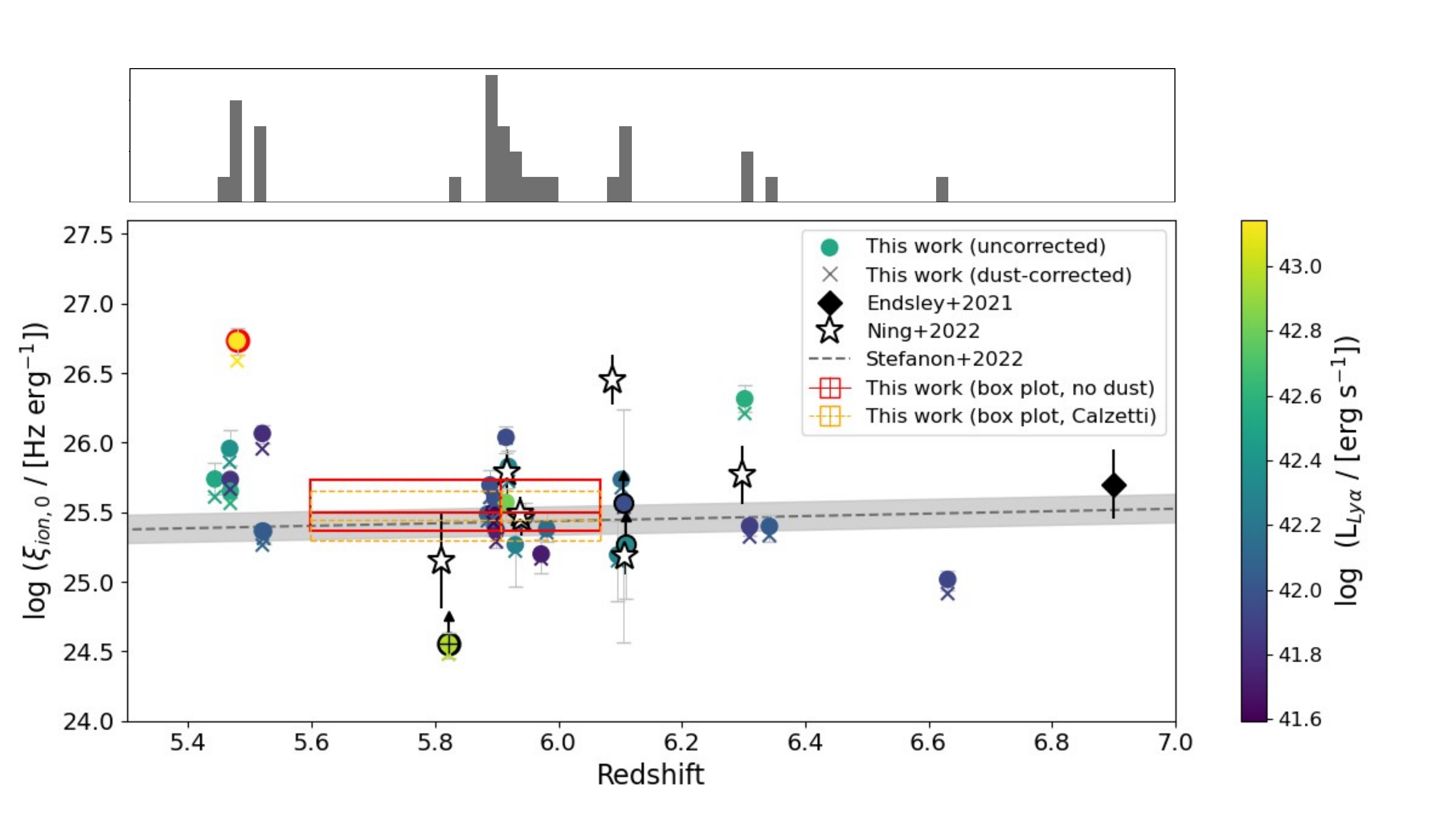}
   \caption{$\xi_{ion,0}$ uncorrected for dust as a function of redshift for 30 individual LAEs (circles), colour-coded by \lya\ luminosity. If we assume a Calzetti dust law instead, the points are lowered by $\sim 0.1$ dex for the few sources where $\beta$ suggests non-negligible dust attenuation (crosses). The source with the highest $\xi_{ion,0}$ (JADES-GS+53.13859-27.79024; circle with red edge) is also the strongest \lya\ emitter, however, this galaxy shows evidence of AGN activity, which would boost the production of ionising photons. Lower limits are shown as circles with black edges, and JADES-GS+53.16674-27.80424 is additionally marked with a black plus sign. The linear fit to observations of $\xi_{ion,0}$ provided in \citet{Stefanon2022} is shown as a dashed line, with its corresponding 68\% confidence interval of the fit. For comparison, we have included measurements from literature at similar redshifts from \citet{Endsley2021} (black diamond) and the LAEs from \citet{Ning2023} (stars). The empty red (orange) rectangle is a box plot summary of this work, showing the median $\xi_{ion,0}$ and redshift, along with the 25\% and 75\% percentiles of the sample uncorrected (corrected) for dust attenuation. The top panel shows the redshift distribution of the sample, where two possible overdensities of LAEs are observed (at $z \sim 5.4$ and $z \sim 5.9$).}
              \label{fig:xi_ion}%
    \end{figure*}

Given the uncertainty in the effects of dust on the \ha\ emission line and the UV continuum, in addition to the blue continuum slopes found in our sample (median values of $\beta = -2.09^{+0.23}_{-0.21}$, corresponding to $L_{UV} = 1.58^{+1.09}_{-0.60} \times 10^{43}$ erg s$^{-1}$), in this work we present the $\xi_{ion,0}$ values uncorrected for dust in  Table~\ref{tab:LAEs_derived_properties}. We show both corrected (crosses) and uncorrected (circles) $\xi_{ion,0}$ estimations in Figure~\ref{fig:xi_ion}, and note that if a Calzetti law is applied, $\xi_{ion,0}$ is reduced by $\sim 0.1$ dex for the few cases where dust is non-negligible.

We find that the median value of $\xi_{ion,0}$ (assuming negligible dust correction for \ha\ and the UV continuum) for our sample is log$\frac{\xi_{ion,0}}{\text{Hz erg}^{-1}} = 25.44^{+0.21}_{-0.15}$. The individual estimations can be found in Table~\ref{tab:LAEs_derived_properties}. This value is in broad agreement with those found for other LAEs at similar redshifts \citep[e.g.][]
{Ning2023, Prieto-Lyon2023}. A compilation of results from the literature is given in \cite{Stefanon2022}, in particular, they provide a linear fit (in logarithmic space) of the $\xi_{ion}$ observations up to $z = 9$ from  
\cite{Stark2015,Marmol-Queralto2016,Bouwens2016,Nakajima2016,Matthee2017,Stark2017,Harikane2018,Shivaei2018,DeBarros2019,Faisst2019,Lam2019,Tang2019,Emami2020,Nanayakkara2020,Endsley2021,Atek2022,Naidu2020}. This extensive compilation illustrates a clear trend of increasing $\xi_{ion,0}$ with redshift \citep[see Figure 4 of ][]{Stefanon2022}.  As shown in Figure~\ref{fig:xi_ion}, our estimations scatter around this relation. It must be noted that this relation is not specific to LAEs, and is mostly tracing the ionising photon production of LBGs. Thus, the increased ionising photon production in a few of our sources is likely linked to the population being studied. \cite{Matthee2022eiger} investigate this further by comparing $\xi_{ion}$ for different galaxy populations, finding non-LAEs tend to have lower $\xi_{ion}$ than LAEs. The ionising photon production efficiency has not been extensively studied in the redshift range of this work, for comparison, we show the median value obtained for 22 massive galaxies at $z \sim 7$ from \citep[black diamond; ][]{Endsley2021}, and the 7 LAEs from \cite[stars; ][]{Ning2023}.

\begin{figure}
        \centering
   \includegraphics[width=0.5\textwidth]{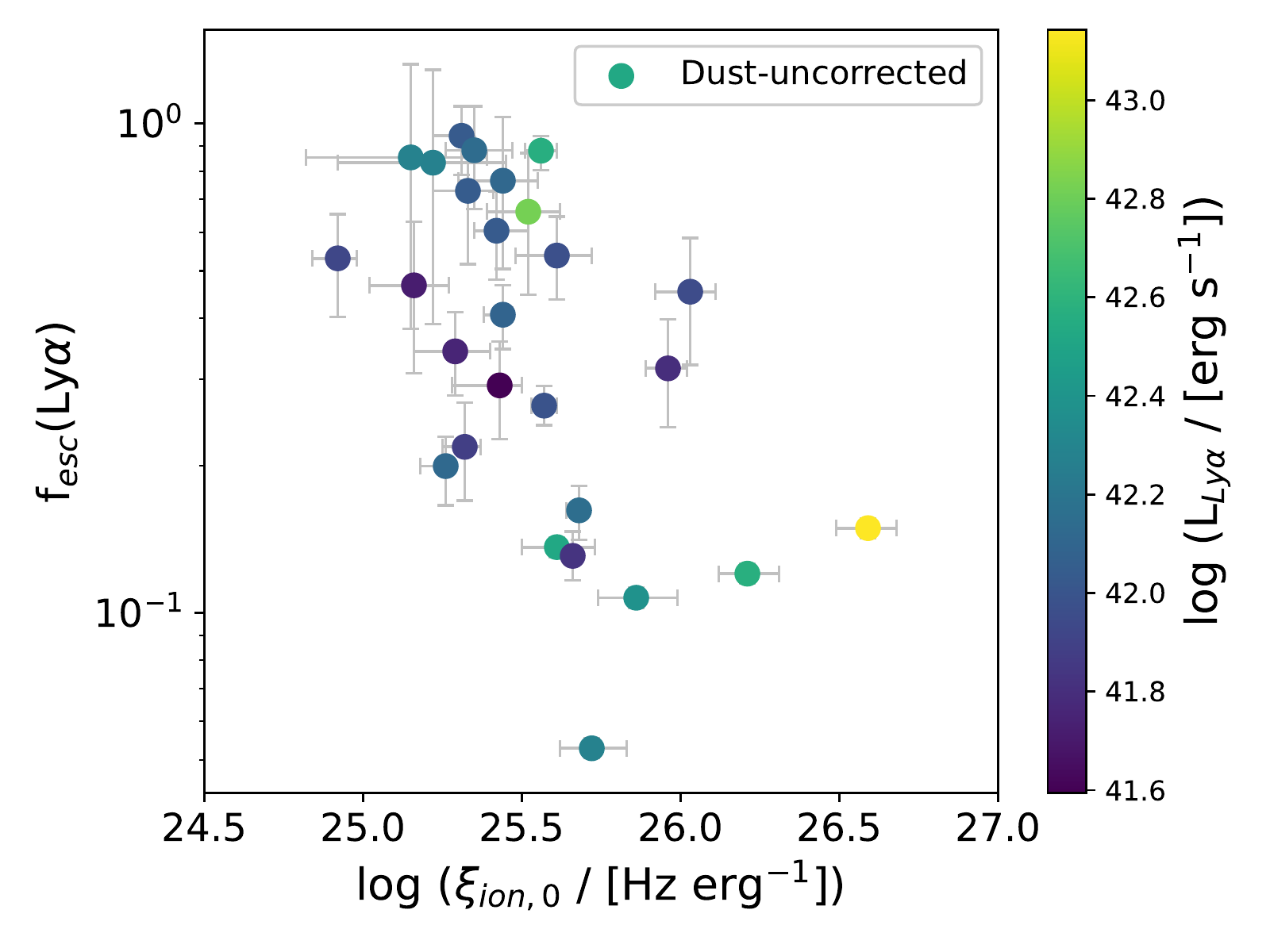}
   \caption{\lya\ escape fractions versus $\xi_{ion,0}$ for our sample, colour-coded by \lya\ luminosity. The LAEs where lower limit measurements of \ha\ fluxes were provided have been excluded. There is a tentative anti-correlation between f$_{\rm{esc}}$(\lya\/) and $\xi_{ion,0}$ for our sample, which we mainly attribute to: (1) the porosity of the ISM and, (2) a time delay between the creation of ionising photons and their escape.}
              \label{fig:xi_Lya}%
    \end{figure}

Figure~\ref{fig:xi_Lya} shows a comparison of $\xi_{ion,0}$ and \lya\ escape fractions, excluding the cases where the measured \ha\ flux is a lower limit (i.e. JADES-GS+53.16674-27.80424, JADES-GS+53.17829-27.77728 and JADES-GS+53.18744-27.77804). We observe a large variation of escape fractions for a given $\xi_{ion,0}$, with a tentative anti-correlation between f$_{\rm{esc}}$(\lya\/) and $\xi_{ion,0}$. The f$_{\rm{esc}}$(\lya\/) values are in agreement with other studies at similar redshifts \citep[][Saxena in prep.]{Ning2023,Tang2023,Jung2023}.

\section{Discussion}
Studying $\xi_{ion,0}$ at the EoR is interesting to understand the escape fractions required for galaxies to be the main sources responsible for the reionisation of the Universe. The redshift range investigated here ($z \sim 5.4 - 6.6$) has not been extensively studied in the past. This study has been made possible due to the deep imaging provided by NIRCam onboard JWST, in combination with the ground-based multi-unit spectrograph MUSE. The latter provided reliable systemic redshifts, that were used to find \ha\ through flux excesses in the JEMS bands F430M, F460M and F480M. 

One of the caveats in this work arises from the \ha\ measurements. As discussed in Section~\ref{section:nii_contamination}, we expect these galaxies to be metal poor, and thus, that contamination in the JEMS filters from other emission lines (especially \nii\/) is negligible. This might not be the case for all the objects in our sample, for example, \cite{Killi2022} suggest evidence of a solar metallicity galaxy at $z \sim 7$. Thus, it is possible that some of the \ha\ luminosities here provided are upper limits. The \ha\ luminosities directly relate to the ionising photon production through equation~\ref{eq:xi_ion}, where the conversion between \ha\ luminosity and ionising photons produced ($N(H^0)$) assumes Case B recombination. In other words, an escape fraction of zero since all the ionising photons produced are being used to power the nebular emission. This assumption is adopted for simplicity, but it has to be noted that we expect non-zero escape fractions for these objects.

A second important point to keep in mind is the dust corrections applied to the \ha\ emission line and the UV stellar continuum at rest-frame $\lambda = 1500$ \AA\/. Attenuation laws are highly uncertain at high redshift, as well as how much their relative contribution is to nebular lines versus to continuum attenuation. We show both our corrected and uncorrected results based on this uncertainty, we note that the blue slopes found with \texttt{Prospector} indicate young objects with little-to-no dust attenuation ($\beta \sim -2.2$). In studies such as \cite{Bouwens2016}, galaxies with UV slopes of $\beta < -2.23$ are left uncorrected for dust \citep[$A_{UV} = 0$ according to a ][law]{Calzett2000}. This is a reasonable approach since a steep $\beta$ slope indicates a young stellar population with low dust attenuation.

We find slightly enhanced ionising photon production efficiency values for some of the LAEs in our sample, in comparison to the average $\xi_{ion,0}$ at the redshift range studied. We again note that after correction for dust attenuation, assuming a Calzetti law and local relations between nebular and continuum attenuation, our values are lowered by $\sim 0.1$ dex in the few cases where $\beta$ suggests non-negligible dust attenuation. However, even after dust correction some of these objects remain above average for their redshift. This is not unheard of in LAEs, for example, \cite{Maseda2020} find enhanced ionising photon production in a sample of 30 LAEs at $z \sim 4-5$, with high \lya\ EW (median value of 250 \AA\/). Moreover, enhanced $\xi_{ion,0}$ values have been observed in LyC leakers at $z \sim 0.3$, thought to be local analogues to high-redshift ($z \sim 7$) galaxies \citep{Schaerer2016}. All these studies point towards LAEs being key to the reionisation of the Universe. Future observations of indirect tracers of LyC escape for this galaxy population at high redshift will allow us to understand how $\xi_{ion}$ relates to LyC (and \lya\/) escape. Regarding the latter, we find a tentative anti-correlation between f$_{\rm{esc}}$(\lya\/) and $\xi_{ion,0}$, which has two interesting implications. Firstly, the escape of \lya\ photons has been shown to have a similar behaviour with \lya\ peak separations \citep[e.g.][]{Gazagnes2020}, which are related to the porosity of the ISM. A more porous ISM favours low density channels through which \lya\ (and LyC) photons can escape. Secondly, the fact that a higher ionising photon production is not linked to higher \lya\ escape fractions, points toward a time delay between the production and (potentially) escape of ionising photons, in agreement with simulations \citep{Barrow2020,Katz2020}. This time delay is related to the creation of low-density channels in galaxies (for example, through SNe feedback).

Somewhat interestingly, we do not find any significant correlation between $\xi_{ion,0}$ and the galactic properties derived with \texttt{Prospector} (table provided in Appendix~\ref{appendix:properties}), however, the median values for our sample indicate it is dominated by metal-poor (log$_{10}$(Z/Z$_{\odot}$) $= -2.2^{+0.4}_{-0.4}$), low-mass (log$_{10}$(M$_*$/M$_{\odot}$) $= 8.3^{+0.7}_{-0.6}$) galaxies. We find that the lookback time at which half of their mass was assembled is given by $t_{50} = 0.2^{+0.1}_{-0.1}$ Gyr, and that they have relatively high ionisation parameters (log<U> $= -2.4^{+0.4}_{-0.3}$). Interestingly, the LAE with the highest $\xi_{ion,0}$, JADES-GS+53.13859-27.79024, also has the highest stellar mass (log$_{10}$(M$_*$/M$_{\odot}$) $\sim 10$), shown as the circle with a red edge in Figure~\ref{fig:xi_ion}. This LAE was flagged as an AGN candidate due to its high stellar mass in \cite{Helton2023}. Moreover, slitless spectroscopy reveals broad \ha\ emission, indicative of AGN activity. This object will be discussed in detail in a future work.

The redshift distribution of the sources in this work is shown on the top panel of Figure~\ref{fig:xi_ion}. Given the small footprint of the MUSE coverage, groups of sources close in redshift space suggest clustering. In particular, we can distinguish two such groups, at $z \sim 5.4$ and $z \sim 5.9$, respectively. The former was identified and presented in \cite{Helton2023}. It is not surprising that the LAEs we observe at the tail end of the EoR seem to be clustered, since there is growing evidence for these kind of objects to lie in overdense environments \citep[e.g. ][]{Jung2020,Tang2023}. Moreover, a recent study by \cite{Witten2023}, which combines the power of simulations with observational data, finds that LAEs deep into the EoR necessarily reside in clustered environments. The clustering acts as a mechanism that aids in the creation of ionised bubbles, which favour \lya\ escape as the IGM becomes more opaque. It is possible that the environment of these LAEs also has an effect on $\xi_{ion,0}$, but further studies are needed in order to draw solid conclusions.

Finally, we call attention to the biased nature of our sample. The IGM transmission makes \lya\ detection at high redshifts a difficult task. Here we study the brightest LAEs at the end of the EoR, in other words, they are bright enough that we can reliably identify the \lya\ emission line, which is later used to infer systemic redshifts. LAEs have been shown to be LyC leakers \cite{Gazagnes2020,Nakajima2020,Izotov2022}, making them interesting objects to analyse. Our sample is not necessarily representative of the entire galaxy population at the EoR; this is a first exploratory work to showcase the science that can be performed using JEMS. However, as previously mentioned, there is increasing evidence that LAEs might be key drivers of the reionisation of the Universe. A follow-up study in the future will focus on a more complete sample using the entire JADES database.  Of particular interest, a recent study by \cite{Ning2023} focuses on the same class of objects at $z \sim 6$, shown as black arrows in Figure~\ref{fig:xi_ion}. Their sample was constructed by selecting spectroscopically confirmed LAEs observed with the fiber-fed, multi-object spectrograph Michigan/Magellan Fiber System \citep[M2FS; ][]{Mateo2012}, in a program that aims to build a sample of high-z galaxies \citep{Jiang2017}. Therefore, even though it was built in an analogous way to the sample presented here, it differs in the instrumentation used. Promisingly, our findings are broadly consistent with theirs.

\section{Summary and conclusions}
We study a sample of 30 bright LAEs at the end of the EoR ($z \sim 5.4-6.6$), with spectroscopic redshifts provided by the MUSE team \citep{Bacon2022}. The \ha\ recombination line is shifted into wavelengths observed by JEMS in this redshift range. We aim to retrieve the \ha\ intrinsic luminosity using a combination of the F430M, F460M and F480M bands, while using Prospector to fit JEMS and selected HST photometry to constrain the UV continuum slope. We finally use our measurements to estimate the ionising photon production efficiency $\xi_{ion,0}$. The blue UV continuum slopes of our sample indicate little-to-no dust attenuation. Additionally, due to the uncertainties in dust corrections of nebular emission lines and stellar continuum, we present uncorrected values in Table~\ref{tab:LAEs_derived_properties}. In summary
\begin{itemize}
    \item We use photoionisation models to quantify the importance of \nii\ contamination in our photometric bands, and find that this effect would only be significant for cases where log<U> $\leq -3.0$, which is not expected from galaxies at $z \sim 6$ \citep{Sugahara2022}. Moreover, \texttt{Prospector} predicts the median log<U> of the sample to be log<U> $= -2.3^{+0.2}_{-0.3}$.
    \item We divide our sample into four redshift bins and estimate the \ha\ flux using combinations of JEMS and find the median luminosity of the sample is log$\frac{L(\text{\ha\/})}{\text{erg s}^{-1}}$ $= 41.41^{+0.44}_{-0.15}$
    \item We use \texttt{Prospector} to fit the SEDs of our sample and find blue UV slopes obtained by fitting a straight line in the rest-frame $\lambda = 1250 - 2600$ \AA\/ spectral region. The median of the sample is given by $\beta = -2.09^{+0.23}_{-0.21}$, and is consistent with young stellar populations with little-to-no dust. We use the fitted lines to estimate the monochromatic luminosity density at rest-frame 1500 \AA\
    \item We use our measurements to estimate the ionising photon production efficiency assuming no dust attenuation and find log$\frac{\xi_{ion,0}}{\text{Hz erg}^{-1}} = 25.44^{+0.21}_{-0.15}$. If a Calzetti attenuation law is assumed instead, with local relations between nebular and continuum relative attenuation, this value is reduced by $\sim 0.1$ dex for the few cases where dust attenuation is expected to be non-negligible. Our $\xi_{ion,0}$ estimations are in broad agreement with other LAEs at similar redshifts \citep{Ning2023}, and are slightly enhanced compared to their expected value based on all the measurements from literature (from $z \sim 0 - 9$). The highest $\xi_{ion,0}$ is found in JADES-GS+53.13859-27.79024, which has the highest \lya\ emission and shows indications of possible AGN activity, this galaxy will be analysed in detail in a future work. 
\end{itemize}
In this work we focus on LAEs, making us biased towards higher $\xi_{ion,0}$ values. This is an interesting galaxy population to study because LAEs are likely to be the main drivers of the reionisation of the Universe. In the future we will produce a similar study on a larger sample, through a stellar mass selection in the JADES data.

\section*{Data Availability Statement}
The data underlying this article will be shared on reasonable request to the corresponding author.

\section*{Acknowledgements}
These observations are associated with JWST Cycle 1 GO program \#1963. Support for program JWST-GO-1963 was provided in part by NASA through a grant from the Space Telescope Science Institute, which is operated by the Associations of Universities for Research in Astronomy, Incorporated, under NASA contract NAS 5-26555. The authors acknowledge the FRESCO team led by PI Pascal Oesch for developing their observing program with a zero-exclusive-access period. The authors acknowledge use of the lux supercomputer at UC Santa Cruz, funded by NSF MRI grant AST 1828315. The research of CCW is supported by NOIRLab, which is managed by the Association of Universities for Research in Astronomy (AURA) under a cooperative agreement with the National Science Foundation. WB, JW and LS acknowledge support from the ERC Advanced Grant 695671, "QUENCH", and the Fondation MERAC. AJB and AS acknowledge funding from the "FirstGalaxies" Advanced Grant from the European Research Council (ERC) under the European Union’s Horizon 2020 research and innovation programme (Grant agreement No. 789056). BER acknowledges support from the NIRCam Science Team contract to the University of Arizona, NAS5-02015. CW thanks the Science and Technology Facilities Council (STFC) for a PhD studentship, funded by UKRI grant 2602262. ECL acknowledges support of an STFC Webb Fellowship (ST/W001438/1). EE, BDJ and FS acknowledge support from the JWST/NIRCam contract to the University of Arizona NAS5-02015. DJE is supported as a Simons Investigator and by JWST/NIRCam contract to the University of Arizona, NAS5-02015. KB acknowledges support from the Australian Research Council Centre of Excellence for All Sky Astrophysics in 3 Dimensions (ASTRO 3D), through project number CE170100013. R.M. acknowledges support by the Science and Technology Facilities Council (STFC) and by the ERC through Advanced Grant 695671 "QUENCH". RM also acknowledges funding from a research professorship from the Royal Society. H{\"U} gratefully acknowledges support by the Isaac Newton Trust and by the Kavli Foundation through a Newton-Kavli Junior Fellowship. RS acknowledges support from a STFC Ernest Rutherford Fellowship (ST/S004831/1).

\bibliographystyle{mnras}
\bibliography{bib} 



 \appendix

\section{Prospector fits}
\label{appendix}
As in Figure~\ref{fig:bestfit}, we show the \texttt{Prospector} best fit SED and selected photometry for the remainder of the sample, including JEMS $30 \times 30$ pixel$^2$ ($0.9 \times 0.9$ arcsec$^2$) cutouts. For clarity, a Gaussian interpolation has been applied to the cutouts. The information relevant to the UV continuum slope fitting is also provided.

   \begin{figure*}
        \centering
   \includegraphics[width=0.65\textwidth,trim={0 6cm 0 0}]{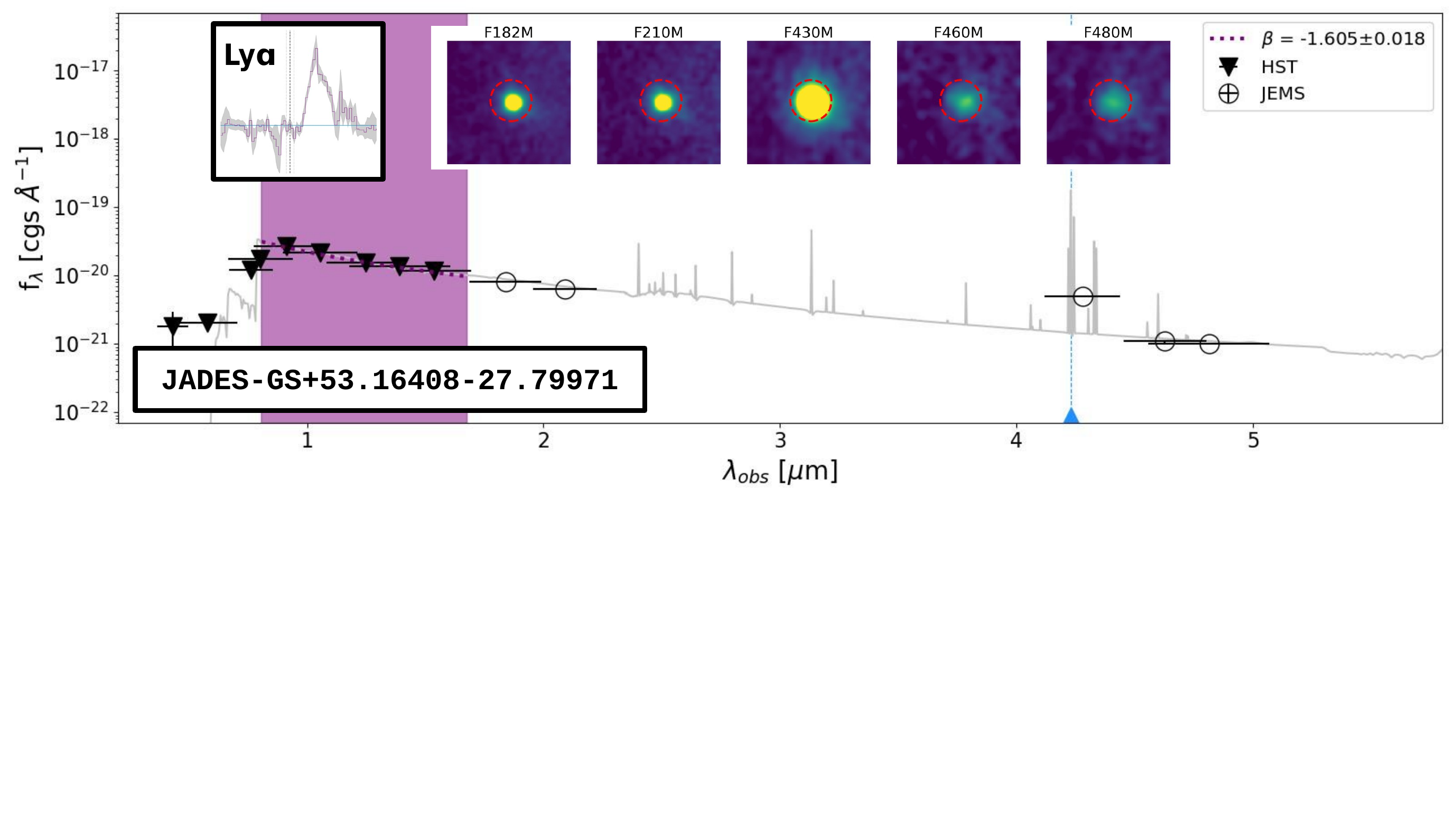}
   \includegraphics[width=0.65\textwidth,trim={0 6cm 0 0}]{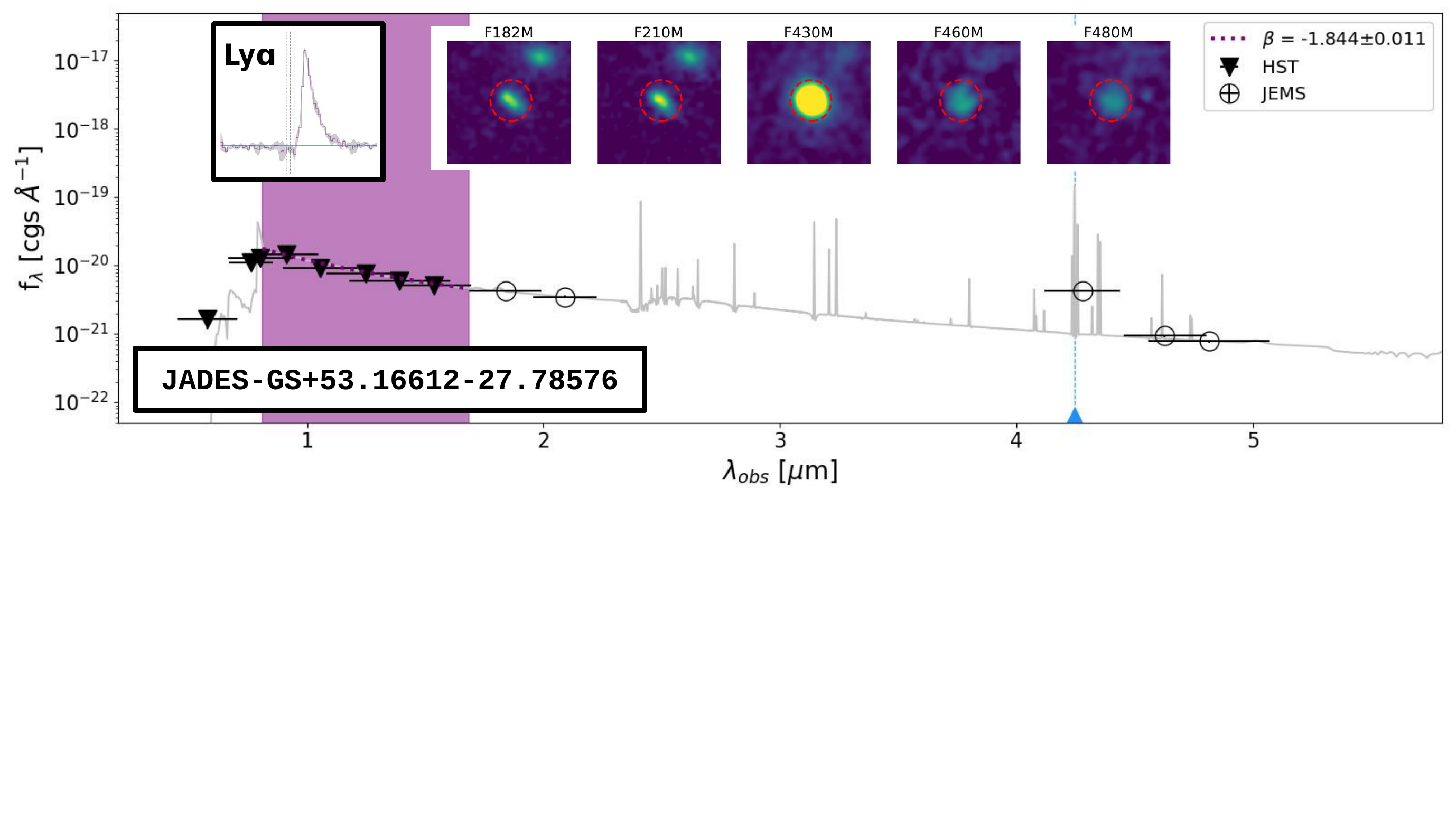}
   \includegraphics[width=0.65\textwidth,trim={0 6cm 0 0}]{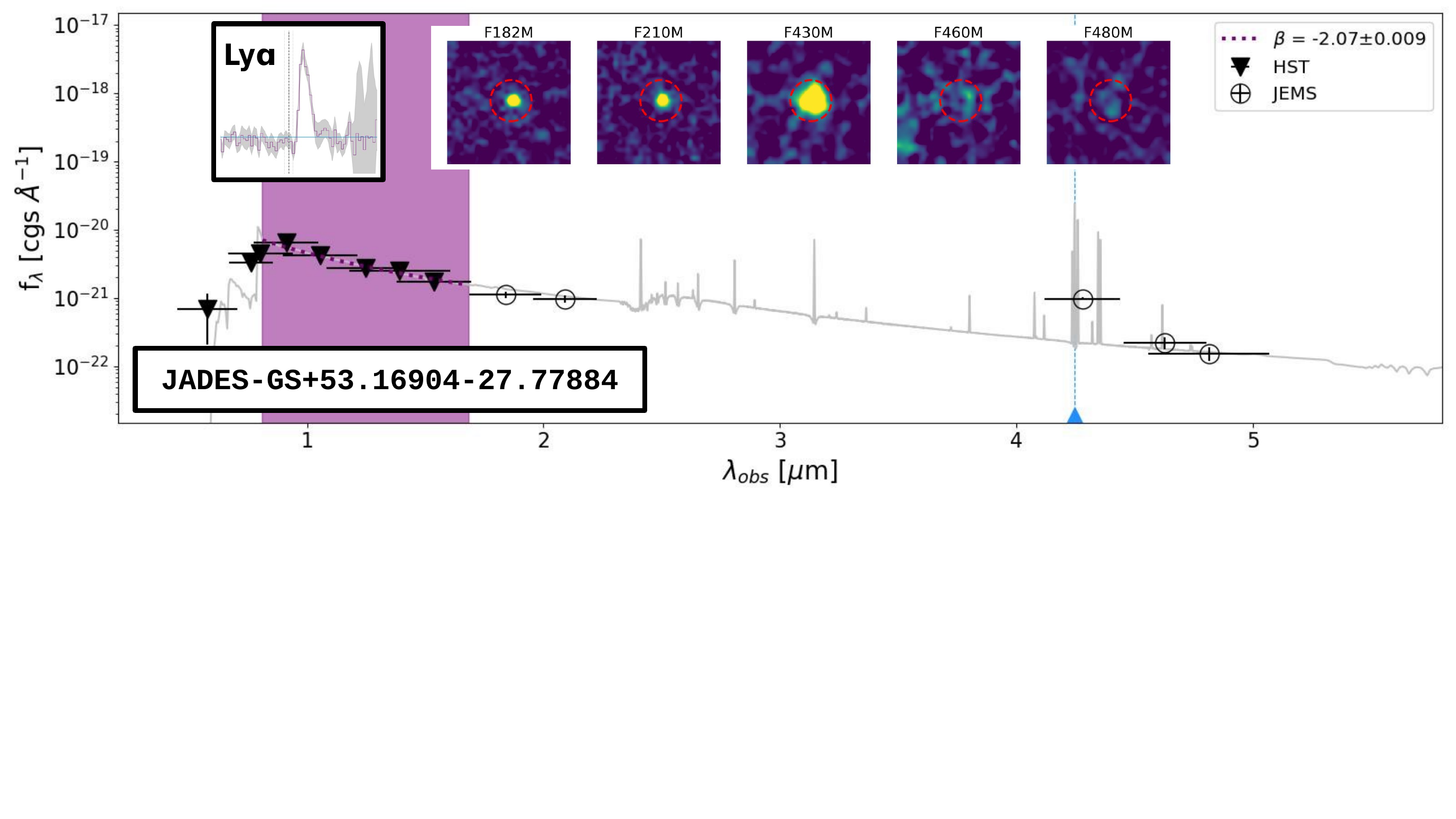}
   \includegraphics[width=0.65\textwidth,trim={0 6cm 0 0}]{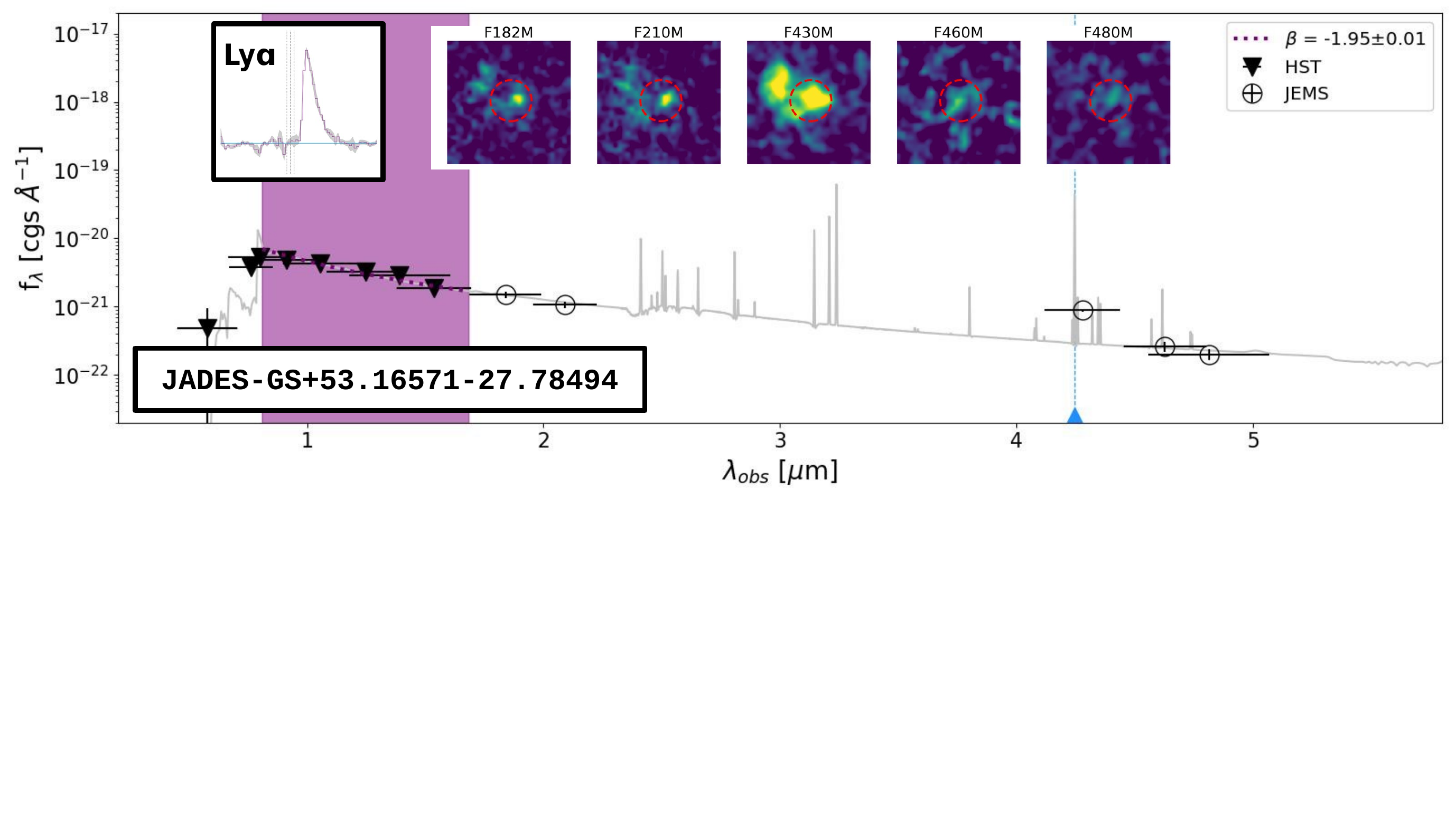}
   \includegraphics[width=0.65\textwidth,trim={0 6cm 0 0}]{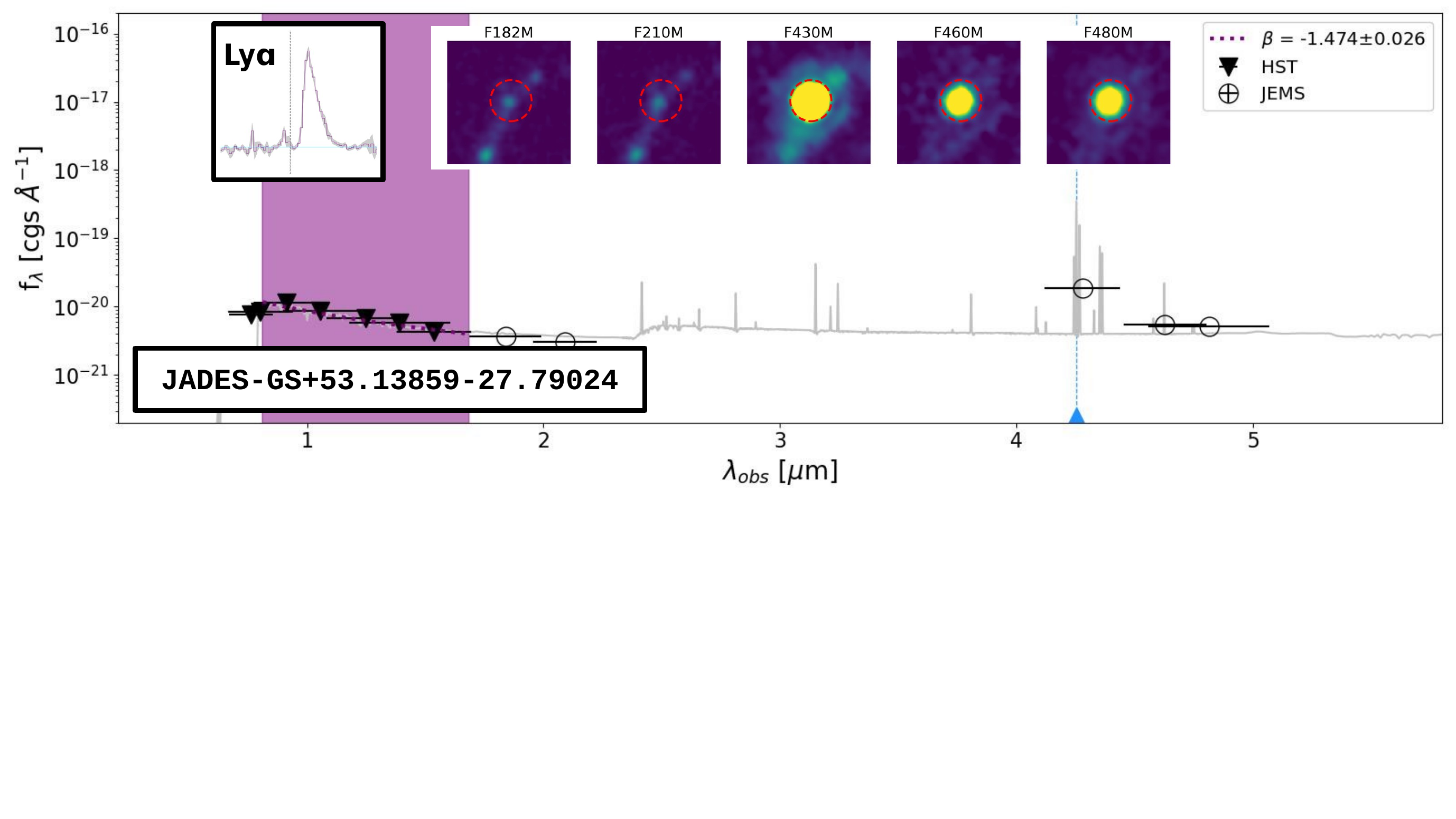}
   \caption{As in figure~\ref{fig:bestfit}, we show the \texttt{Prospector} best fit SED (grey curve) and photometric points (HST; triangles, JEMS; circles) for the remainder of the sample. The JEMS identifier is given in the bottom left of each figure. $30 \times 30$ pixel$^2$ cutouts ($0.9 \times 0.9$ arcsec$^2$) are shown for JEMS, along with the size of the aperture used for the photometry extraction (red dashed circle). The spectral region used for the $\beta$ estimation is shaded in purple. For reference, the shape of the \lya\ profile is also included. The location of the redshifted \ha\ is shown as a dotted vertical line.}
   \label{fig:app}
    \end{figure*}

    \begin{figure*}
        \centering
   \includegraphics[width=0.7\textwidth,trim={0 6cm 0 0}]{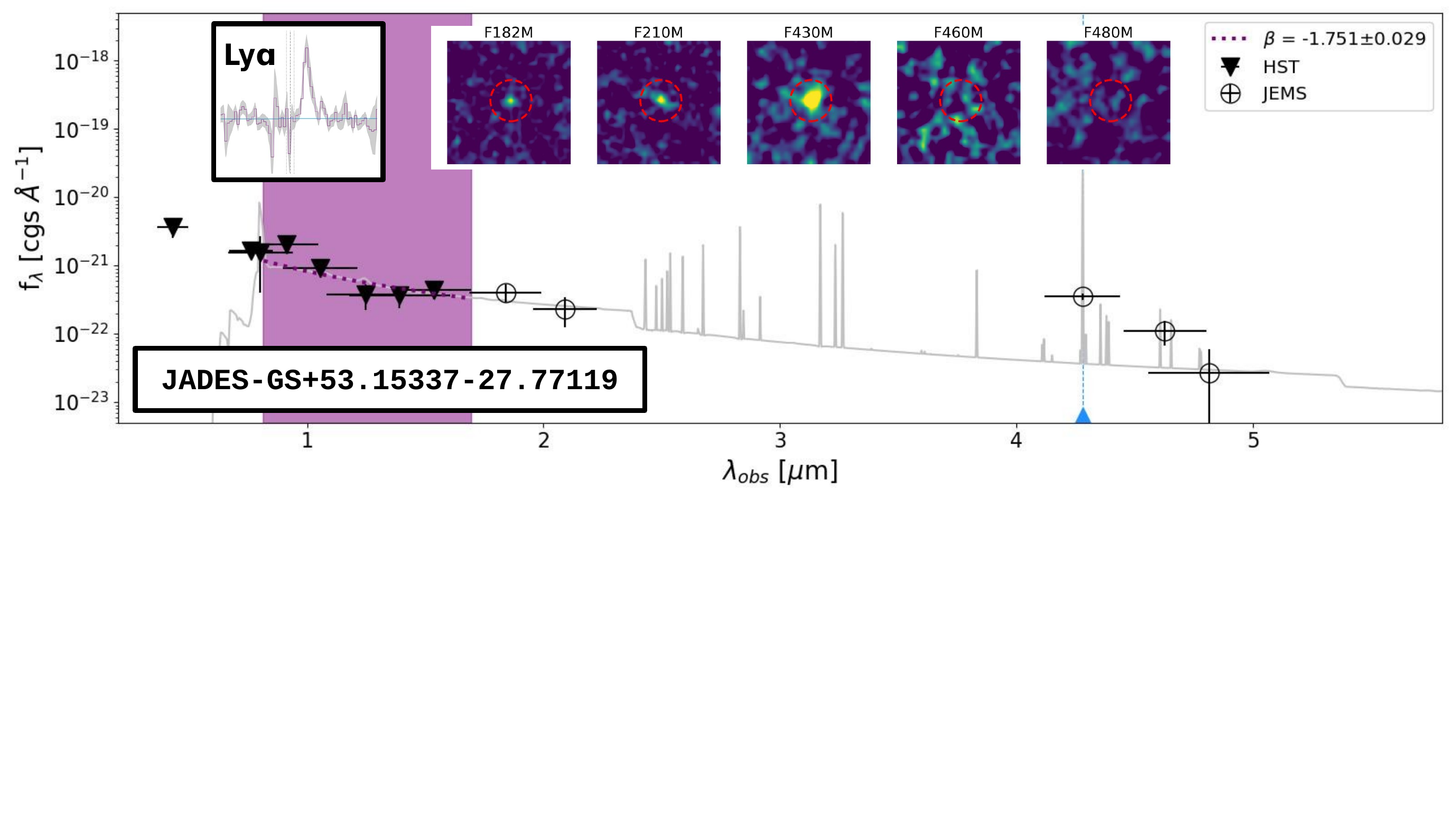}
   \includegraphics[width=0.7\textwidth,trim={0 6cm 0 0}]{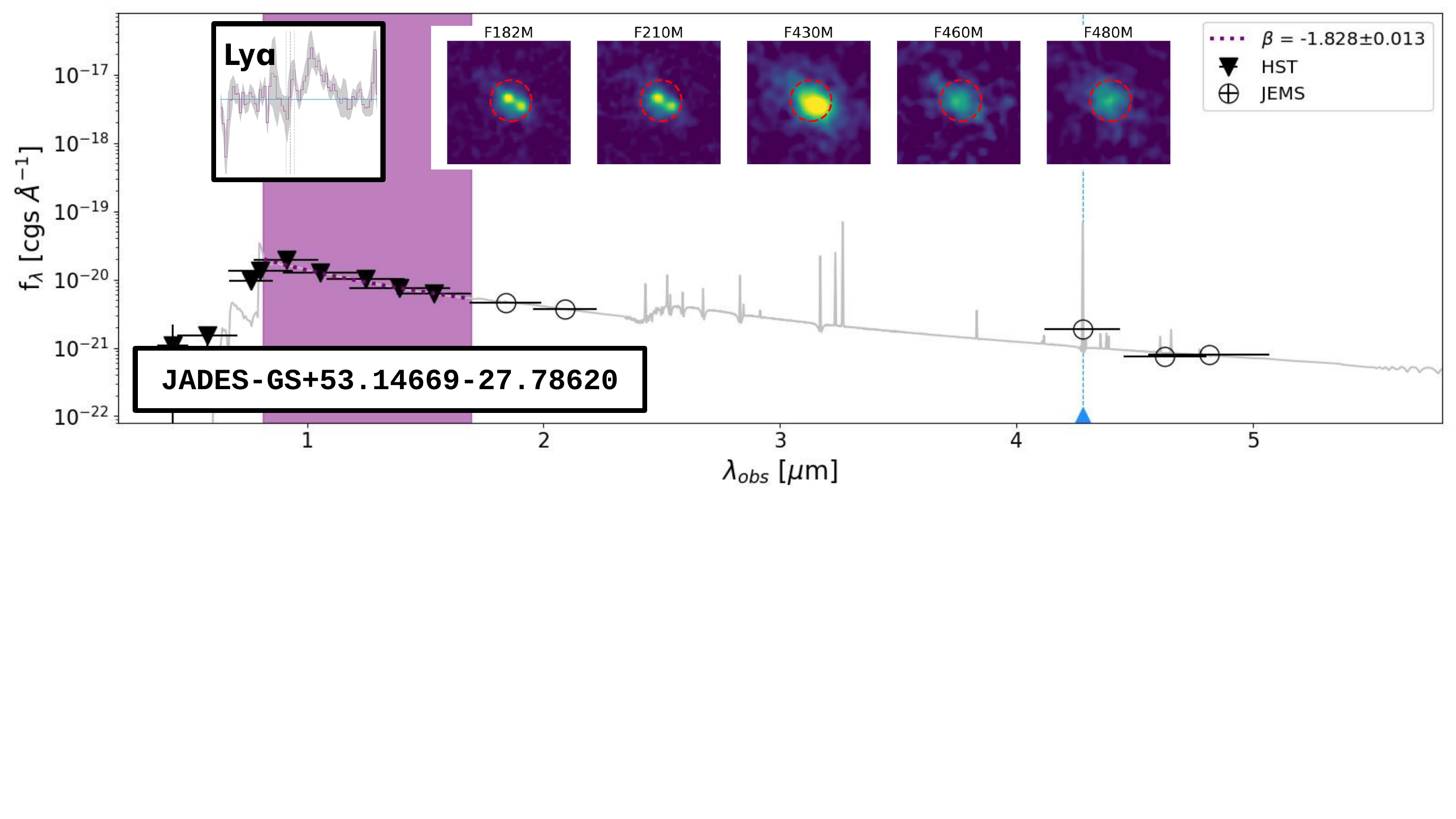}
   \includegraphics[width=0.7\textwidth,trim={0 6cm 0 0}]{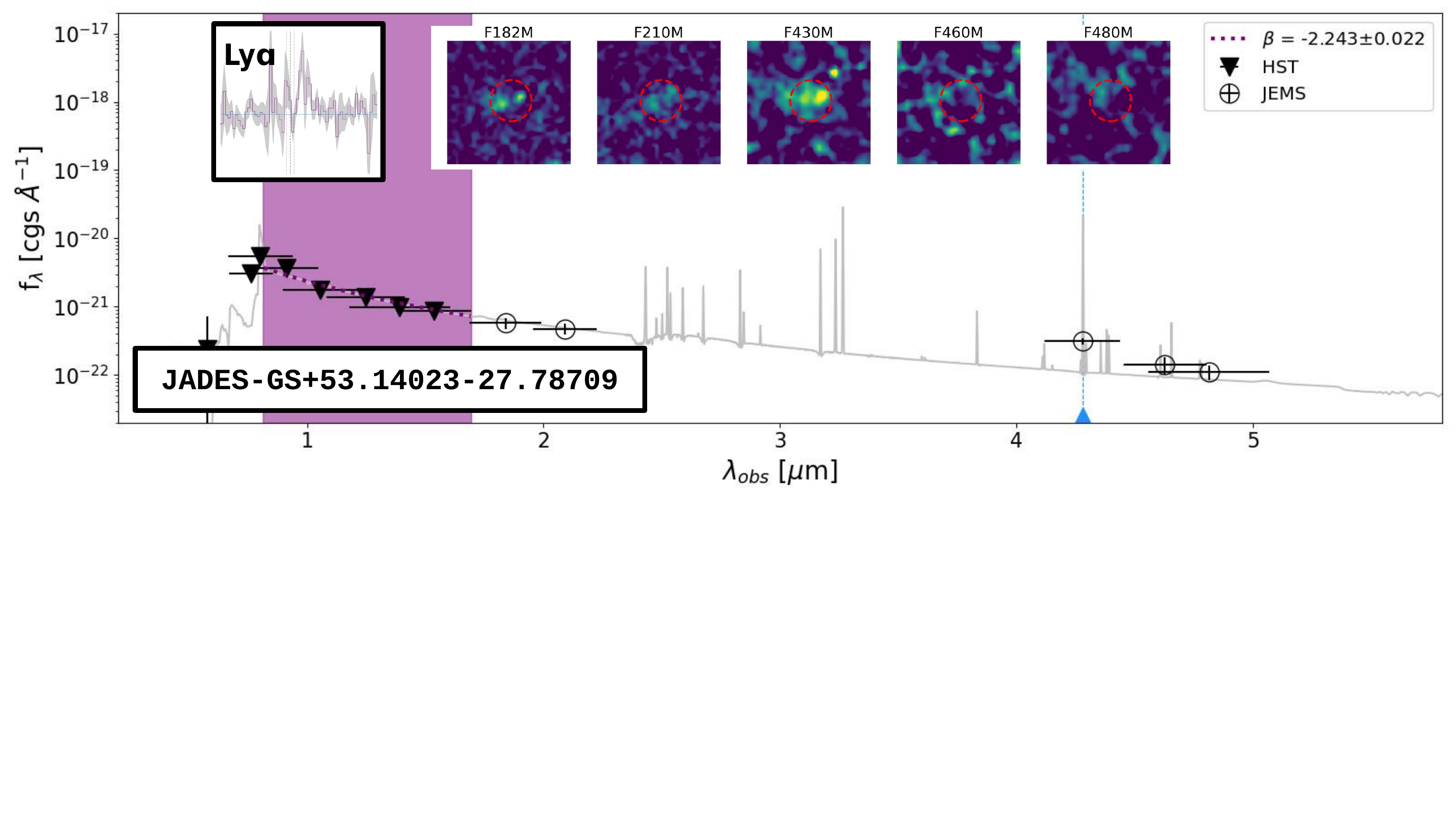}
   \includegraphics[width=0.7\textwidth,trim={0 6cm 0 0}]{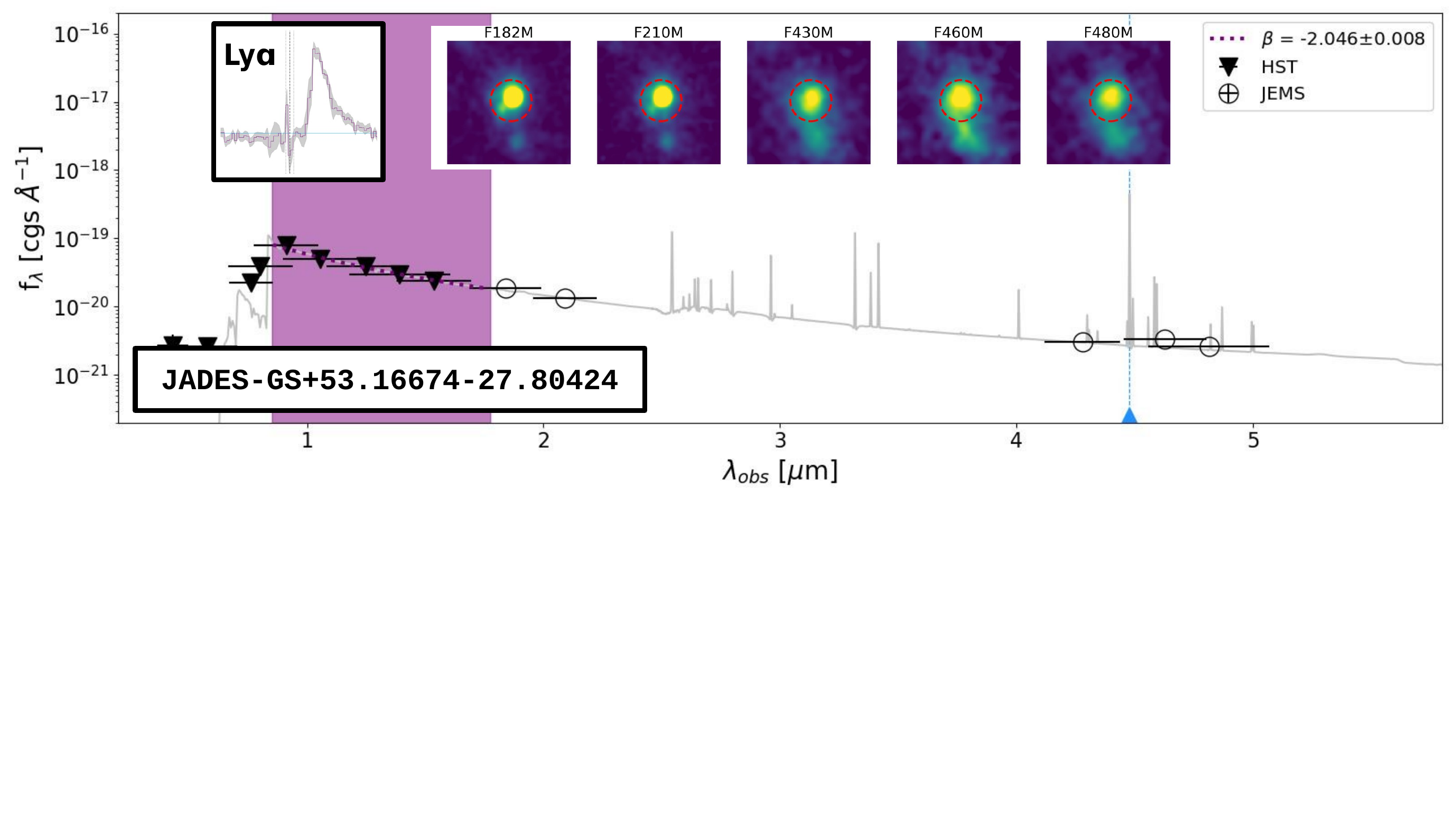}
   \includegraphics[width=0.7\textwidth,trim={0 6cm 0 0}]{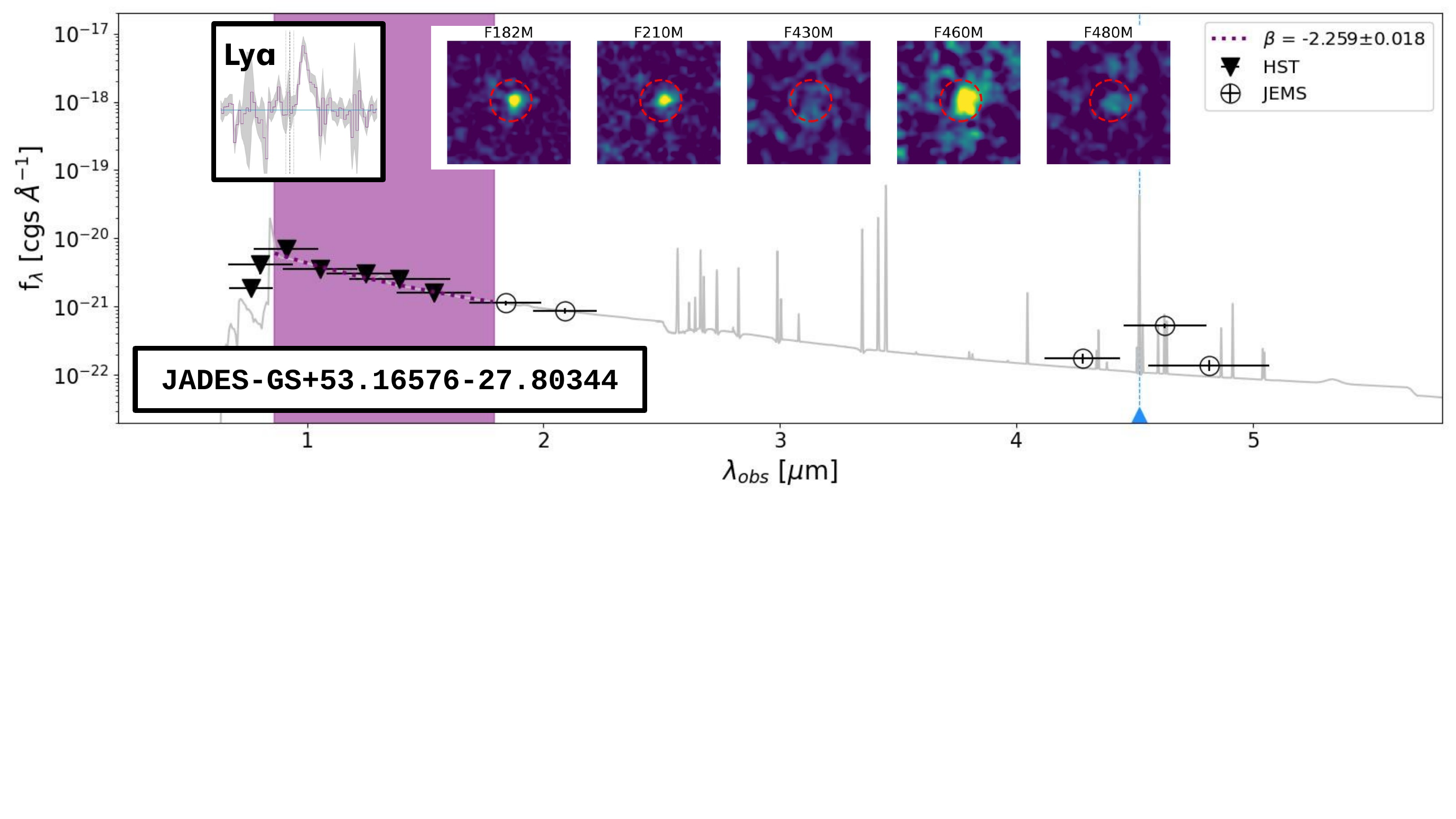}
   \caption{Continuation of Figure~\ref{fig:app}}
    \end{figure*}

    \begin{figure*}
        \centering
   \includegraphics[width=0.7\textwidth,trim={0 6cm 0 0}]{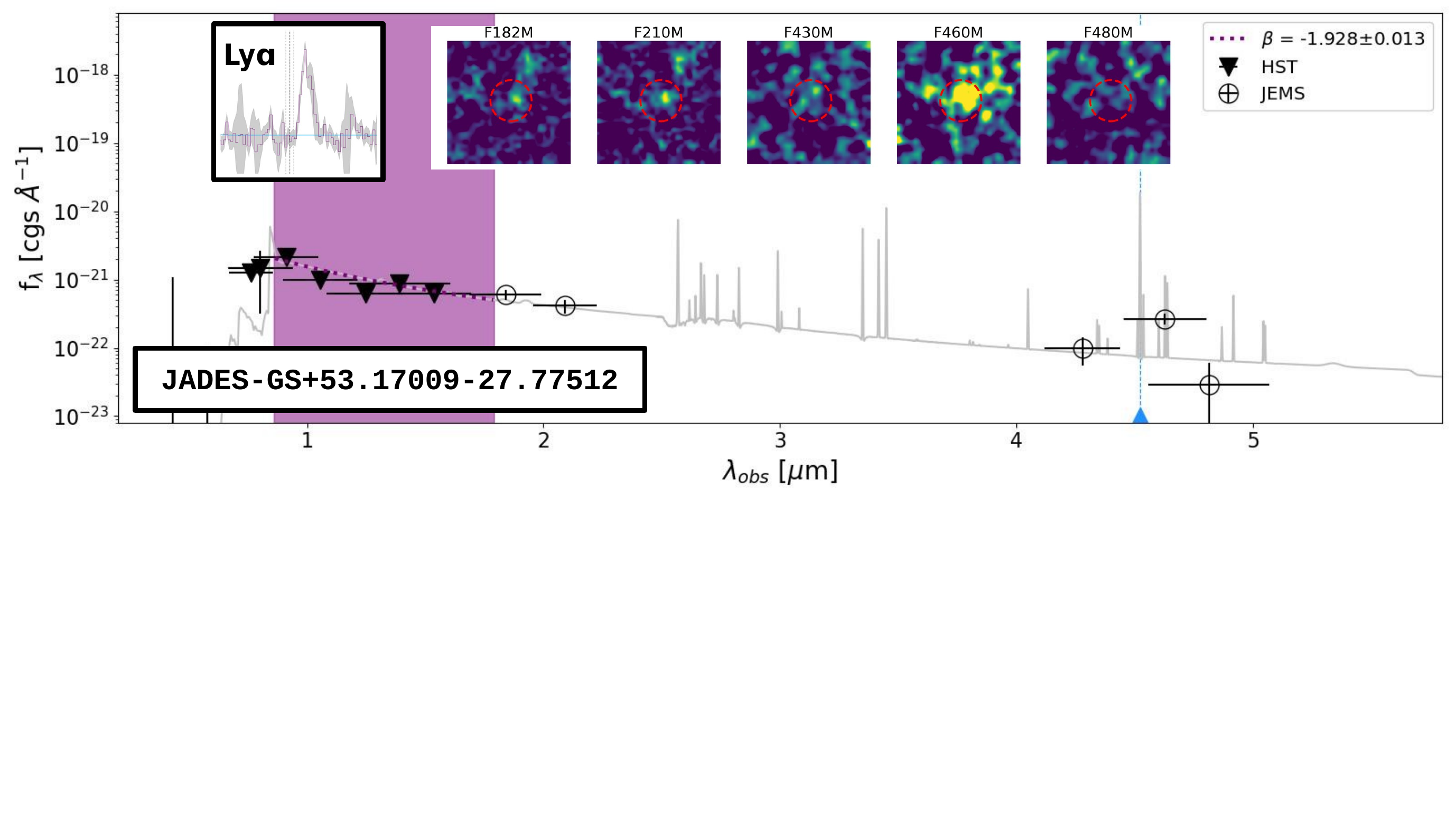}
   \includegraphics[width=0.7\textwidth,trim={0 6cm 0 0}]{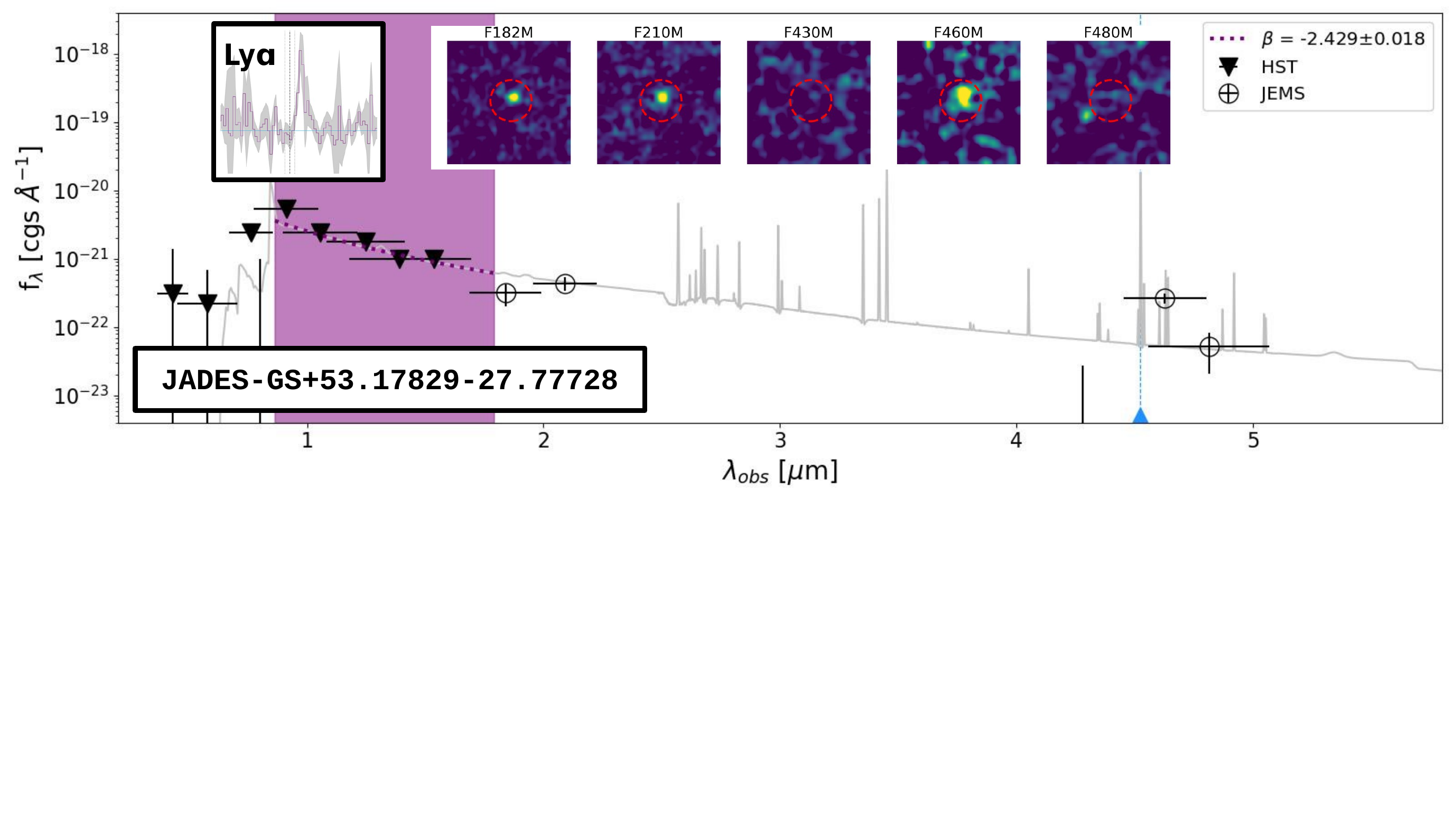}
   \includegraphics[width=0.7\textwidth,trim={0 6cm 0 0}]{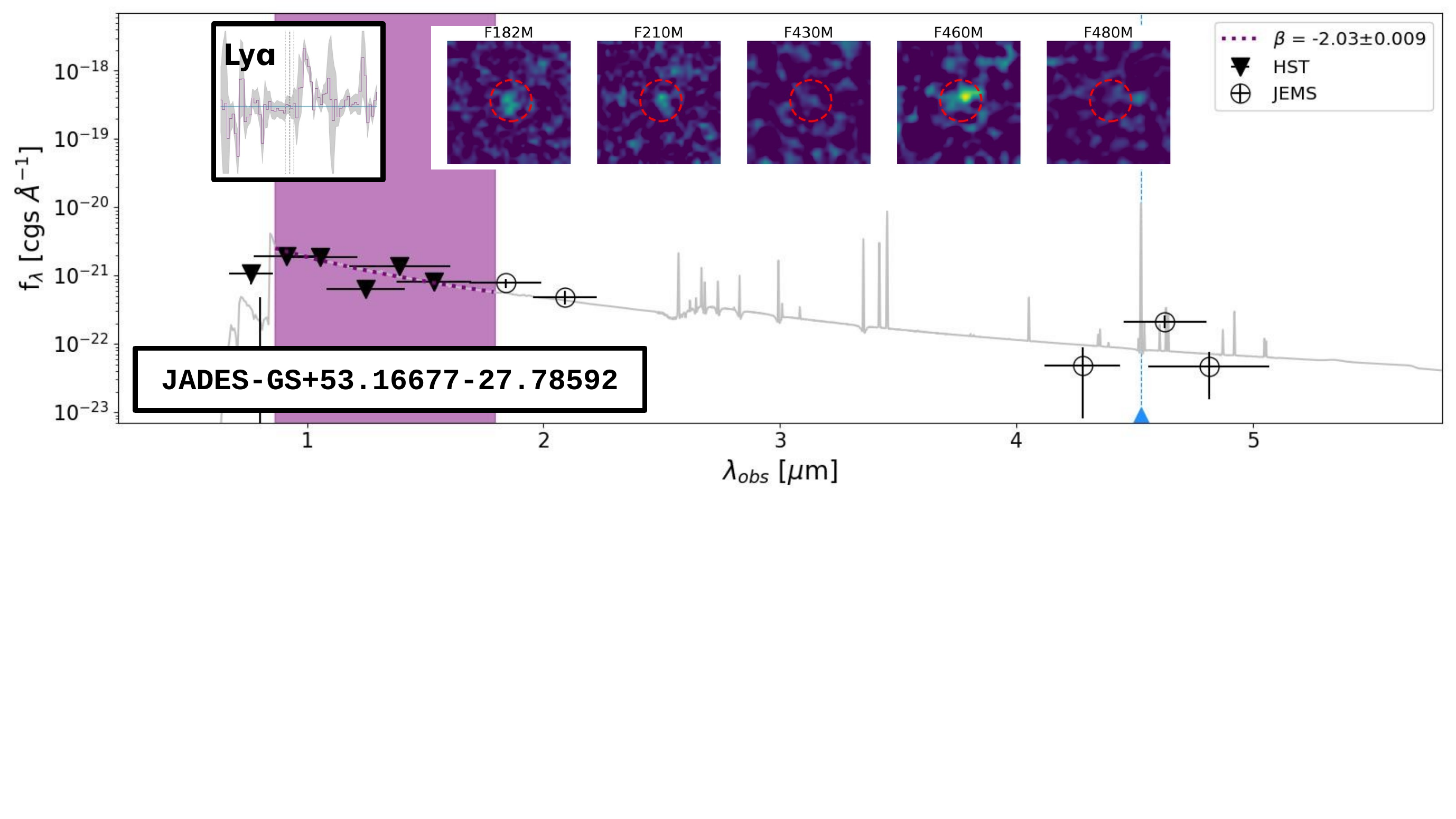}
   \includegraphics[width=0.7\textwidth,trim={0 6cm 0 0}]{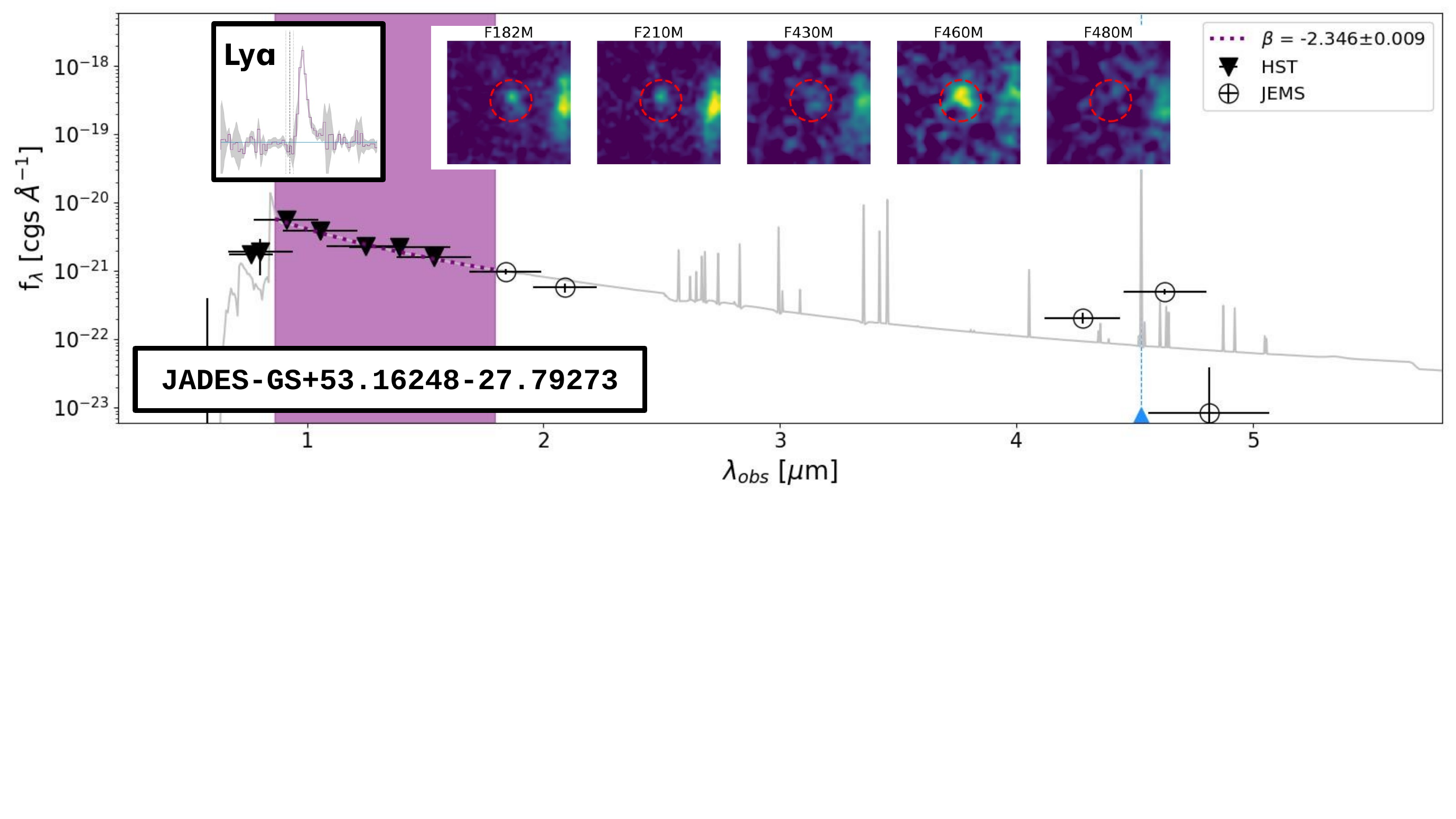}
   \includegraphics[width=0.7\textwidth,trim={0 6cm 0 0}]{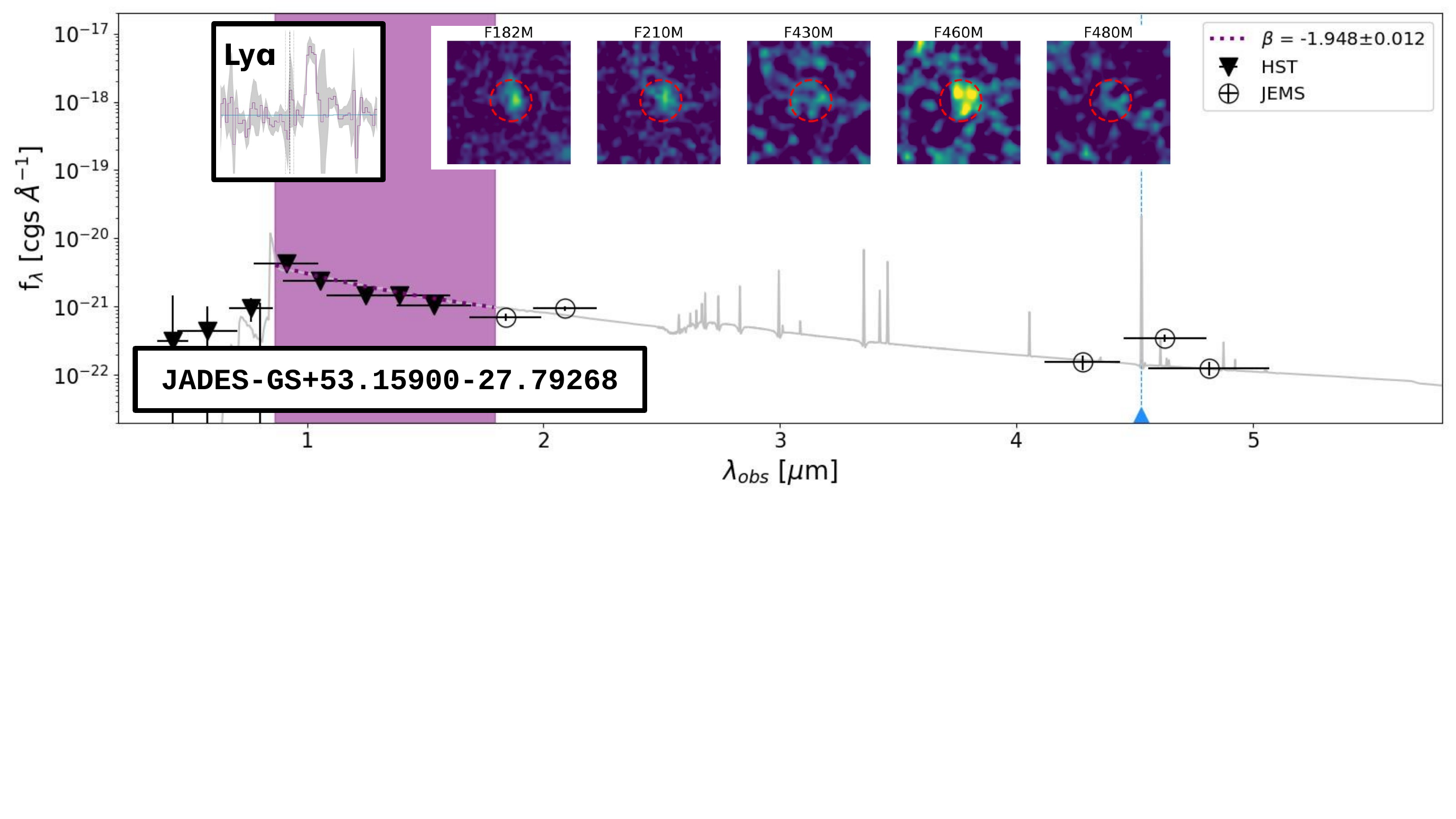}
   \caption{Continuation of Figure~\ref{fig:app}}
    \end{figure*}

    \begin{figure*}
        \centering
   \includegraphics[width=0.7\textwidth,trim={0 6cm 0 0}]{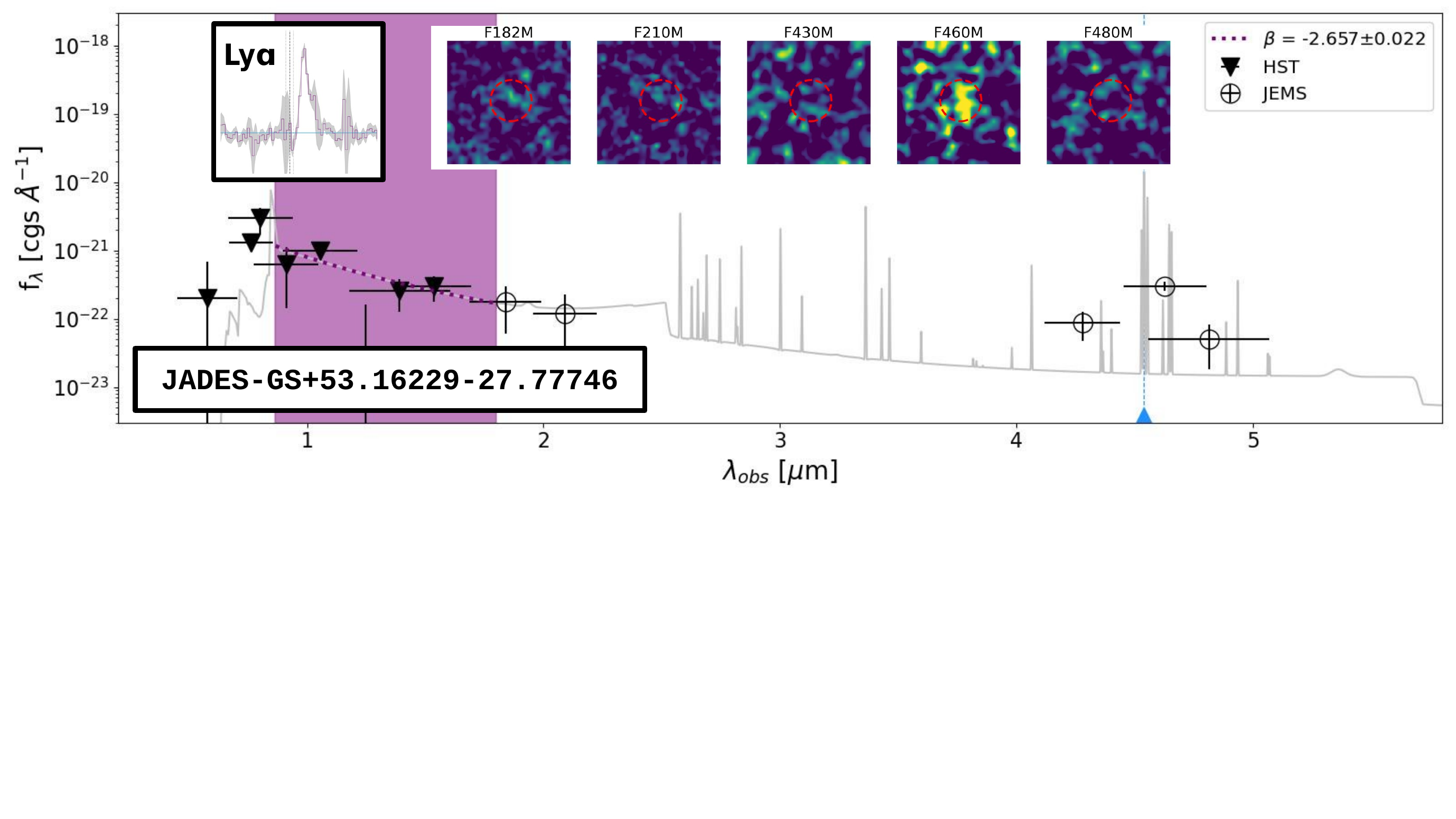}
   \includegraphics[width=0.7\textwidth,trim={0 6cm 0 0}]{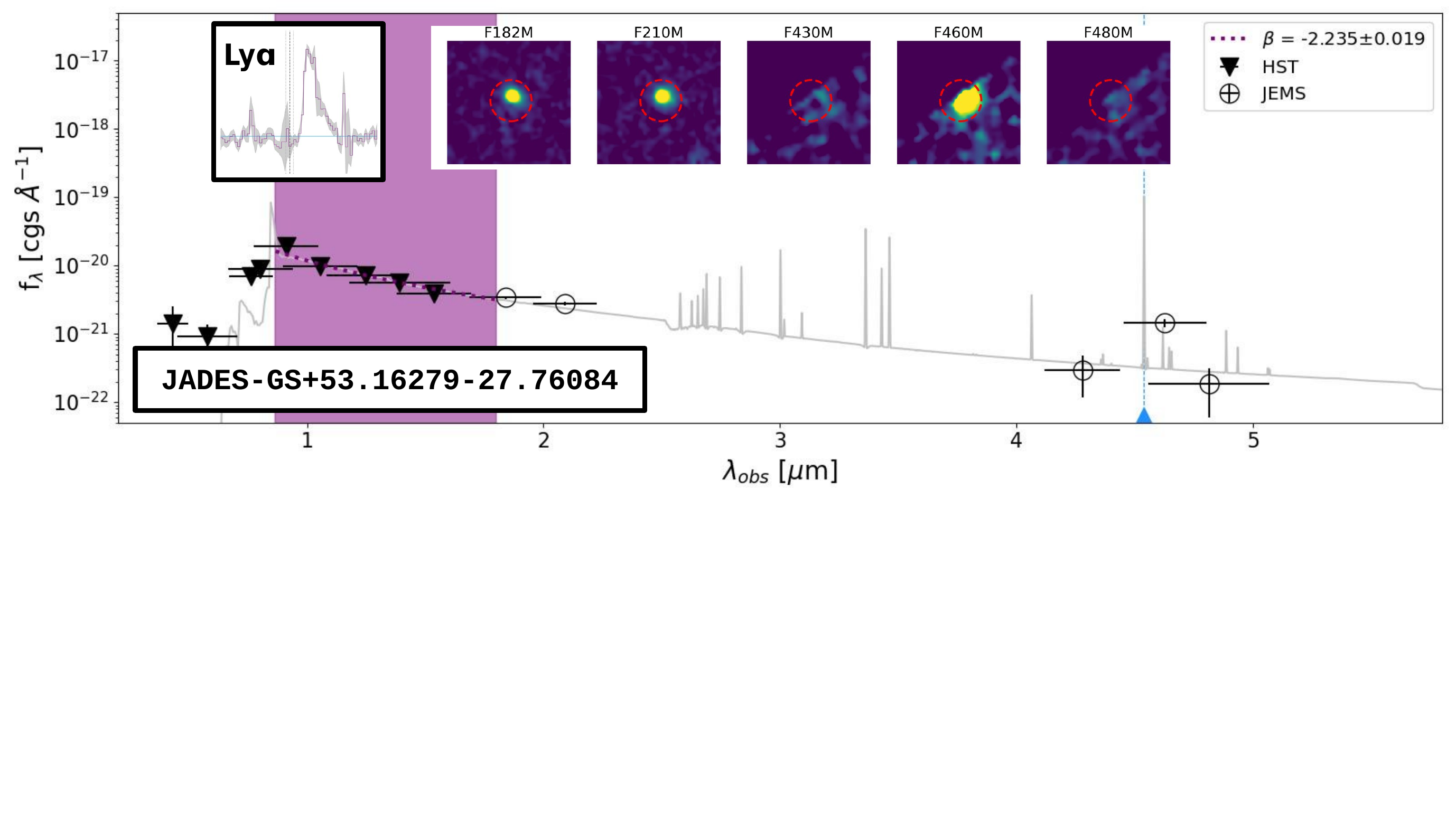}
   \includegraphics[width=0.7\textwidth,trim={0 6cm 0 0}]{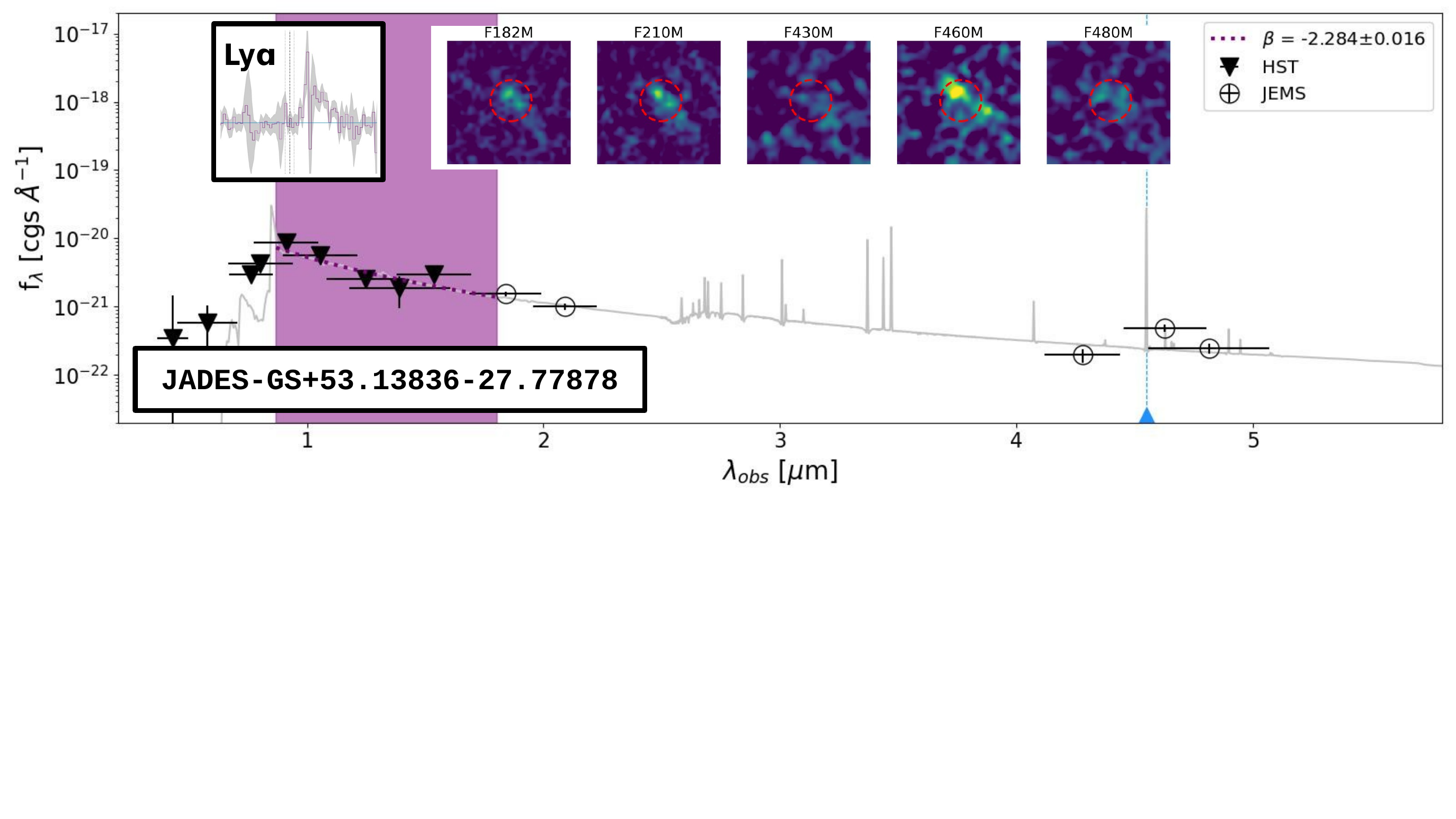}
   \includegraphics[width=0.7\textwidth,trim={0 6cm 0 0}]{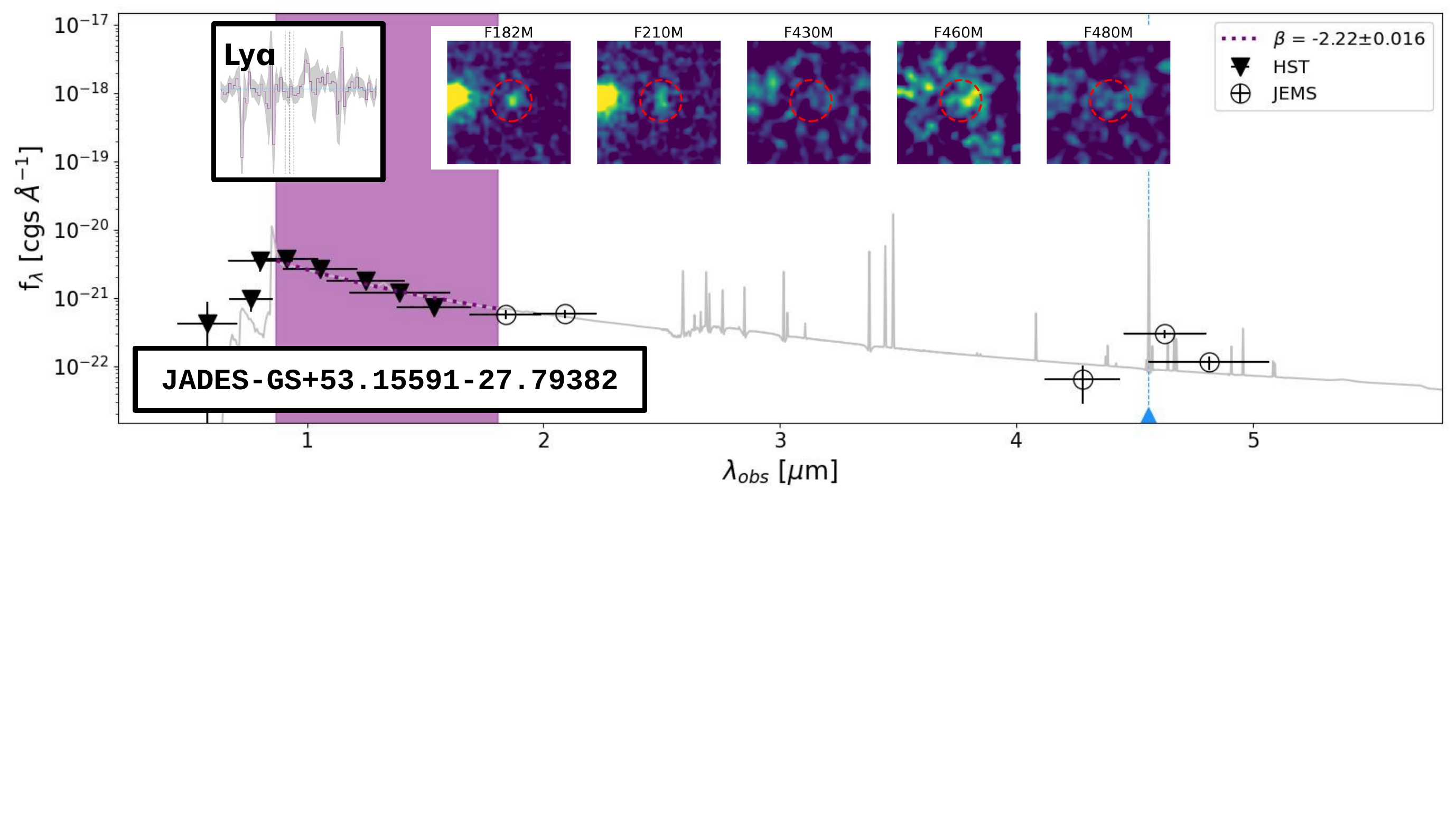}
   \includegraphics[width=0.7\textwidth,trim={0 6cm 0 0}]{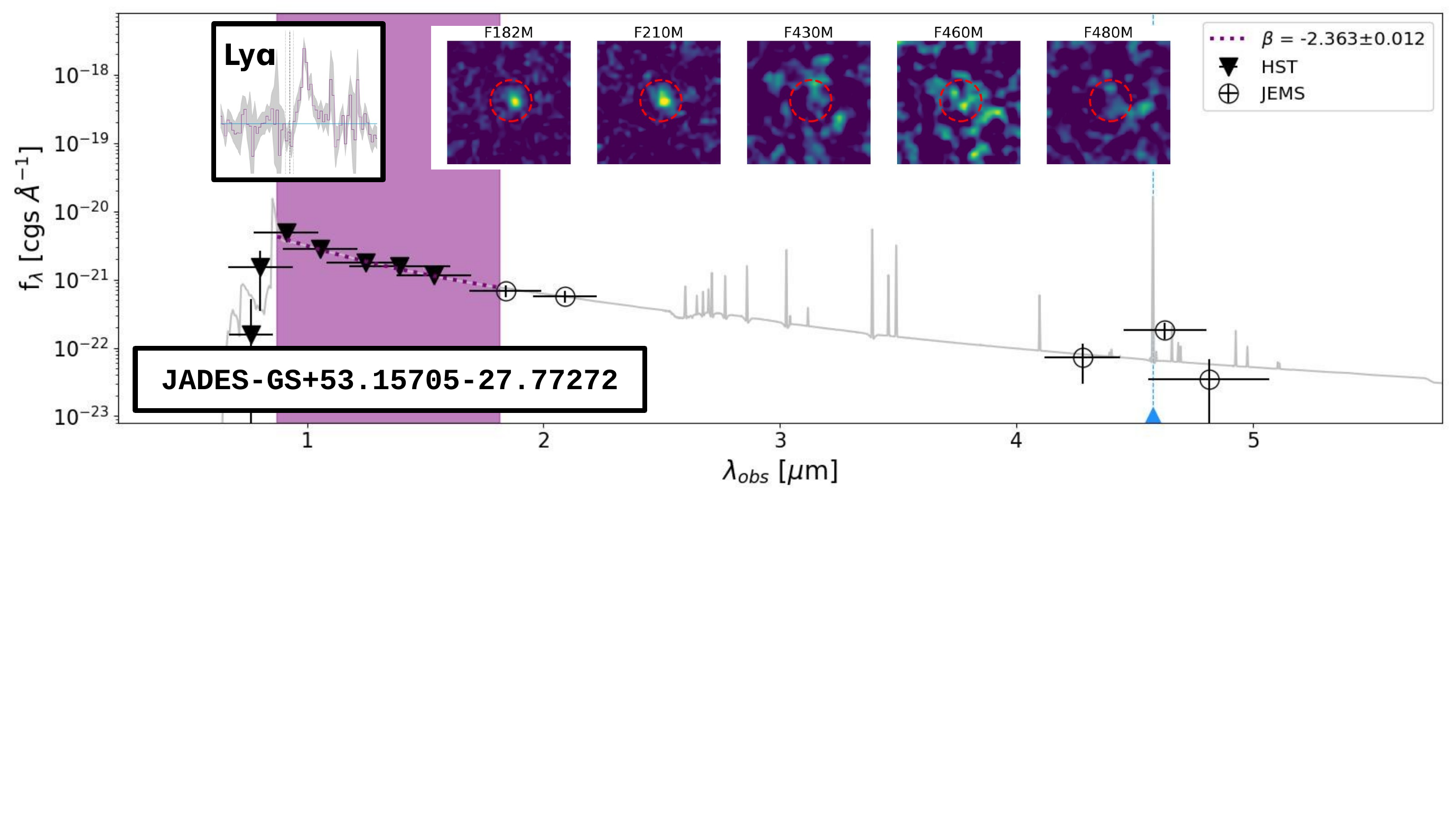}
   \caption{Continuation of Figure~\ref{fig:app}}
    \end{figure*}

    \begin{figure*}
        \centering
   \includegraphics[width=0.7\textwidth,trim={0 6cm 0 0}]{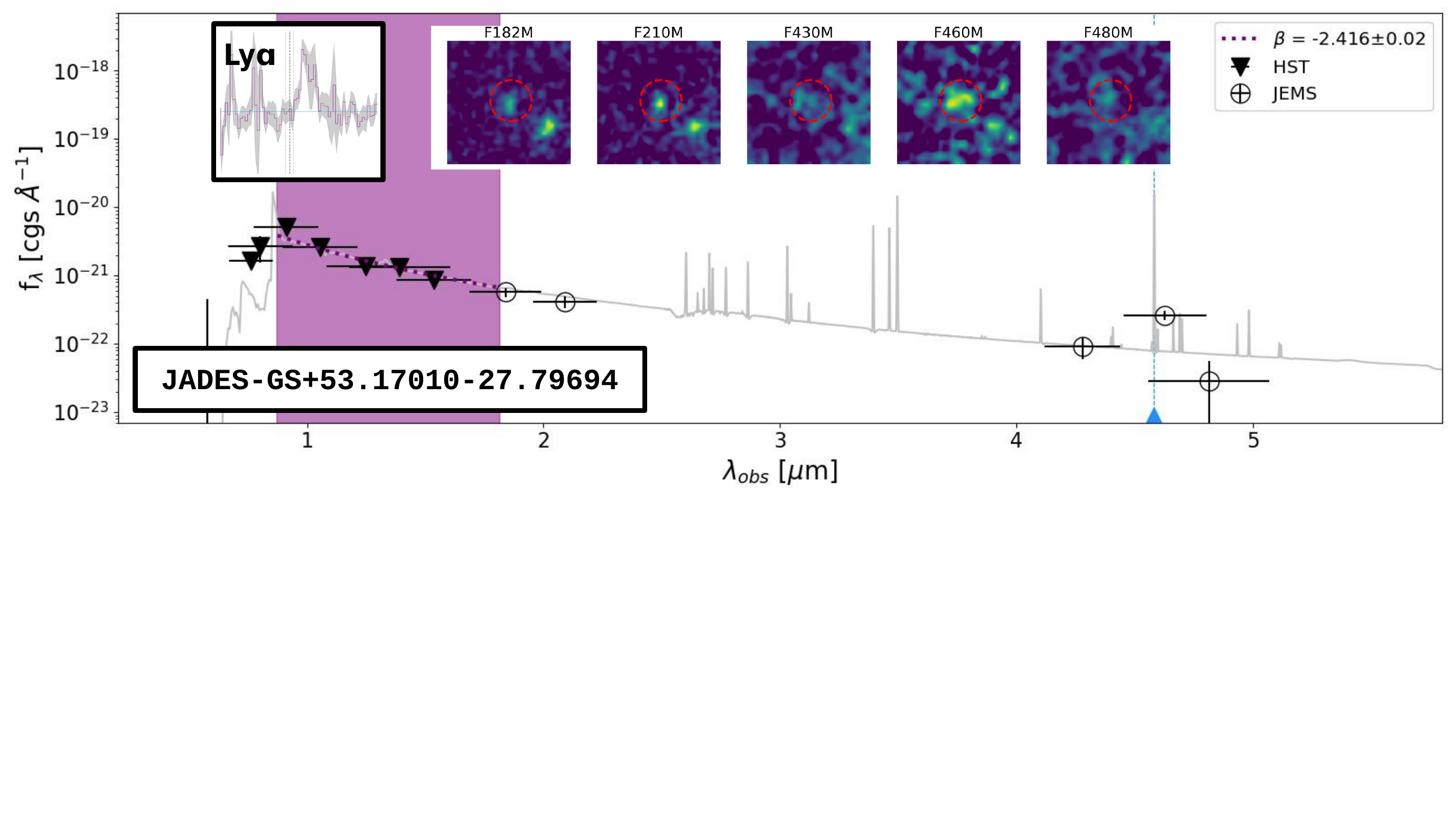}
   \includegraphics[width=0.7\textwidth,trim={0 6cm 0 0}]{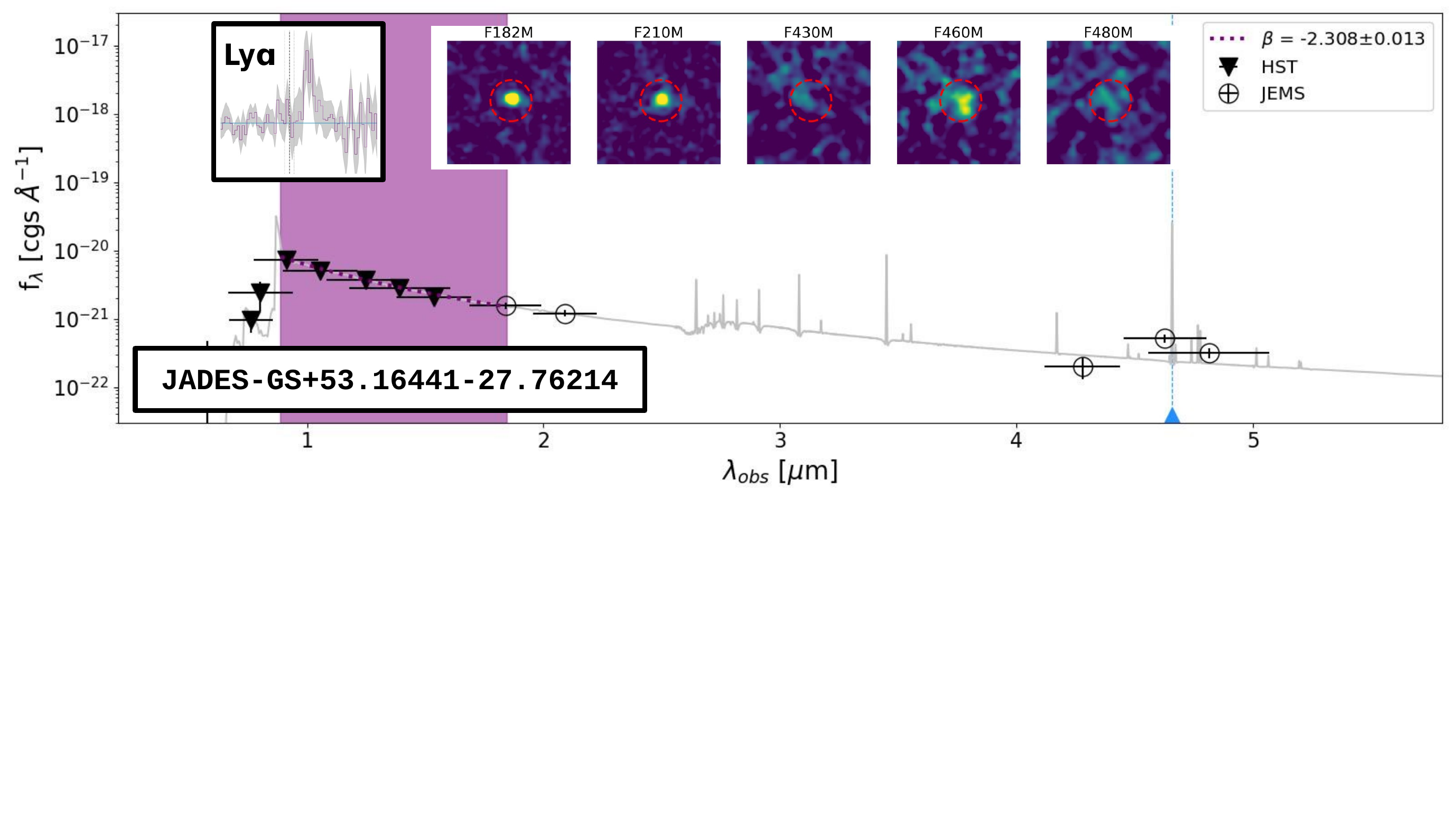}
   \includegraphics[width=0.7\textwidth,trim={0 6cm 0 0}]{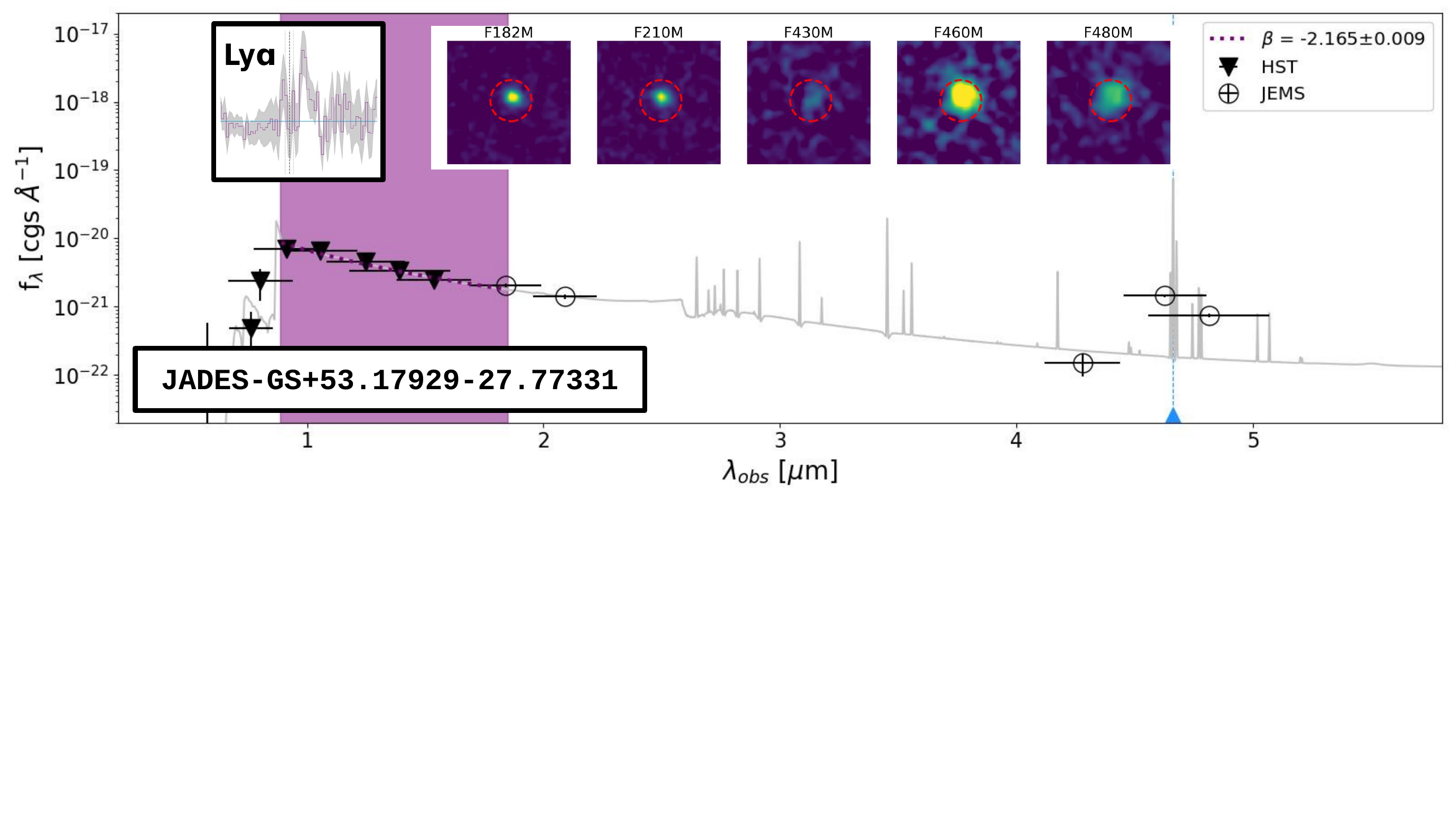}
   \includegraphics[width=0.7\textwidth,trim={0 6cm 0 0}]{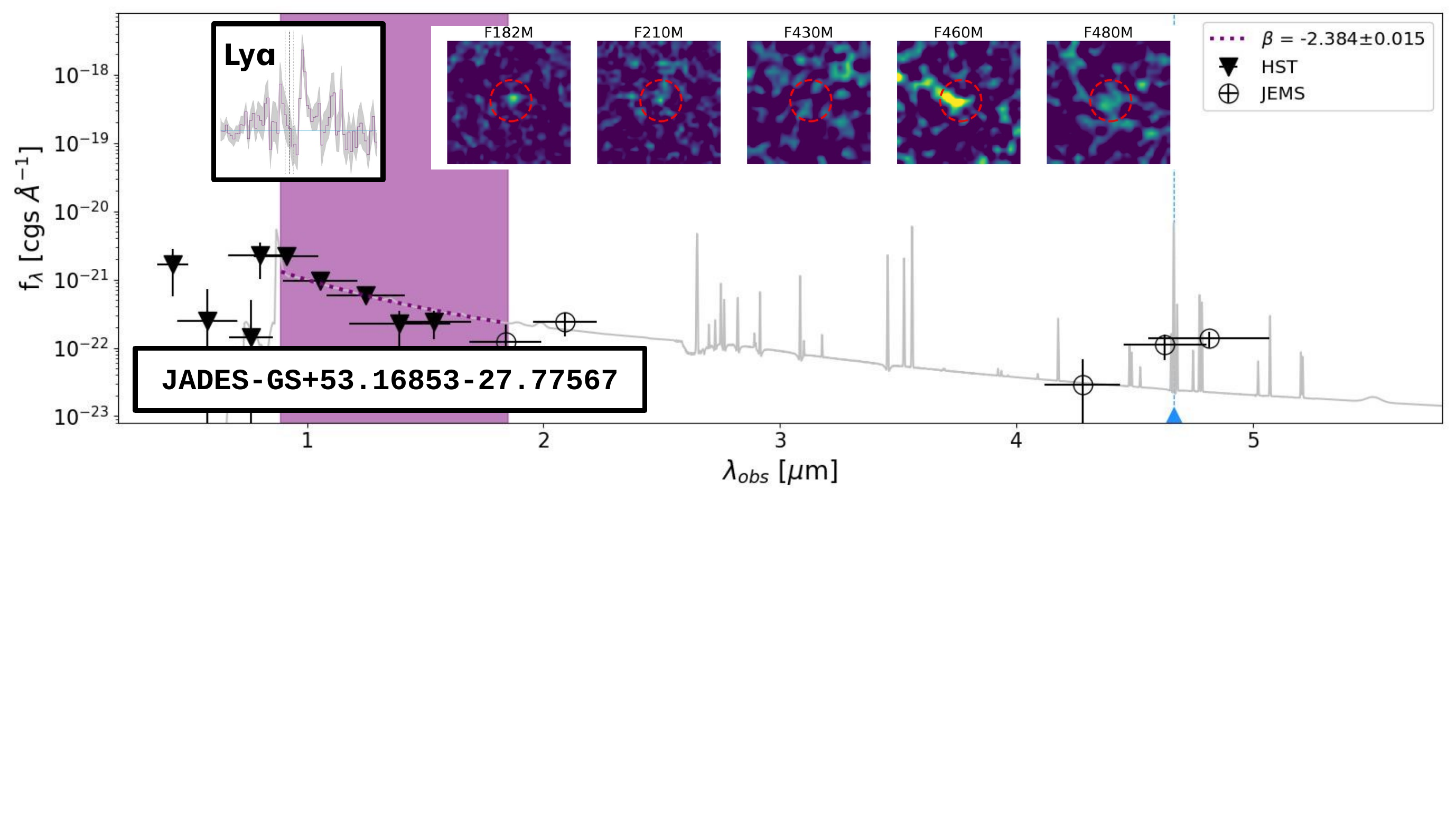}
   \includegraphics[width=0.7\textwidth,trim={0 6cm 0 0}]{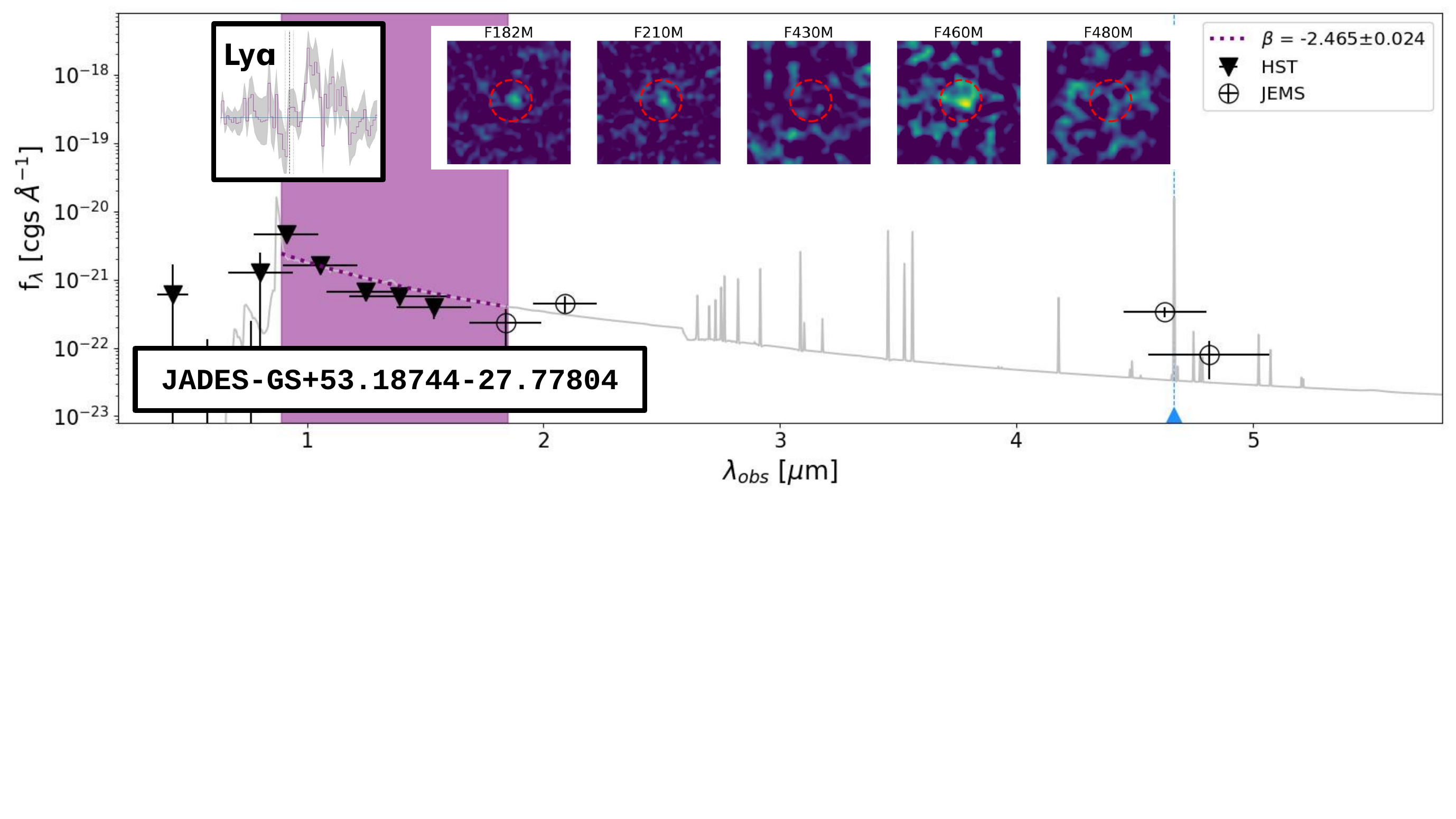}
   \caption{Continuation of Figure~\ref{fig:app}}
    \end{figure*}

    \begin{figure*}
        \centering
   \includegraphics[width=0.7\textwidth,trim={0 6cm 0 0}]{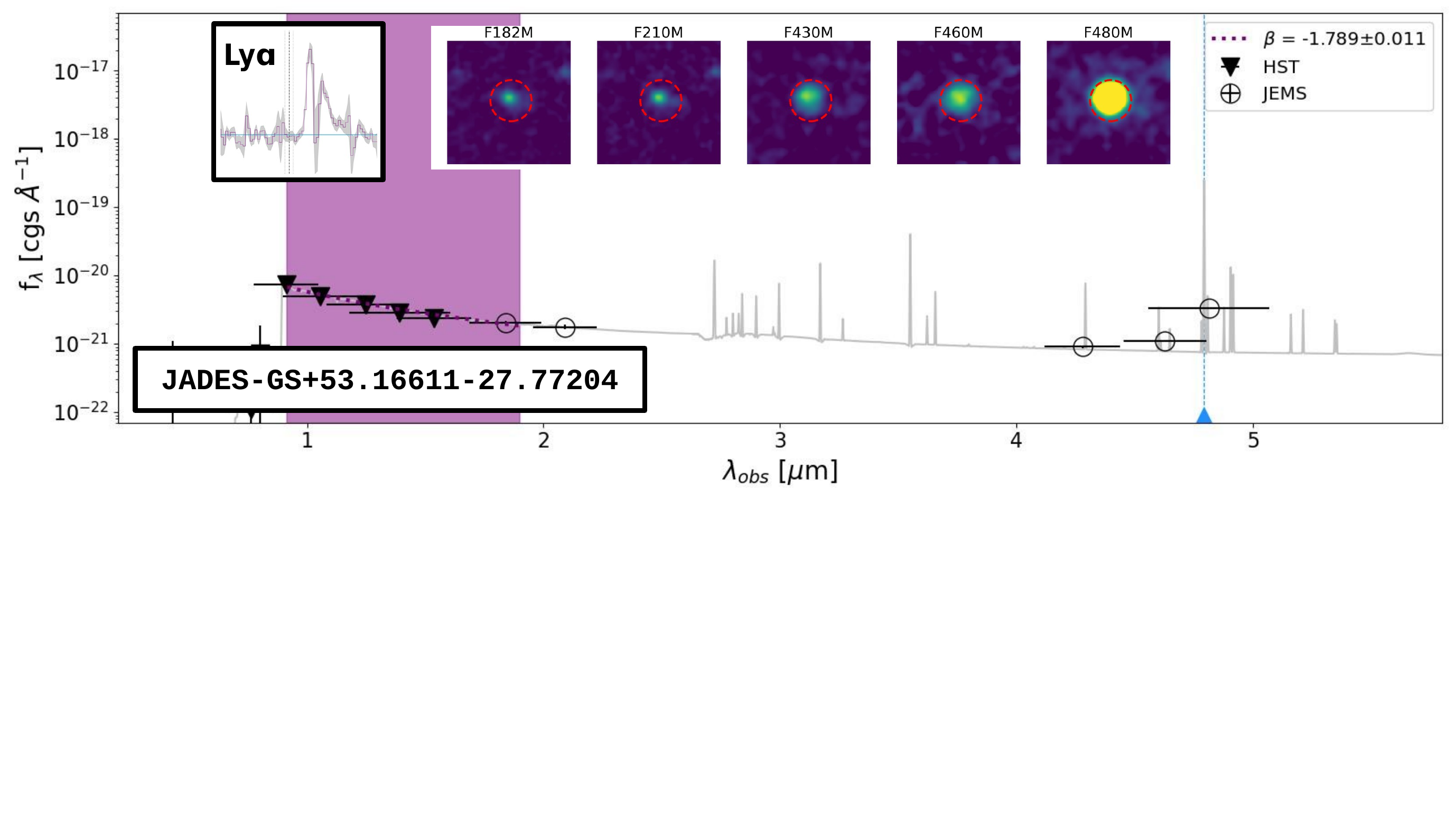}
   \includegraphics[width=0.7\textwidth,trim={0 6cm 0 0}]{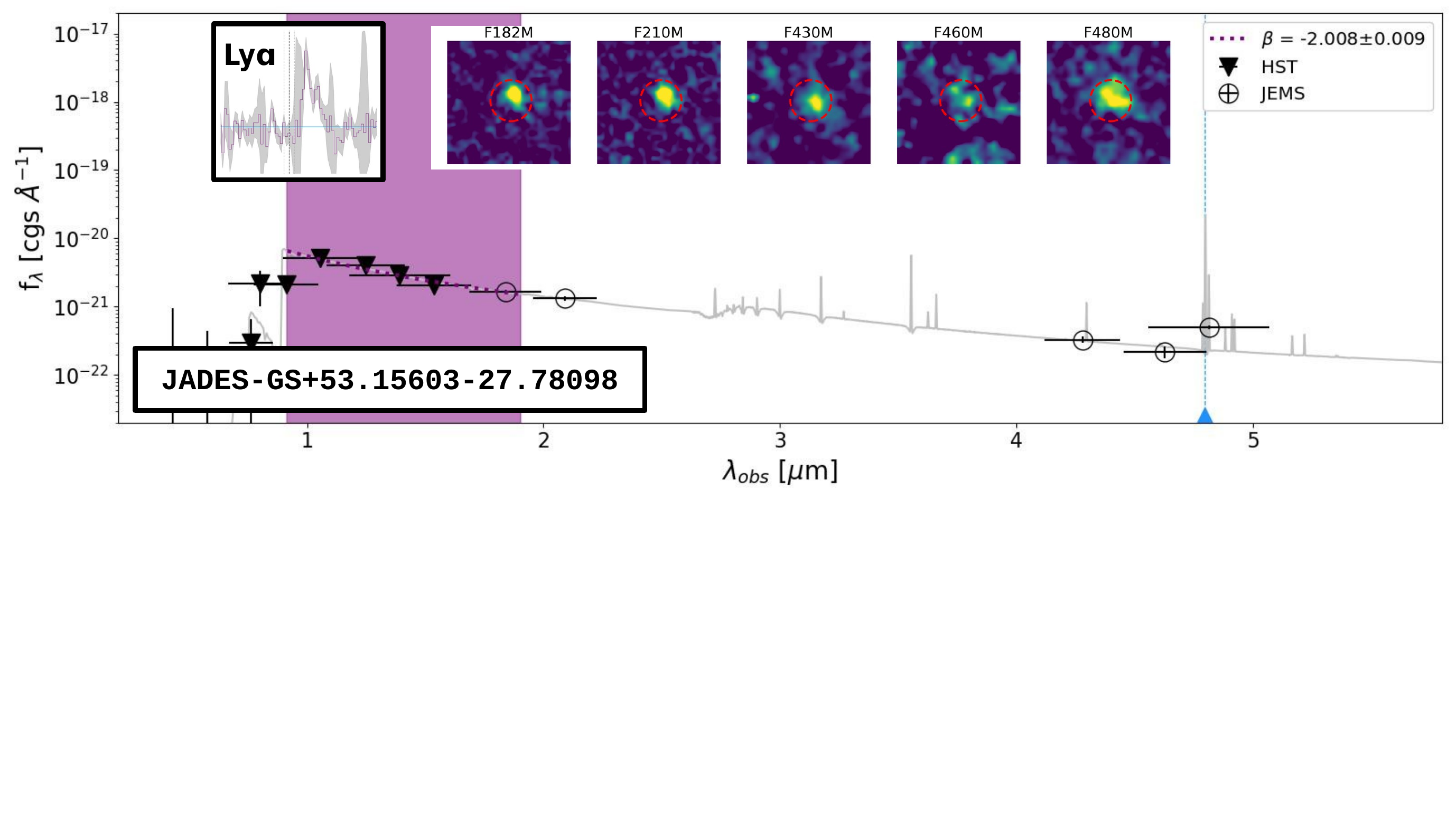}
   \includegraphics[width=0.7\textwidth,trim={0 6cm 0 0}]{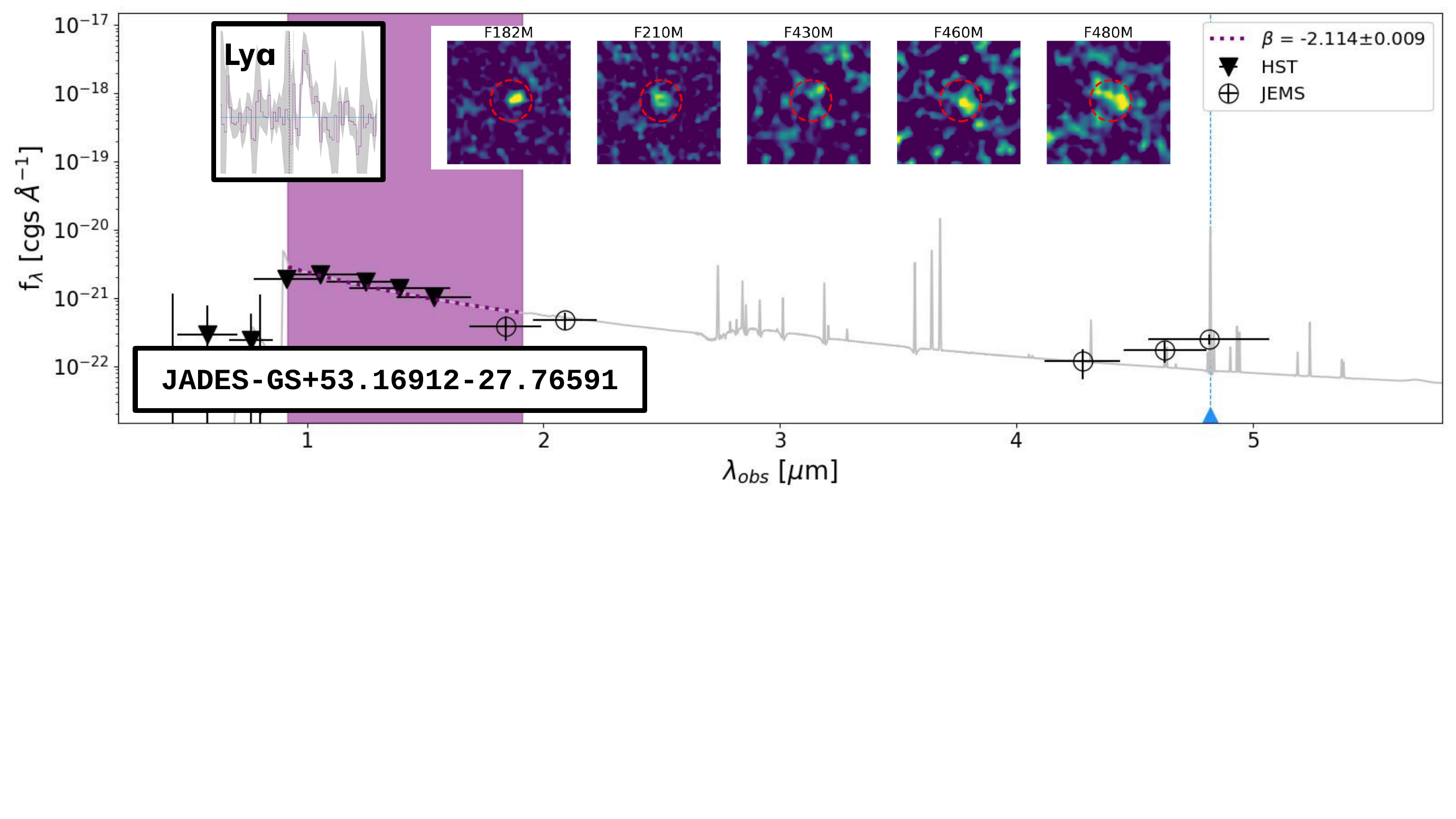}
   \includegraphics[width=0.7\textwidth,trim={0 6cm 0 0}]{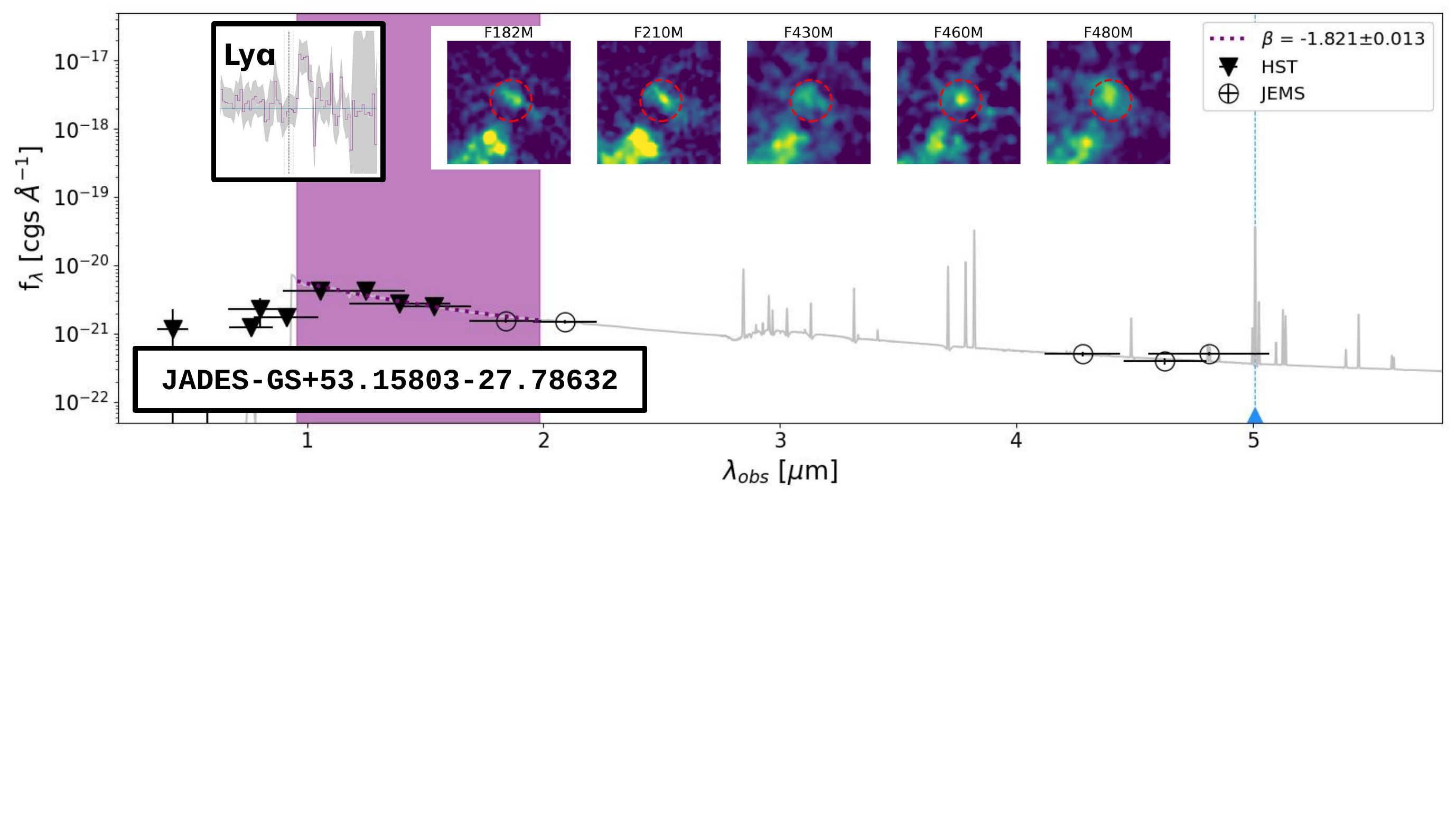}
   \caption{Continuation of Figure~\ref{fig:app}}
    \end{figure*}

\newpage

\section{Prospector derived properties}
\label{appendix:properties}
Here we present the galactic properties derived with \texttt{Prospector}. All errors represent the 25\% and 75\% percentiles of each parameter.

   \begin{table*}
    \selectfont
        \centering
        \begin{tabular}{ccccccc}
        \hline
        \noalign{\smallskip}
        N & log(M$_*$) & SFR$_{10}$ & SFR$_{100}$ & log(Z$_*$) & log<U> & t$_{50}$ \\
         & [M$_{\odot}$]& [M$_{\odot}$ yr$^{-1}$] & [M$_{\odot}$ yr$^{-1}$] & [Z$_{\odot}$] & & [Gyr]\\
        \noalign{\smallskip}
        \hline
        \noalign{\smallskip}
        1 & 9.22 (-0.00,+0.00) & 3.76 (-0.02,+0.02) & 15.80 (-0.26,+0.26) & -2.17 (-0.04,+0.04) & -3.08 (-0.06,+0.06) & 0.06 (-0.00,+0.00) \\
2 & 8.92 (-0.00,+0.24) & 3.00 (-0.26,+0.38) & 1.78 (-0.01,+0.14) & -1.04 (-1.06,+0.01) & -3.41 (-0.01,+0.42) & 0.36 (-0.00,+0.14) \\
3 & 8.42 (-0.04,+0.16) & 1.74 (-0.66,+0.09) & 1.58 (-0.29,+0.15) & -2.83 (-0.15,+0.08) & -2.14 (-1.22,+0.11) & 0.10 (-0.01,+0.17) \\
4 & 8.61 (-0.15,+0.07) & 0.96 (-0.21,+0.16) & 1.08 (-0.17,+0.34) & -2.09 (-0.39,+0.49) & -2.47 (-0.68,+0.93) & 0.28 (-0.09,+0.16) \\
5 & 10.00 (-0.00,+0.00) & 305.98 (-1.95,+1.95) & 87.07 (-1.64,+1.64) & -0.68 (-0.00,+0.00) & -3.09 (-0.00,+0.00) & 0.02 (-0.00,+0.00) \\
6 & 7.03 (-0.40,+0.44) & 0.38 (-0.10,+0.06) & 0.06 (-0.02,+0.04) & -1.95 (-0.91,+0.52) & -3.26 (-0.27,+0.48) & 0.14 (-0.14,+0.35) \\
7 & 9.33 (-0.17,+0.01) & 1.26 (-0.61,+2.36) & 3.19 (-0.99,+8.69) & -2.20 (-0.29,+0.18) & -2.63 (-0.66,+1.07) & 0.22 (-0.09,+0.28) \\
8 & 7.57 (-0.48,+0.38) & 0.45 (-0.13,+0.06) & 0.14 (-0.04,+0.09) & -0.27 (-0.26,+0.16) & -1.83 (-1.67,+0.63) & 0.29 (-0.26,+0.14) \\
9 & 9.66 (-0.06,+0.11) & 12.64 (-6.48,+20.18) & 18.56 (-5.52,+10.94) & -2.21 (-0.20,+0.38) & -2.71 (-1.21,+0.69) & 0.33 (-0.27,+0.07) \\
10 & 7.86 (-0.17,+0.38) & 1.07 (-0.21,+0.22) & 0.48 (-0.18,+0.11) & -1.41 (-0.68,+0.47) & -1.99 (-0.90,+0.60) & 0.16 (-0.11,+0.11) \\
11 & 7.33 (-0.49,+0.44) & 0.71 (-0.25,+0.23) & 0.10 (-0.03,+0.10) & -0.77 (-1.15,+0.43) & -1.65 (-1.68,+0.42) & 0.23 (-0.22,+0.07) \\
12 & 7.63 (-0.32,+0.05) & 0.33 (-0.00,+0.10) & 0.26 (-0.12,+0.02) & -2.58 (-0.33,+0.80) & -2.78 (-0.44,+1.12) & 0.09 (-0.03,+0.11) \\
13 & 8.28 (-0.23,+0.19) & 0.26 (-0.09,+0.22) & 0.48 (-0.18,+0.68) & -2.79 (-0.13,+0.44) & -2.23 (-0.93,+0.71) & 0.25 (-0.14,+0.13) \\
14 & 8.12 (-0.19,+0.19) & 0.53 (-0.13,+0.27) & 0.73 (-0.18,+0.36) & -2.70 (-0.23,+0.24) & -2.29 (-0.88,+0.61) & 0.14 (-0.10,+0.20) \\
15 & 8.44 (-0.29,+0.11) & 0.36 (-0.08,+0.11) & 0.32 (-0.09,+0.20) & -2.33 (-0.34,+1.13) & -1.64 (-1.09,+0.35) & 0.41 (-0.17,+0.09) \\
16 & 7.44 (-0.58,+0.54) & 0.27 (-0.09,+0.31) & 0.08 (-0.05,+0.20) & -2.23 (-0.62,+0.75) & -2.99 (-0.77,+0.93) & 0.30 (-0.28,+0.16) \\
17 & 8.41 (-0.24,+0.21) & 2.53 (-0.68,+0.13) & 0.93 (-0.13,+0.12) & -1.90 (-0.56,+0.72) & -2.67 (-0.84,+0.66) & 0.31 (-0.10,+0.08) \\
18 & 9.84 (-0.06,+0.12) & 27.24 (-18.42,+47.68) & 27.99 (-16.39,+27.38) & -2.58 (-0.12,+0.66) & -2.25 (-1.05,+0.98) & 0.31 (-0.25,+0.04) \\
19 & 8.24 (-0.36,+0.18) & 0.64 (-0.27,+0.58) & 0.49 (-0.07,+0.15) & -1.38 (-0.54,+1.55) & -2.52 (-1.19,+0.47) & 0.24 (-0.12,+0.20) \\
20 & 7.99 (-0.14,+0.15) & 0.32 (-0.07,+0.05) & 0.31 (-0.06,+0.11) & -1.76 (-0.79,+0.27) & -2.00 (-0.61,+0.70) & 0.23 (-0.09,+0.15) \\
21 & 7.99 (-0.24,+0.15) & 0.23 (-0.10,+0.05) & 0.37 (-0.05,+0.07) & -2.52 (-0.25,+1.31) & -2.67 (-0.63,+0.50) & 0.26 (-0.13,+0.07) \\
22 & 7.60 (-0.27,+0.25) & 0.35 (-0.10,+0.04) & 0.21 (-0.04,+0.08) & -2.17 (-0.59,+1.12) & -1.26 (-0.76,+0.20) & 0.16 (-0.13,+0.18) \\
23 & 7.75 (-0.09,+0.72) & 0.87 (-0.36,+0.11) & 0.48 (-0.12,+0.26) & -0.93 (-1.67,+0.22) & -1.39 (-1.26,+0.33) & 0.04 (-0.02,+0.35) \\
24 & 8.74 (-0.45,+0.00) & 2.08 (-0.11,+0.28) & 1.55 (-0.47,+0.03) & -2.86 (-0.01,+1.11) & -1.53 (-1.05,+0.00) & 0.23 (-0.18,+0.02) \\
25 & 7.41 (-0.33,+0.25) & 0.16 (-0.04,+0.05) & 0.07 (-0.02,+0.04) & -2.14 (-0.57,+0.80) & -2.40 (-0.73,+0.99) & 0.34 (-0.22,+0.08) \\
26 & 7.26 (-0.11,+0.16) & 0.34 (-0.06,+0.04) & 0.08 (-0.02,+0.02) & -2.70 (-0.14,+0.41) & -2.17 (-0.57,+0.44) & 0.23 (-0.18,+0.09) \\
27 & 9.41 (-0.66,+0.06) & 39.24 (-24.04,+33.32) & 14.46 (-8.42,+2.88) & -1.82 (-0.03,+1.07) & -3.74 (-0.11,+0.07) & 0.08 (-0.06,+0.20) \\
28 & 9.06 (-0.30,+0.13) & 2.17 (-1.05,+2.07) & 5.11 (-2.35,+4.81) & -2.73 (-0.18,+0.48) & -2.19 (-0.46,+0.58) & 0.13 (-0.04,+0.13) \\
29 & 8.26 (-0.36,+0.19) & 0.39 (-0.14,+0.22) & 0.46 (-0.17,+0.31) & -1.71 (-0.79,+1.19) & -2.63 (-0.79,+0.71) & 0.28 (-0.21,+0.10) \\
30 & 9.04 (-0.23,+0.26) & 1.55 (-0.49,+1.92) & 2.58 (-1.13,+1.80) & -1.85 (-0.66,+0.66) & -2.85 (-0.79,+0.92) & 0.28 (-0.11,+0.08) \\

        \end{tabular}
        \caption{Galactic properties derived with \texttt{Prospector}, rounded to the second decimal.  All errors represent the 25\% and 75\% percentiles of each parameter.\textsl{Column 1:} sequential identifier from this work. \textsl{Column 2:} logarithm of the stellar mass in solar units. \textsl{Column 3:} star formation rate averaged over 10 Myr, in units of solar masses per year. \textsl{Column 4:} star formation rate averaged over 100 Myr, in units of solar masses per year. \textsl{Column 5:} logarithm of metallicity in solar units. \textsl{Column 6:} dimensionless ionisation parameter. \textsl{Column 7:} time of half mass assembly.}  
        \label{tab:prospector}
    \end{table*}

\bsp	
\label{lastpage}
\end{document}